\documentclass[a4paper]{article} 
\usepackage{amssymb,amscd,mathrsfs,latexsym}
\usepackage[all]{xy}
\input amssym.def       
\input amssym.tex

\def\goth{\frak}          
\def\double{\Bbb}

\def\cc{{\double C}}     
\def\nn{{\double N}}       
\def\zz{{\double Z}}

\def\rr{{\double R}}

\newtheorem{theorem}{Theorem}[section]
\newtheorem{lemma}[theorem]{Lemma}
\newtheorem{corollary}[theorem]{Corollary}
\newtheorem{definition}[theorem]{Definition}
\newtheorem{proposition}[theorem]{Proposition}
\newtheorem{remark}[theorem]{Remark}

\newtheorem{example}[theorem]{Example}

\def\limind{\mathop{\mathrm{lim}}\limits_{\longrightarrow}}

\def\si{\sigma}

\def\cinf{C^{\infty}}

\newcommand{\be}{\begin{equation}}
\newcommand{\ee}{\end{equation}}
\newcommand{\beq}{\begin{eqnarray}}
\newcommand{\eeq}{\end{eqnarray}}
\newcommand{\om}{\omega}
\newcommand{\Om}{\Omega}
\newcommand{\al}{\alpha}
\def\nat{\natural}
\def\id{\mbox{\textup{Id}}}

\newcommand{\la}{\lambda}
\newcommand{\Ec}{{\cal E}}
\newcommand{\Vc}{{\cal V}}
\newcommand{\Lc}{{\cal L}}
\newcommand{\non}{\nonumber}
\newcommand{\eps}{\varepsilon}
\newcommand{\Sc}{{\cal S}}
\newcommand{\Wc}{{\cal W}}
\newcommand{\Rc}{{\cal R}}
\newcommand{\Mc}{{\cal M}}

\newcommand{\ch}{\mathop{\mathrm{ch}}}

\newcommand{\Tr}{{\mathop{\mathrm{Tr}}}}
\newcommand{\tr}{{\mathop{\mathrm{tr}}}}
\newcommand{\Ac}{{\cal A}}

\newcommand{\te}{\theta}

\newcommand{\Te}{\Theta}

\def\Id{\mbox{\textup{Id}}}

\def\d{\partial}
\def\dd{\textup{\bf d}}

\def\Hc{{\cal H}}

\def\Bc{{\cal B}}
\def\Cc{{\cal C}}
\def\Jc{{\cal J}}
\def\Kc{{\cal K}}
\def\Fc{{\cal F}}
\def\im{\mbox{\textup{Im}}}
\def\ker{\mbox{\textup{Ker}}}
\def\Bb{\overline{B}}
\def\Bbe{\overline{B}_{\epsilon}}
\def\bb{\overline{b}}
\def\mor{\mathop{\mathrm{Mor}}}
\def\hom{\mathop{\mathrm{Hom}}}
\def\dom{\mathop{\mathrm{Dom}}}

\def\End{\mathop{\mathrm{End}}}
\def\ev{\mathop{\mathrm{ev}}}
\def\hotimes{\hat{\otimes}}

\def\St{\widetilde{S}}
\def\Tt{\widetilde{T}}
\def\Sg{{\goth S}}
\def\Act{\widetilde{\cal A}}
\def\Tct{\widetilde{\cal T}}
\def\Bct{\widetilde{\cal B}}
\def\Cct{\widetilde{\cal C}}
\def\at{\widetilde{a}}
\def\cct{\widetilde{\cc}}
\def\Sge{{\goth S_{\epsilon}}}
\def\Sgan{{\goth S_{\textup{\scriptsize an}}}}
\def\Sgd{{\goth S_{\delta}}}

\def\Ome{\Omega_{\epsilon}}
\def\Omd{\Omega_{\delta}}
\def\Omtd{\widetilde{\Omega}_{\delta}}
\def\Oman{\Omega_{\textup{\scriptsize an}}}
\def\Omtan{\widetilde{\Omega}_{\textup{\scriptsize an}}}

\def\Tc{{\cal T}}
\def\Dc{{\cal D}}

\def\Kc{{\cal K}}

\begin{document}

\begin{center}

{ \Large A BIVARIANT CHERN CHARACTER FOR\\
\vskip 0.2cm
 FAMILIES OF SPECTRAL TRIPLES}
\vskip 1cm
{\bf Denis PERROT}
\vskip 0.5cm
SISSA, via Beirut 2-4, 34014 Trieste, Italy \\[2mm]
{\tt perrot@fm.sissa.it}\\[2mm]
\today 
\end{center}
\vskip 0.5cm
\begin{abstract} 
In this paper we construct a bivariant Chern character defined on ``families of spectral triples''. Such families should be viewed as a version of unbounded Kasparov bimodules adapted to the category of bornological algebras. The Chern character then takes its values in the bivariant entire cyclic cohomology of Meyer. The basic idea is to work within Quillen's algebra cochains formalism, and construct the Chern character from the exponential of the curvature of a superconnection, leading to a heat kernel regularization of traces. The obtained formula is a bivariant generalization of the JLO cocycle. 
\end{abstract}

\vskip 0.5cm

\noindent {\bf Keywords:} Bivariant entire cyclic cohomology, bornological algebras.\\


\section{Introduction}

Recall that according to Connes \cite{C1}, a noncommutative space is described by a spectral triple $(\Ac,\Hc,D)$, where $\Hc$ is a separable Hilbert space, $\Ac$ an associative algebra represented by bounded operators on $\Hc$, and $D$ is a self-adjoint unbounded (Dirac) operator with compact resolvent, such that the commutator $[D,a] $ is densely defined for any $a\in \Ac$ and extends to a bounded operator. The triple $(\Ac,\Hc,D)$ carries a nontrivial homological information as a $K$-homology class of $\Ac$. The major motivation leading Connes to introduce periodic cyclic cohomology \cite{C0} is that the latter is the natural receptacle for a Chern character defined on {\it finitely summable} representatives of $K$-homology. This finiteness condition was removed later and replaced by the weaker condition of $\te$-summability, i.e. the heat kernel $\exp(-tD^2)$ associated to the laplacian of the Dirac operator has to be trace-class for any $t>0$ \cite{C2}. In that case, the algebra $\Ac$ has to be endowed with a norm and the Chern character of the spectral triple is expressed as an {\it infinite-dimensional cocycle} in the entire cyclic cohomology $HE^*(\Ac)$. Except the original construction of Connes, one of the interesting explicit formulas for such a Chern character is provided by the so-called JLO cocycle \cite{JLO}. Here the heat kernel plays the role of a {\it regulator} in the algebra of operators on $\Hc$, and the JLO formula incorporates the data of the spectral triple in a rather simple way. This led Connes and Moscovici to use the powerful machinery of asymptotic expansions of the heat kernel, giving rise to {\it local} expressions extending the classical index theorems of Atiyah-Singer to very interesting non-commutative situations \cite{CM95,CM98}.\\
In this paper we want to generalize the construction of a Chern character to families of spectral triples ``over a noncommutative space'' described by a second associative algebra $\Bc$. In the context of $C^*$-algebras, such objects correspond to the unbounded version of Kasparov's bivariant $K$-theory \cite{Bl}. In this picture, an element of the group $K\!K(\Ac,\Bc)$ is represented by a triple $(\Ec,\rho,D)$, where $\Ec$ is an Hilbert $\Bc$-module. $D$ should be viewed as a family of Dirac operators over $\Bc$, acting by unbounded endomorphisms on $\Ec$, and $\rho$ is a representation of $\Ac$ as bounded endomorphisms of $\Ec$ commuting with $D$ modulo bounded endomorphisms. In the particular case $\Bc=\cc$, this description just reduces to spectral triples over $\Ac$. The construction of a general bivariant Chern character as a transformation from an algebraic version of $K\!K(\Ac,\Bc)$ (for $\Ac$ and $\Bc$ not necessarily $C^*$-algebras) to a bivariant cyclic cohomology has already been considered by several authors. For example Nistor \cite{Ni1,Ni2} constructed a bivariant Chern character for $p$-summable quasihomomorphisms \cite{Cu1}, with values in the Jones-Kassel bivariant cyclic cohomology groups. Cuntz and Quillen also constructed a bivariant Chern character under some summability assumptions,  with values in their own description of the bivariant periodic cyclic theory \cite{CQ1,CQ2}. On the other hand, Puschnigg constructed a well-behaved cyclic cohomology theory for $C^*$-algebras, namely the {\it local cyclic cohomology} \cite{Pu1,Pu2}. Upon generalization of a previous work of Cuntz \cite{Cu2}, the local cyclic theory appears to be the suitable target for a completely general bivariant Chern character (without summability assumptions) defined of Kasparov's $K$-theory. However, the existence and properties of such constructions are often based on excision in cyclic cohomology and the universal properties of bivariant $K$-theory. By considering unbounded bimodules we will follow a different way, involving heat kernel regularization in the spirit of the JLO cocycle, keeping in mind that we are interested in {\it explicit formulas} for a bivariant Chern character incorporating the data $\rho$ and $D$. Our motivation mainly comes from the potential applications to mathematical physics, especially quantum field theory and string/brane theory, where such objects arise naturally:
\vskip 2mm
\noindent $\bullet$ The heat kernel method admits a functional integral representation. The quantities under investigation then correspond to expectation values of observables corresponding to some quantum-mechanical system. This was first used by Alvarez-Gaum\'e and Witten in their study of mixed-gravitational anomalies \cite{AW}, and led to the asymptotic symbol calculus of Getzler \cite{BGV}.
\vskip 2mm
\noindent $\bullet$ The basic idea of introducing a heat kernel regularization of Chern characters in classical differential geometry is due to Quillen \cite{Q1}. Bismut then succesfully applied this method in his approach of the Atiyah-Singer index theorem for families of elliptic operators on submersions \cite{B}. It is worth mentioning that Bismut also uses a stochastic representation of the heat kernel.
\vskip 2mm
\noindent $\bullet$ The Bismut-Quillen approach is essential for the analytic and topological understanding of anomalies (both chiral and gravitational) in quantum field theory \cite{P1,P2}. A bivariant Chern character designed in an equivariant setting may shed some light on the interplay between BRS cohomology and the recently discovered cyclic cohomology of Hopf algebras \cite{CM98,CM99}.
\vskip 2mm
\noindent $\bullet$ Twisted $K$-theory and $K$-homology recently appeared in the physics literature through the classification of $D$-branes \cite{MS,W}. This also falls into the scope of a bivariant Chern character.\\

First we have to consider the right category of algebras. For our purpose, it turns out that {\it bornological associative algebras} are exactly what we need. These are associative algebras endowed with an additional structure describing the notion of a {\it bounded subset}. Complete bornological algebras provide the general framework for entire cyclic cohomology. This theory has been developed in detail by Meyer in \cite{Me}. The interesting feature of the bivariant entire cyclic cohomology is that it contains infinite-dimensional cocycles and thus can be used as the receptacle of a bivariant Chern character for our families of spectral triples carrying some properties of $\te$-summability. Given two complete bornological algebras $\Ac$ and $\Bc$, we will consider the $\zz_2$-graded semigroup $\Psi_*(\Ac,\Bc)$, $*=0,1$, of {\it unbounded $\Ac$-$\Bc$-bimodules}. The latter is an adaptation of Kasparov's unbounded bimodules to the realm of bornological algebras. In our geometric picture, such a bimodule represents a family of spectral triples over the non-commutative space $\Bc$. Our aim is to construct an explicit formula for a Chern character defined on the subsemigroup of $\te$-summable bimodules,
\be
\ch: \Psi_*^{\te}(\Ac,\Bct)\to HE_*(\Ac,\Bc)\ ,\qquad *=0,1\ , \label{ch}
\ee
carrying suitable properties of additivity, differentiable homotopy invariance and functoriality. Here $HE_*(\Ac,\Bc)$ is the bivariant entire cyclic cohomology of $\Ac$ and $\Bc$, and $\Bct$ is the unitalization of $\Bc$. On the technical side, we will use both the $X$-complex description of cyclic cohomology due to Cuntz-Quillen \cite{CQ1,CQ2}, and the usual $(b,B)$-complex of Connes. The $X$-complex is useful for some conceptual explanations of the abstract properties of cyclic (co)homology. Given a complete bornological algebra $\Ac$, its entire cyclic homology is computed by the supercomplex \cite{Me}
\be
X(\Tc\Ac): \Tc\Ac \rightleftarrows \Om^1\Tc\Ac_{\nat}\ ,
\ee
where $\Tc\Ac$ is the {\it analytic tensor algebra of $\Ac$}, obtained by a certain bornological completion of the tensor algebra over $\Ac$, and $\Om^1\Tc\Ac_{\nat}=\Om^1\Tc\Ac/[\Tc\Ac,\Om^1\Tc\Ac]$ is the commutator quotient space of the universal one-forms over $\Tc\Ac$. This means that the entire cyclic homology of $\Ac$ is completely described through the homological properties of its analytic tensor algebra in dimension 0 and 1. Furthermore, taking the analytic tensor algebra of $\Tc\Ac$ is harmless: indeed $X(\Tc\Ac)$ and $X(\Tc\Tc\Ac)$ are homotopically equivalent complexes. In other words, entire cyclic homology does not distinguish between a complete bornological algebra and its successive nested analytic tensor algebras. This is a particular case of the analytic version \cite{Me} of Goodwillie's theorem \cite{G}. This result is a key point of our bivariant Chern character. The construction of (\ref{ch}) will follow two steps:
\vskip 2mm
\noindent a) Using the Goodwillie theorem, we first construct an invertible bivariant class $[\gamma]\in HE_0(\Ac,\Tc\Ac)$ realizing the equivalence between the entire cyclic homologies of $\Ac$ and $\Tc\Ac$.
\vskip 2mm
\noindent b) We consider a bimodule in $\Psi_*(\Ac,\Bct)$. Then under certain $\te$-summability conditions, we construct an element $[\chi]\in HE_*(\Tc\Ac,\Bc)$ involving the exponential of the curvature of a superconnection, which automatically incorporates the desired heat kernel regularization. This step uses Quillen's theory of algebra cochains as an essential tool \cite{Q2,Q3}. Then the composition $[\gamma]\cdot[\chi]\in HE_*(\Ac,\Bc)$ is the bivariant Chern character (\ref{ch}).\\

The paper is organized as follows. In sections \ref{born}, \ref{ent} we recall the basic definitions and properties of bornological spaces and entire cyclic cohomology. In section \ref{good} we present our construction of the Goodwillie equivalence $[\gamma]\in HE_0(\Ac,\Tc\Ac)$. The semigroup of unbounded bimodules $\Psi_*(\Ac,\Bc)$ is introduced in section \ref{bim}. Sections \ref{super} and \ref{biv} are devoted to the fundamental construction of the element $\chi\in HE_*(\Tc\Ac,\Bc)$. Finally, we end the paper with an application of our Chern character to the non-bivariant cases, namely ordinary $K$-theory and $K$-homology in section \ref{ex}. In particular we check that the composition product on $HE$ describes correctly the index pairing between idempotents and spectral triples. Besides, the study of the Bott class allows to normalize the bivariant Chern character. The appendix contains a straightforward adaptation, to bornological algebras, of Quillen's algebra cochains formalism.\\

We would like to mention a last point. There is a priori no obvious intersection product $\Psi(\Ac,\Bc)\times\Psi(\Bc,\Cc)\to\Psi(\Ac,\Cc)$ as in Kasparov theory. Also we will never ask if our construction is compatible with such a product. In fact, it is possible to show that the bivariant Chern character is compatible with the Kasparov product on $p$-summable quasihomomorphisms, but this involves a retraction of our entire cocycles onto periodic ones. These matters will be treated elsewhere.\\

All algebras are supposed to be based on the ground field $\cc$. We work in the non-unital graded category, i.e. homomorphisms between algebras do not necessarily preserve units, and all operations like commutators, tensor products, etc... involving graded objects are automatically graded.

\section{Bornology}\label{born}

This section is intended to give a short introduction to {\it bornological vector spaces} \cite{HN}. These are vector spaces with an additional structure describing abstractly the notion of boundedness. Concrete examples of bornological spaces are provided by normed or locally convex spaces. Bornology is the correct framework allowing the development of {\it entire} cyclic cohomology in full generality; this has been done by Meyer in \cite{Me}. Since this topic is not so familiar to mathematical physicists, we feel the need to recall the definitions and basic properties. Our sketch is by no means supposed to give a sufficient knowledge about bornology; we refer to \cite{HN,Me} for details.\\

Let $\Vc$ be a vector space over $\cc$. A subset $S\subset \Vc$ is a disk iff it is circled and convex. Given any subset $S$, we denote by $S^{\Diamond}$ its circled convex hull: it is the smallest disk containing $S$. If $S$ is a disk, its linear span $\Vc_S$ is endowed with a semi-norm $||\cdot ||_S$ whose unit ball is the closure of $S$. $S$ is called completant iff $\Vc_S$ is a Banach space.
\begin{definition}
Let $\Vc$ be a vector space. A (convex) bornology $\Sg(\Vc)$ is a collection of subsets of $\Vc$ verifying the following axioms:
\begin{itemize}
\item $\{x\}\in\Sg(\Vc)$ for any vector $x\in\Vc$.
\item $S_1+S_2\in\Sg(\Vc)$ for any $S_1,S_2\in\Sg(\Vc)$.
\item If $S\in \Sg(\Vc)$, then $T\in\Sg(\Vc)$ for any $T\subset S$.
\item $S^{\Diamond}\in\Sg(\Vc)$ for any $S\in\Sg(\Vc)$.
\end{itemize}
Any $S\in\Sg(\Vc)$ is called a \emph{small subset} of the bornological space $\Vc$.
\end{definition}
The bornology $\Sg(\Vc)$ is called completant iff any small subset $S\in\Sg(\Vc)$ is contained in a completant small disk. In that case, $(\Vc,\Sg(\Vc))$ is a {\it complete bornological vector space}.
\begin{example}\textup{
If $\Vc$ is a locally convex space, then the {\it bounded bornology} $\goth{Bound}(\Vc)$ is the collection of subsets $S$ bounded for all seminorms on $\Vc$. If $\Vc$ is complete for the locally convex topology, then it is complete as a bornological space. Fr\'echet spaces endowed with the bounded bornology are important examples of complete bornological spaces. }
\end{example}
\begin{example}\textup{
If $\Vc$ is any vector space, the fine bornology $\goth{Fine}(\Vc)$ is the smallest admissible bornology: a subset is small iff it is contained in the disked hull of a finite number of points of $\Vc$. In particular, any small subset is contained in a finite-dimensional subspace of $\Vc$. A bornological space with fine bornology is always complete because finite-dimensional spaces are complete.}
\end{example}
\begin{example}\textup{
A useful way to construct a bornology on $\Vc$ is to start from a collection $\goth{U}$ of subsets not satisfying the axioms of a bornology, and then to consider the smallest bornology $\Sg(\Vc)$ containing $\goth{U}$. We say that $\Sg(\Vc)$ is generated by $\goth{U}$. }
\end{example}

\noindent {\bf Bornological convergence:} A sequence $\{x_n\}_{n\in\nn}$ of points in a bornological space $\Vc$ is said to converge bornologically to the limit $x_{\infty}\in\Vc$ iff there is a small disk $S\in\Sg(\Vc)$ such that $x_n-x_{\infty}\in S$ for any $n$ and $\lim_{n\to\infty}||x_n-x_{\infty}||_S=0$. A set is said to be closed for the bornology iff it is sequentially closed for bornologically convergent sequences. The closed sets for the bornology fulfill the axioms of a topology, hence a bornological space has also a topology (though in general not a vector space topology).\\

\noindent {\bf Bounded maps:} Let $\Vc$ and $\Wc$ be two bornological vector spaces. A linear map $l:\Vc\to\Wc$ is {\it bounded} iff $l(S)\in \Sg(\Wc)$ for any small $S\in\Sg(\Vc)$. An arbitrary set $\{l_j\}_{j\in J}$ of linear maps is {\it equibounded} iff $\{l_j(x)|j\in J,x\in S\}$ is a small subset of $\Wc$ for any $S\in\Sg(\Vc)$. We denote by $\hom(\Vc,\Wc)$ the vector space of bounded linear maps between $\Vc$ and $\Wc$. The sets of equibounded maps form a bornology called the {\it equibounded bornology} on $\hom(\Vc,\Wc)$. It is complete if $\Wc$ is complete. We will always endow the spaces of bounded linear maps with the equibounded bornology. \\

\noindent {\bf Completions:} Let $\Vc$ be a bornological vector space. Its {\it bornological completion} $\Vc^c$ is the complete bornological vector space defined as the solution of the following universal problem: there is a bounded linear map $u:\Vc\to\Vc^c$ such that, for any complete bornological space $\Wc$ and any bounded linear map $l:\Vc\to\Wc$, there is a unique bounded linear map from $\Vc^c$ to $\Wc$ factorizing $l$. The completion always exists, and can be explicitly realized as the inductive limit of a system of Banach spaces (see \cite{HN} and the appendix of \cite{Me}). It is of course unique by universality. If $\Vc$ is a normed space endowed with the bounded bornology, then its bornological completion coincides with its Hausdorff completion. However, it should be stressed that the universal map $\Vc\to\Vc^c$ may fail to be injective for an arbitrary bornological space $\Vc$.\\

\noindent {\bf Multilinear maps:} An $n$-linear map $l:\Vc_1\times\ldots\times\Vc_n\to\Wc$ between bornological spaces is bounded iff $l(S_1,\ldots,S_n)\in\Sg(\Wc)$ for any small sets $S_i\in\Sg(\Vc_i)$. If $\Wc$ is complete, then there is a unique bounded $n$-linear map $\Vc_1^c\times\ldots\times\Vc_n^c\to \Wc$ factorizing $l$. \\

\noindent {\bf Completed tensor products:} Let $\Vc_1$ and $\Vc_2$ be two bornological vector spaces. We endow their algebraic tensor product $\Vc_1\otimes\Vc_2$ with the bornology generated by the subsets $S_1\otimes S_2$, for any $S_i\in\Sg(\Vc_i)$. The completion of $\Vc_1\otimes\Vc_2$ with respect to this bornology is the {\it completed tensor product} $\Vc_1\hotimes\Vc_2$. The completed tensor product is associative, whence the definition of the $n$-fold tensor product $\Vc_1\hotimes\ldots\hotimes\Vc_n$ of $n$ bornological spaces. The latter is universal for the bounded $n$-linear maps $\Vc_1\times\ldots\times\Vc_n\to \Wc$ with complete range $\Wc$.\\

\noindent {\bf Algebras:} A bornological algebra is a bornological space $\Ac$ endowed with a bounded bilinear map (product) $\Ac\times\Ac\to \Ac$. The algebra $\Ac$ is complete iff it is complete as a vector space. In this paper we will be concerned only with associative bornological algebras. \\

\noindent {\bf Subspaces, quotients:} Let $\Vc$ be a bornological vector space and $\Wc\subset\Vc$ a vector subspace. There is a canonical bornology on $\Wc$: a subset $S\in\Wc$ is small iff it is small for $\Vc$. On the other hand, the quotient space $\Vc/\Wc$ has also a bornology: $S\in\Sg(\Vc/\Wc)$ iff there is a small $T\in\Sg(\Vc)$ such that $S=T$ mod $\Wc$. When $\Vc$ is complete, then the subspace $\Wc\subset\Vc$ and quotient $\Vc/\Wc$ are complete iff $\Wc$ is bornologically closed in $\Vc$.\\

\noindent {\bf Bornological complexes:} A bornological space $\Vc$ with a bounded linear map $\d:\Vc\to\Vc$ satisfying $\d^2=0$ is a bornological complex. Its homology is as usual the bornological vector space $H_*(\Vc)=\ker \d/\im\d$.

\section{Entire cyclic cohomology}\label{ent}

Here we recall the formulation of cyclic (co)homology within the $X$-complex framework of Cuntz and Quillen \cite{CQ1}. The analytic adaptation of that theory presented by Meyer in \cite{Me} allows to define elegantly the {\it entire} cyclic homology, cohomology and bivariant cohomology for {\it bornological} algebras. There are in fact two equivalent ways to describe the entire cyclic cohomology of a complete bornological algebra $\Ac$. The first one is to use Connes' $(b,B)$ complex of non-commutative forms completed with respect to a certain bornology; we call this completion $\Ome \Ac$. The second one is the $X$-complex of the completed tensor algebra $\Tc\Ac$. These complexes are homotopy equivalent \cite{Me}, and give rise to the definition of entire cyclic cohomology. The construction of the bivariant Chern character proposed in our paper uses simultaneously the $(b,B)$-complex and $X$-complex approaches. Also, a third complex will be needed as an intermediate step; we call it the completed de Rham-Karoubi complex $\Omd\Ac_{\nat}$. \\

\subsection{Non-commutative differential forms}

Let $\Ac$ be a complete bornological algebra. The algebra of non-commutative differential forms over $\Ac$ is the direct sum $\Om\Ac=\bigoplus_{n\ge 0}\Om^n\Ac$ of the $n$-dimensional subspaces $\Om^n\Ac=\Act\hotimes\Ac^{\hotimes n}$ for $n\ge 1$ and $\Om^0\Ac=\Ac$, where $\Act=\Ac\oplus \cc$ is the unitalization of $\Ac$. It is customary to use the differential notation $a_0da_1\ldots da_n$ (resp. $da_1\ldots da_n$) for the string $a_0\otimes a_1\ldots\otimes a_n$ (resp. $1\otimes a_1\ldots\otimes a_n)$. The differential $d: \Om^n\Ac\to\Om^{n+1}\Ac$ is uniquely specified by $d(a_0da_1\ldots da_n)=da_0da_1\ldots da_n$ and $d^2=0$. The multiplication in $\Om\Ac$ is defined as usual and fulfills the Leibniz rule $d(\om_1\om_2)=d\om_1\om_2 +(-)^{|\om_1|}\om_1d\om_2$, where $|\om_1|$ is the degree of $\om_1$. Each $\Om^n\Ac$ is a complete bornological space by construction, and we endow $\Om\Ac$ with the direct sum bornology. This turns $\Om\Ac$ into a complete bornological differential graded (DG) algebra, i.e. the multiplication map and $d$ are bounded. It is the universal complete bornological DG algebra generated by $\Ac$. \\

On $\Om\Ac$ are defined various operators. First of all, the Hochschild boundary $b: \Om^{n+1}\Ac\to \Om^n\Ac$ is $b(\om da)=(-)^n[\om,a]$ for $\om\in\Om^n\Ac$, and $b=0$ on $\Om^0\Ac=\Ac$. One easily shows that $b$ is bounded and $b^2=0$. Then the Karoubi operator $\kappa:\Om^n\Ac\to \Om^n\Ac$ is constructed out of $b$ and $d$:
\be
1-\kappa=db+bd\ .
\ee
Therefore $\kappa$ is bounded and commutes with $b$ and $d$. The last operator is Connes' $B:\Om^n\Ac\to \Om^{n+1}\Ac$,
\be
B=(1+\kappa+\ldots+\kappa^n)d\quad \mbox{on} \ \Om^n\Ac\ ,
\ee
which is bounded and verifies $B^2=0=Bb+bB$ and $B\kappa=\kappa B=B$. \\

We now define three other bornologies on $\Om\Ac$, leading to the notion of {\it entire} cyclic cohomology:
\vskip 2mm
\noindent $\bullet$ The {\bf entire bornology} $\Sge(\Om\Ac)$ is generated by the sets
\be
\bigcup_{n\ge 0} [n/2]!\,\St (dS)^n\quad ,\ S\in\Sg(\Ac)\ ,\label{sge}
\ee 
where $[n/2]=k$ if $n=2k$ or $n=2k+1$, and $\St=S+\cc$. That is, a subset of $\Om\Ac$ is small iff it is contained in the circled convex hull of a set like (\ref{sge}). We write $\Ome\Ac$ for the completion of $\Om\Ac$ with respect to this bornology. $\Ome\Ac$ will be the complex of entire chains.
\vskip 2mm
\noindent $\bullet$  The {\bf analytic bornology} $\Sgan(\Om\Ac)$ is generated by the sets $\bigcup_{n\ge 0} \St(dS)^n$, $S\in\Sg(\Ac)$. The corresponding completion of $\Om \Ac$ is $\Oman\Ac$. It is related to the $X$-complex description of entire cyclic homology (see below).
\vskip 2mm
\noindent $\bullet$ The {\bf de Rham-Karoubi bornology} $\Sgd(\Om\Ac)$ is generated by the collection of sets $\bigcup_{n\ge 0} \frac{1}{[n/2]!}\,\St(dS)^n$,  $S\in\Sg(\Ac)$, with completion $\Omd\Ac$. This will give rise to the de Rham-Karoubi complex.
\vskip 2mm
The multiplication in $\Om\Ac$ is bounded for the three bornologies above, as well as all the operators $d,b,\kappa,B$. Moreover, the $\zz_2$-graduation of $\Om\Ac$ given by even and odd forms is preserved by the completion process, so that $\Ome\Ac, \Oman\Ac$ and $\Omd\Ac$ are $\zz_2$-graded differential algebras, endowed with the operators $b,\kappa,B$ fulfilling the usual relations. In particular, $\Ome\Ac$ is called the $(b,B)$-complex of {\it entire chains}. Note also that the multiplication or division of $n$-forms by $[n/2]!$ obviously provide linear bornological isomorphisms between $\Ome\Ac, \Oman\Ac$ and $\Omd\Ac$.

\subsection{The analytic tensor algebra}

Let $\Ac$ be a complete bornological algebra, $\Om\Ac=\Om^+\Ac\oplus\Om^-\Ac$ the $\zz_2$-graded algebra of differential forms. The even part $\Om^+\Ac$ is a trivialy graded subalgebra. We endow $\Om^+\Ac$ with a new associative product, the {\it Fedosov product} \cite{CQ1}
\be
\om_1\odot\om_2 =\om_1\om_2 -d\om_1d\om_2\ ,\quad \om_{1,2}\in\Om^+\Ac\ .
\ee
Associativity is easy to check. In fact the algebra $(\Om^+\Ac,\odot)$ is isomorphic to the non-unital tensor algebra $T\Ac=\bigoplus_{n\ge 1}\Ac^{\hotimes n}$, under the correspondence
\be
\Om^+\Ac\ni a_0da_1\ldots da_{2n} \longleftrightarrow a_0\otimes\om(a_1,a_2)\otimes\ldots \otimes\om(a_{2n-1},a_{2n}) \in T\Ac\ ,\label{cor}
\ee
where $\om(a_i,a_j):=a_ia_j- a_i\otimes a_j \in \Ac\oplus \Ac^{\hotimes 2}$ is the {\it curvature} of $(a_i,a_j)$. It turns out that the Fedosov product $\odot$ is bounded for the bornology $\Sgan$ restricted to $\Om^+\Ac$ \cite{Me}, and thus extends to the analytic completion $\Oman^+\Ac$. The complete bornological algebra $(\Oman^+\Ac,\odot)$ is also denoted by $\Tc\Ac$ and called the {\it analytic tensor algebra} of $\Ac$ in \cite{Me}.

\subsection{$X$-complex}\label{X}

The $X$-complex first appeared in Quillen's work on algebra cochains \cite{Q2}, and then was used by Cuntz-Quillen in their formulation of cyclic homology \cite{CQ1,CQ2}. Here we recall the $X$-complex construction for bornological algebras, following Meyer \cite{Me}. \\
Let $\Ac$ be a complete bornological algebra. The $X$-complex of $\Ac$ is the $\zz_2$-graded complex
\be
X(\Ac):\quad \Ac\ \xymatrix@1{\ar@<0.5ex>[r]^{\nat d} &  \ar@<0.5ex>[l]^{\bb}}\ \Om^1\Ac_{\nat}\ ,
\ee
where $\Om^1\Ac_{\nat}$ is the completion of the commutator quotient space $\Om^1\Ac/b\Om^2\Ac=\Om^1\Ac/[\Ac,\Om^1\Ac]$ endowed with the quotient bornology. The class of the generic element $(a_0da_1\, \mbox{mod}\, [,])\in \Om^1\Ac_{\nat}$ is usually denoted by $\nat a_0da_1$. The map $\nat d:\Ac\to \Om^1\Ac_{\nat}$ thus sends $a\in\Ac$ to $\nat da$. Also, the Hochschild boundary $b:\Om^1\Ac\to \Ac$ vanishes on the commutators $[\Ac,\Om^1\Ac]$, hence passes to a well-defined map $\bb:\Om^1\Ac_{\nat}\to\Ac$. Explicitly the image of $\nat a_0da_1$ by $\bb$ is the commutator $[a_0,a_1]$. These maps are bounded and satisfy $\nat d\circ \bb=0$ and $\bb\circ\nat d=0$, so that $X(\Ac)$ indeed defines a complete $\zz_2$-graded bornological  complex.\\

We now focus on the $X$-complex of the analytic tensor algebra $\Tc\Ac$. In that case, $\Om^1\Tc\Ac_{\nat}=\Om^1\Tc\Ac/[\Tc\Ac,\Om^1\Tc\Ac]$ is always complete, and as a bornological vector space $X(\Tc\Ac)$ is canonically isomorphic to the analytic completion $\Oman\Ac$. Here we must take care of a notational problem. Since the symbol $d$ is already used for the differential on $\Om\Ac$, we always choose the bold print $\dd$ for the differential on $\Om\Tc\Ac$. Then the correspondence between $X(\Tc\Ac)$ and $\Oman\Ac$ goes as follows \cite{CQ1,Me}: first, one has a $\Tc\Ac$-bimodule isomorphism 
\beq
\Om^1\Tc\Ac &\simeq& \Tct\Ac\hotimes \Ac\hotimes\Tct\Ac\\
x\, \dd a\, y &\leftrightarrow& x\otimes a\otimes y\qquad \mbox{for}\ a\in\Ac\ ,\ x,y\in\Tct\Ac\ , \non
\eeq
where $\Tct\Ac:=\cc\oplus\Tc\Ac$ is the unitalization of $\Tc\Ac$. This implies that the bornological space $\Om^1\Tc\Ac_{\nat}$ is isomorphic to $\Tct\Ac\hotimes\Ac$, which can further be identified with the analytic completion of odd forms $\Oman^-\Ac$, through the correspondence $x\otimes a\leftrightarrow xda$, $\forall a\in \Ac, x\in \Tct\Ac$. Thus collecting the even part $X_0(\Tc\Ac)=\Tc\Ac$ and the odd part $X_1(\Tc\Ac)=\Om^1\Tc\Ac_{\nat}$ together, yields a linear bornological isomorphism $X(\Tc\Ac)\simeq \Oman\Ac$. We still denote by $(\nat \dd,\bb)$ the boundaries induced on $\Oman\Ac$ through this isomorphism; Cuntz and Quillen explicitly computed them in terms of the usual operators on differential forms \cite{CQ1}:
\beq
\bb&=&b-(1+\kappa)d\quad \mbox{on}\ \Om^{2n+1}\Ac\ ,\\
\nat \dd&=&\sum_{i=0}^{2n}\kappa^id -\sum_{i=0}^{n-1}\kappa^{2i}b\quad \mbox{on}\ \Om^{2n}\Ac\ .\non
\eeq

The crucial result \cite{CQ1,Me} is that the complex $(\Oman\Ac,\nat \dd,\bb)=X(\Tc\Ac)$ is homotopy equivalent to the complex of entire chains $\Ome\Ac$ endowed with the differential $(b+B)$. Let us recall briefly the job \cite{CQ1,Me}. The Karoubi operator $\kappa$ verifies the polynomial identity $(\kappa^n-1)(\kappa^{n+1}-1)=0$ on $\Om^n\Ac$, hence $\kappa^2$ also verifies a polynomial identity. It follows that $\kappa^2$ has a discrete spectrum $\si$, and $\Om\Ac$ decomposes into the direct sum of the generalized eigenspaces $V_{\la}$ for any $\la\in\si$. One of the eigenvalues of $\kappa^2$ is $1$, with multiplicity $2$. Let $P$ be the projection of $\Om\Ac$ onto $V_1$, vanishing on the other eigenspaces. Since $P$ and its orthogonal projection $P^{\bot}$ commute with all operators commuting with $\kappa$, the subspaces $P\Om\Ac$ and $P^{\bot}\Om\Ac$ are stable with respect to $d,b$ and $B$. One shows \cite{Me} that $P,P^{\bot}$ are bounded for the bornologies $\Sgan(\Om\Ac)$ and $\Sge(\Om\Ac)$, hence extend to the completions $\Oman\Ac$ and $\Ome\Ac$. Moreover, the subcomplex $(P^{\bot}\Oman\Ac,\nat \dd,\bb)$ is contractible, hence $\Oman\Ac$ retracts on $P\Oman\Ac$ for the differential $(\nat \dd,\bb)$. Also, $(P^{\bot}\Ome\Ac,b+B)$ is contractible and $\Ome\Ac$ retracts on $P\Ome\Ac$ for the differential $(b+B)$. Let $c:\Oman\Ac\to\Ome\Ac$ be the bornological vector space isomorphism
\be
c(a_0da_1\ldots da_n)=(-)^{[n/2]}[n/2]!\, a_0da_1\ldots da_n\quad \forall n\in\nn\ .\label{res}
\ee
Then $c$ maps isomorphically $P\Oman\Ac$ onto $P\Ome\Ac$, and under this correspondence, the boundaries $(\nat \dd,\bb)$ and $b+B$ coincide: $c^{-1}(b+B)c=(\nat \dd,\bb)$ on $P\Oman\Ac$. It follows that the $X$-complex $X(\Tc\Ac)$ is homotopy equivalent to the $(b+B)$-complex of entire chains $\Ome\Ac$. This leads to the definition of entire cyclic (co)homology:
\begin{definition}
Let $\Ac$ be a complete bornological algebra. \\
i) The entire cyclic homology of $\Ac$ is the homology of the $X$-complex of the analytic tensor algebra $\Tc\Ac$:
\be
HE_*(\Ac)=H_*(X(\Tc\Ac))\ ,\qquad *=0,1\ ,
\ee
or equivalently, the $(b+B)$-homology of the $\zz_2$-graded complex of entire chains $\Ome\Ac$.\\
ii) Let $(X(\Tc\Ac))'$ be the $\zz_2$-graded complex of \emph{bounded} maps from $X(\Tc\Ac)$ to $\cc$, with differential the transposed of $(\nat \dd,\bb)$. Then the entire cyclic cohomology of $\Ac$ is the cohomology of this dual complex:
\be
HE^*(\Ac)=H^*((X(\Tc\Ac))')\ ,\qquad *=0,1\ .
\ee
iii) If $\Ac$ and $\Bc$ are complete bornological algebras, then $\hom(X(\Tc\Ac),X(\Tc\Bc))$ denotes the space of \emph{bounded} linear maps from $X(\Tc\Ac)$ to $X(\Tc\Bc)$. It is naturally a complete $\zz_2$-graded bornological complex, the differential of a map $f$ corresponding to the commutator $(\nat \dd,\bb)\circ f-(-)^{|f|}f\circ(\nat \dd,\bb)$. The bivariant entire cyclic cohomology of $\Ac$ and $\Bc$ is then the cohomology of this complex:
\be
HE_*(\Ac,\Bc)=H_*(\hom(X(\Tc\Ac),X(\Tc\Bc)))\ ,\qquad *=0,1\ .
\ee
\end{definition}
In the case $\Ac=\cc$, one shows \cite{Me} that $X(\Tc\cc)$ is homotopically equivalent to $X(\cc): \cc\rightleftarrows 0$, thus the entire cyclic homology of $\cc$ is simply $HE_0(\cc)=\cc$ and $HE_1(\cc)=0$. This implies that for any complete bornological algebra $\Ac$, we get the usual isomorphisms $HE_*(\cc,\Ac)\simeq HE_*(\Ac)$ and $HE_*(\Ac,\cc)\simeq HE^*(\Ac)$. Furthermore, since the composition of bounded maps is bounded, there is a well-defined composition product on bivariant entire cyclic cohomology:
\be
HE_i(\Ac,\Bc)\times HE_j(\Bc,\Cc)\to HE_{i+j +2\zz}(\Ac,\Cc)\ ,\quad i,j=0,1
\ee
for complete bornological algebras $\Ac,\Bc,\Cc$. Any bounded homomorphism $\rho: \Ac\to \Bc$ extends to a bounded homomorphism $\rho_*:\Tc\Ac\to \Tc\Bc$ by setting $\rho_*(a_1\otimes\ldots\otimes a_n)=\rho(a_1)\otimes\ldots\otimes\rho(a_n)$. The boundedness of $\rho_*$ becomes obvious once we rewrite it using the isomorphism $\Tc\Ac\simeq(\Oman^+\Ac,\odot)$, since $\rho_*(a_0da_1\ldots da_{2n})= \rho(a_1)d\rho(a_1)\ldots d\rho(a_{2n})$. The homomorphism $\rho_*$ gives rise to a bounded $X$-complex morphism $X(\rho_*): X(\Tc\Ac)\to X(\Tc\Bc)$:
\beq
x&\mapsto& \rho_*(x)\\
\nat x\dd y &\mapsto& \nat \rho_*(x)\dd\rho_*(y)\qquad \forall x,y\in\Tc\Ac\ .\non
\eeq
We write $\ch(\rho)$ for the class of $X(\rho_*)$ in $HE_0(\Ac,\Bc)$. It is the simplest example of bivariant Chern character. Last, remark that $HE_*(\Ac,\Ac)$ is a $\zz_2$-graded unital ring, the unit corresponding to the Chern character of the identity homomorphism of $\Ac$.

\subsection{The entire de Rham-Karoubi complex}

There is still another complex related to cyclic homology, namely the de Rham-Karoubi complex \cite{Kar}. In our context of bornological algebras, we have to consider its completed version. So let $\Ac$ be a complete bornological algebra. Recall that the de Rham-Karoubi bornology $\Sgd$ on $\Om\Ac$ is generated by the subsets $\bigcup_{n\ge 0} \frac{1}{[n/2]!}\,\St\otimes S^{\otimes n}$, for any small set $S\in\Sg(\Ac)$. The completion of $\Om\Ac$ with respect to this bornology is $\Omd\Ac$. Let $\Omd\Ac_{\nat}$ be the {\it completion} of $\Omd\Ac/[\Omd\Ac,\Omd\Ac]$ with respect to the quotient bornology, and $\nat:\Omd\Ac\to\Omd\Ac_{\nat}$ be the natural bounded map. The composition $\nat d: \Omd\Ac\to \Omd\Ac_{\nat}$ is bounded and vanishes on the commutator subspace $[\Omd\Ac,\Omd\Ac]$, thus it factors through a well-defined bounded map $\nat d:\Omd\Ac_{\nat}\to \Omd\Ac_{\nat}$. One obviously has $(\nat d)^2=0$, whence a bornological complex. 
\begin{definition}
Let $\Ac$ be a complete bornological algebra. The \emph{entire de Rham-Karoubi} cohomology of $\Ac$ is the cohomology of the $\zz_2$-graded complex $(\Omd\Ac_{\nat},\nat d)$:
\be
H^*_{\mathrm{dR}}(\Ac)=H^*(\Omd\Ac_{\nat},\nat d)\ , \quad *=0,1\ .
\ee
\end{definition}
There is a direct relation between the entire cyclic homology and the entire de Rham-Karoubi cohomology. Let $c':\Om\Ac\to\Om\Ac$ be the linear isomorphism sending the $n$-form $a_0da_1\ldots da_n$ to $\frac{1}{n!}a_0da_1\ldots da_n$. It is also a {\it bornological isomorphism} between $(\Om\Ac,\Sge)$ and $(\Om\Ac,\Sgd)$, and thus extends to an isomorphism of complete bornological spaces $c':\Ome\Ac\to\Omd\Ac$. It is easy to show that the composition $\nat c':\Ome\Ac\to \Omd\Ac_{\nat}$ is a bounded morphism from the $(b+B)$-complex of entire chains to the de Rham-Karoubi complex, whence a natural (covariant) map
\be
HE_*(\Ac)\to H^*_{\mathrm{dR}}(\Ac)\ ,\quad *=0,1\ .
\ee
The entire de Rham-Karoubi complex arises in differential geometry, when characteristic classes of vector bundles are constructed from connections and curvatures \cite{Kar}. If we let $A\in \Om^1\Ac$ be a one-form with curvature $F=dA+A^2\in\Om^2\Ac$, then the {\it Chern character} form (here we omit some irrelevant $2\pi i$ factors)
\be
\ch(A)=\nat\exp F \ \in \Omd\Ac_{\nat}
\ee 
indeed defines an entire de Rham cocycle whose class lies in $H^0_{\mathrm{dR}}(\Ac)$. The use of exponentials will be crucial in our bivariant Chern character construction, because it allows to combine heat kernel regularization with characteristic classes \cite{Q1,B}. Also the entire de Rham-Karoubi complex will be an important intermediate step.

\section{A Goodwillie theorem}\label{good}

One of the main properties of cyclic homology is the so-called Goodwillie theorem \cite{G}. Roughly, it asserts that periodic cyclic homology is stable when taking nilpotent extensions. In other words, if $0\to N\to E\to A\to 0$ is an extension of an algebra $A$, with $N$ is a nilpotent ideal in $E$, then $A$ and $E$ have the same periodic cyclic (co)homology. This has been generalized by Meyer in the context of bornological algebras and entire cyclic homology \cite{Me}, where algebraic nilpotence has to be replaced by the notion of {\it analytic} nilpotence. However, we don't need the whole theory of analytically nilpotent extensions here. Given a complete bornological algebra $\Ac$, we will only be concerned with the {\it universal analytically nilpotent} extension
\be
0\to \Jc\Ac\to \Tc\Ac\to \Ac\to 0\ .\label{exa}
\ee
Here the bounded projection $\Tc\Ac\to \Ac$ is induced by the multiplication map $m: a_1\otimes\ldots \otimes a_n\mapsto a_1\ldots a_n$ , and $\Jc\Ac$ is its kernel. Using the identification $\Tc\Ac\simeq (\Oman^+\Ac,\odot)$, the multiplication map simply coincides with the projection of an even differential form onto its degree zero component. The canonical linear embedding $\si_{\Ac}:\Ac\hookrightarrow\Tc\Ac$ provides a bounded linear splitting of the exact sequence (\ref{exa}). The Goodwillie theorem then claims that the projection homomorphism $m:\Tc\Ac\to\Ac$ induces an {\it isomorphism} between $HE_*(\Tc\Ac)$ and $HE_*(\Ac)$. Moreover, the Chern character of $m$, corresponding to the class of the chain map $X(m_*):X(\Tc\Tc\Ac)\to X(\Tc\Ac)$ in the entire bivariant cyclic cohomology $HE_0(\Tc\Ac,\Ac)$, has an inverse in $HE_0(\Ac,\Tc\Ac)$. The latter is constructed as follows \cite{Me}. There is a unique bounded homomorphism $v_{\Ac}:\Tc\Ac\to\Tc\Tc\Ac$ such that $v_{\Ac}\circ\si_{\Ac}=\si_{\Tc\Ac}\circ\si_{\Ac}$, and it gives rise to a chain map $X(v_{\Ac}):X(\Tc\Ac)\to X(\Tc\Tc\Ac)$, whose cohomology class is the inverse of $\ch(m)$, i.e. $\ch(m)\cdot [X(v_{\Ac})]=1$ in $HE_0(\Tc\Ac,\Tc\Ac)$ and $[X(v_{\Ac})]\cdot \ch(m)=1$ in $HE_0(\Ac,\Ac)$.\\
In this section we shall present a slightly different construction of the inverse of $\ch(m)$. It will be represented by a bounded chain map of degree zero
\be
\gamma: X(\Tc\Ac)\to \Ome\Tc\Ac
\ee
from the $X$-complex of $\Tc\Ac$ to the $(b+B)$-complex of entire chains over $\Tc\Ac$. Since $(\Ome\Tc\Ac,b+B)$ is homotopy equivalent to $X(\Tc\Tc\Ac)$, the map $\gamma$ indeed defines a bivariant class $[\gamma]\in HE_0(\Ac,\Tc\Ac)$. The explicit expression of $\gamma$ will be an important part of the bivarant Chern character. Our aim is to prove corollaries \ref{cgood} and \ref{ccom} below.\\

Before proceeding, we need some information about the bornology of $\Ome\Tc\Ac$. The latter is the completion of $\Om\Tc\Ac=\bigoplus_{n\ge 0}\Om^n\Tc\Ac$ for the bornology $\Sge(\Om\Tc\Ac)$ generated by the sets $\bigcup_{n\ge 0}[n/2]!\, \Tt(\dd T)^n$ for any small $T\in\Sg(\Tc\Ac)$, where $\Tc\Ac$ is already the completion of the algebra $T\Ac\simeq (\Om^+\Ac,\odot)$ for the bornology $\Sgan(\Om^+\Ac)$ generated by $\bigcup_{n\ge 0} \St(dS)^n$, $S\in\Sg(\Ac)$. We could also obtain $\Ome\Tc\Ac$ after only one completion of a certain bornological space. Let $\Om T\Ac:=\bigoplus_{n\ge 0}\Om^nT\Ac$ be the DG algebra of non-commutative differential forms over the {\it non-complete} tensor algebra $T\Ac$, where $\Om^nT\Ac=\Tt\Ac\otimes (T\Ac)^{\otimes n}$ involves only {\it algebraic} (non-completed) tensor products. Endow $\Om T\Ac$ with the bornology $\Sg(\Om T\Ac)$ generated by the sets $\bigcup_{n\ge 0}[n/2]!\, \Tt(\dd T)^n$, for any small $T\in\Sg(T\Ac)=\Sgan(\Om^+\Ac)$.
\begin{lemma}
The completion of the bornological space $(\Om T\Ac,\Sg)$ is canonically isomorphic to $\Ome\Tc\Ac$.
\end{lemma}
{\it Proof:} It is a direct consequence of the universal property of bornological completions. Let us give some details. The natural map $(\Om T\Ac,\Sg)\to(\Om\Tc\Ac,\Sge)$ extending the arrow $T\Ac\to \Tc\Ac$ is clearly bounded. Composing with the universal map $(\Om\Tc\Ac,\Sge)\to\Ome\Tc\Ac$, we get a bounded map $(\Om T\Ac,\Sg)\to \Ome\Tc\Ac$. We thus have to show that $\Ome\Tc\Ac$ is a solution of the universal problem associated to the bornological space $(\Om T\Ac,\Sg)$. Let $\Wc$ be any complete bornological space, and consider a bounded map $f: \Om T\Ac\to \Wc$. Then the universal property of $\Tc\Ac$ implies that there is a unique bounded map $f':\Om\Tc\Ac\to \Wc$ factorizing $f$. From $f'$ we then get a bounded map $f'':\Ome\Tc\Ac\to \Wc$, also factorizing $f$. Moreover, the universal properties of completions imply that such a $f''$ is necessarily unique, hence the complete space $\Ome\Tc\Ac$ identifies canonically to the bornological completion of $(\Om T\Ac,\Sg)$.\hfill \rule{1ex}{1ex}\\

The space of non-commutative forms $\Om T\Ac$ endowed with the usual boundaries $(b,B)$ is a (non-complete) bornological bicomplex. Consider the following subcomplex:
\be
\Theta=b\Om^2T\Ac\oplus \bigoplus_{n\ge 2}\Om^nT\Ac\ ,
\ee
which is stable by $b$ and $B$. The quotient $\Om T\Ac/\Theta$ corresponds to the $\zz_2$-graded bornological complex
\be
X(T\Ac):\quad T\Ac\ \xymatrix@1{\ar@<0.5ex>[r]^{\nat \textup{\scriptsize \bf d}} &  \ar@<0.5ex>[l]^{\bb}}\ \frac{\Om^1T\Ac}{b\Om^2T\Ac}:=\Om^1T\Ac_{\nat}\ , 
\ee
whose completion identifies with $X(\Tc\Ac)$. The projection $\pi: \Om T\Ac\to \Om T\Ac/\Theta$ being bounded, it extends to a bounded chain map
\be 
\pi: \Ome\Tc\Ac\to X(\Tc\Ac)\ ,
\ee
representing a bivariant entire cohomology class $[\pi]\in HE_0(\Tc\Ac,\Ac)$. It turns out that $[\pi]$ has an inverse $[\gamma]\in HE_0(\Ac,\Tc\Ac)$, that we are going to construct as a bounded chain map
\be
\gamma: X(\Tc\Ac)\to \Ome\Tc\Ac\ .
\ee

According to the terminology of Cuntz and Quillen \cite{CQ0,CQ1}, the non-completed tensor algebra $T\Ac$ is algebraically {\it quasi-free}. This means that there is a {\it right connection} $\nabla: \Om^1T\Ac\to\Om^2T\Ac$, i.e. a linear map satisfying
\be
\nabla(x\om)=x\nabla\om\ ,\quad \nabla(\om x)=\nabla\om x+ \om\dd x\ ,\quad \forall x\in T\Ac\ ,\ \om\in\Om^1T\Ac\ .
\ee
Equivalently, there is a linear map $\phi: T\Ac\to \Om^2T\Ac$ such that $\phi(xy)=\phi(x)y+x\phi(y)+\dd x\dd y$ for any $x,y\in T\Ac$. One obtains $\phi$ from $\nabla$ by setting $\phi(x):=\nabla(\dd x)$. To show that such maps exist, one can use the fact that $T\Ac$ is a free algebra, and construct $\phi$ recursively:
\be
\phi(a):=0\quad \forall a\in\Ac\ ;\quad \phi(a\otimes x)=a\phi(x)+\dd a\dd x\quad \forall a\in\Ac\ ,\ x\in T\Ac\ ,\label{def}
\ee
and then $\nabla (x\dd y)=x\phi(y)$, $\forall x,y\in T\Ac$. In the remainder of the text we will always use these definitions of $\nabla$ and $\phi$. Note that any other admissible map $\phi'$ may be obtained from $\phi$ by adding a derivation from $T\Ac$ to the $T\Ac$-bimodule $\Om^2T\Ac$.\\
Now extend $\nabla$ to a map $\Om^nT\Ac\to \Om^{n+1}T\Ac$ for any $n\ge 1$ by the recursive relation
\be
\nabla(\om\dd x)=(\nabla\om) \dd x\ ,\quad \forall \om\in \Om^nT\Ac\ ,\ x\in T\Ac\ ,
\ee
then put $\phi=\nabla B:\Om^nT\Ac\to\Om^{n+2}T\Ac$ for any $n\ge 0$. The latter extends the previous map $\phi$ to $\Om T\Ac$. Explicitly one has
\beq
\phi(x_0\dd x_1\ldots\dd x_n)&=& \nabla(1+\kappa+\ldots +\kappa^n)(\dd x_0\dd x_1\ldots\dd x_n) \non\\
&=& \sum_{i=0}^n (-)^{ni}\phi(x_i)\dd x_{i+1}\ldots\dd x_{i-1}\ ,
\eeq
for any $x_j\in T\Ac$. One can compute the successive powers of $\phi$. It turns out that given an element $x$ of $T\Ac$, $\phi^k(x)$ vanishes for $k$ sufficiently large (depending on $x$). Indeed, let $x=a_1\otimes a_2\ldots \otimes a_n$. From the definition (\ref{def}), one sees that for $k=[n/2]$, $\phi^k(x)$ contains only terms of the form $\dd a_i\ldots \dd a_n\dd a_1\ldots\dd a_{i-1}$ or $a_i\dd a_{i+1}\ldots \dd a_{i-1}$, and $\phi(a_i)=0$ implies $\phi^{k+1}(x)=0$. More generally, $\phi^k(\om)=0$ for any $\om\in\Om T\Ac$ and $k\gg 0$. Consequently, the operator $1-\phi$ is invertible on $\Om T\Ac$, because the power series
\be
(1-\phi)^{-1}=\sum_{k=0}^{\infty}\phi^k
\ee
only takes a finite number of non-zero terms on any $\om\in\Om T\Ac$.
\begin{proposition}\label{pequi}
i) For any $n\ge 2$, $\nabla b+b\nabla=-\id$ on $\Om^nT\Ac$. Hence $\nabla$ is a contracting homotopy for the $b$-cohomology of the subcomplex $\Theta=b\Om^2T\Ac\oplus \bigoplus_{n\ge 2}\Om^nT\Ac$ of $\Om T\Ac$.\\
ii) The map $\gamma: X(T\Ac)\to \Om T\Ac$ defined by
\beq
T\Ac \ni x &\mapsto & (1-\phi)^{-1}(x) \\
\Om^1T\Ac_{\nat}\ni \nat x\dd y&\mapsto& (1-\phi)^{-1}(x\dd y+b(x\phi(y)))\non
\eeq
is a morphism from the $X$-complex of $T\Ac$ to the $(b+B)$-complex $\Om T\Ac$.\\
iii) Let $\pi:\Om T\Ac\to X(T\Ac)$ be the natural projection. There is a contracting homotopy $h: \Om T\Ac\to \Om T\Ac$ such that 
\beq
\pi\circ \gamma& =& \id \quad\mbox{on} \quad X(T\Ac)\ ,\\
\gamma\circ\pi &=& \id+ (b+B)h+ h(b+B)\quad\mbox{on} \quad \Om T\Ac\ .
\eeq
\end{proposition}
{\it Proof:} i) Let us first show that for any $\om\in\Om^{n}T\Ac$, $n\ge 1$, and $x\in T\Ac$ one has
$$
\nabla(x\om)=x\nabla\om\ ,\qquad \nabla(\om x)=\nabla\om x-(-)^{n}\om \dd x\ .
$$
The first relation is trivial. The second one is proved recursively on $n$. Suppose the relation is realized for $n$, then for any $\om\in\Om^nT\Ac$ and $x,y\in T\Ac$
\beq
&&\nabla(\om \dd x\, y)=\nabla(\om \dd(xy))-\nabla(\om x\dd y)=\nabla\om\, \dd(xy)-\nabla(\om x)\dd y\non\\
&&=\nabla\om\, \dd(xy)-\nabla\om\, x\dd y+(-)^{n}\om \dd x\dd y= \nabla(\om \dd x)y+(-)^{n}\om \dd x\dd y\ ,\non
\eeq
proving the relation for $n+1$. Next, any element of $\Om^nT\Ac$, for $n\ge 2$, is a sum of elements like $\om \dd x$ with $\om \in\Om^{n-1}T\Ac$, $x\in T\Ac$. Thus
\beq
(\nabla b+b\nabla)(\om \dd x)&=&(-)^{n-1}\nabla [\om,x]+b(\nabla\om\, \dd x)\non\\
&=&(-)^{n-1}(\nabla\om\, x+(-)^{n}\om \dd x)-(-)^{n-1}x\nabla\om+(-)^{n} [\nabla\om,x]\non\\
&=& -\om \dd x\ ,\non
\eeq
which concludes the proof.\\
ii) First, the injection $x\dd y\mapsto x\dd y+b(x\phi(y))$ vanishes on $b\Om^2T\Ac=[T\Ac,\Om^1T\Ac]$ (see \cite{CQ1} \S7), hence is well-defined on $\Om^1T\Ac_{\nat}$. Therefore $\gamma$ is well-defined. Next, since $\phi=\nabla B$ one has $\phi B=0$. Also $b\nabla+\nabla b=-\id$ on $\Om^{n\ge 2}T\Ac$, thus the relation $b\phi -\phi b=b\nabla B-\nabla Bb=-(1+\nabla b)B-\nabla Bb=-B$ holds on $\Om^{n\ge 1}T\Ac$. This implies $(1-\phi)(b+B)=b+B-\phi b=b(1-\phi)$, and composing from right and left by $(1-\phi)^{-1}$ (which preserves the subspace $\bigoplus_{n\ge 1}\Om^nT\Ac$) yields
\be
(b+B)(1-\phi)^{-1}=(1-\phi)^{-1}b\quad \mbox{on}\ \Om^{n\ge 1}T\Ac\ .\label{r}
\ee
Let $x\in T\Ac$. Using (\ref{r}) and $\phi B=0$, one has
\beq
\gamma(\nat \dd x) &=&(1-\phi)^{-1}(\dd x+b\phi(x))\non\\
&=& (\sum_{n=0}^{\infty} \phi^n)B(x) +(1-\phi)^{-1}b\phi(x)\non\\
&=& B(x) +(b+B)(1-\phi)^{-1}\phi(x)\ ,\non
\eeq
and since $b(x)=0$ one deduces $\gamma(\nat \dd x) =(b+B)(1-\phi)^{-1}(x)=(b+B)\gamma (x)$. Now let $\nat x\dd y \in \Om^1T\Ac_{\nat}$. One has $\gamma(\bb\nat x\dd y)= \gamma b(x\dd y)= (1-\phi)^{-1}b(x\dd y)$. On the other hand,
\beq
(b+B)\gamma(\nat x\dd y)&=& (b+B)(1-\phi)^{-1}(x\dd y+b(x\phi(y)))\non\\
&=& (1-\phi)^{-1}b(x\dd y)\non
\eeq
by using (\ref{r}). Hence $(b+B)\circ \gamma=\gamma\circ (\nat \dd+ \bb)$, proving that $\gamma$ is a chain map.\\
iii) One obviously has $\pi\circ\gamma=\id$ on $X(T\Ac)$, whence a split exact sequence of complexes
$$
\xymatrix{
0 \ar[r] & \Theta \ar[r] & \Om T\Ac \ar[r]_{\pi} & X(T\Ac) \ar@{.>}@/_1pc/[l]_{\gamma} \ar[r] & 0\ .}
$$
Let $Q:=\gamma\circ \pi$ be the projection of $\Om T\Ac$ onto the image of $X(T\Ac)$. Then $\Theta=\ker Q$, $X(T\Ac)\simeq \im Q$ as complexes and $\Om T\Ac \simeq \Theta\oplus X(T\Ac)$. We now construct a contracting homotopy for $\Theta$. Let $h=(1-\phi)^{-1}\nabla$ on $\Theta$. One has $h\Theta\subset\Theta$ because the image of $h$ are differential forms of degree $\ge 2$. Extend $h$ to $\Om T\Ac$ by setting $h=0$ on $X(T\Ac)$. Then a simple computation using (\ref{r}) and $\nabla b+b\nabla=-1$ on $\Om^{n\ge 2}T\Ac$ shows that $(b+B)h+h(b+B)=-\id$ on $\Theta$. Thus with $h=(1-\phi)^{-1}\nabla(1-Q)$ on $\Om T\Ac$, one gets
$$
\gamma\circ\pi=Q= \id +(b+B)h+h(b+B)\ ,
$$
and the proposition follows.\hfill\rule{1ex}{1ex}\\

Proposition \ref{pequi} shows that $\gamma$ and $\pi$ realize inverse homotopy equivalences between $X(T\Ac)$ and $\Om T\Ac$. This is not very interesting at first sight, since the homology of these complexes is in fact trivial! However, these are bornological complexes and their completions compute the entire cyclic homology of $\Ac$. The key point is that all the maps we have constructed are in fact bounded for the bornology $\Sg(\Om T\Ac)$:
\begin{proposition}
The maps $\nabla,\phi,(1-\phi)^{-1},h$ are bounded on $\Om T\Ac$ and thus extend to bounded linear maps on the completion $\Ome\Tc\Ac$. Also, $\gamma$ is bounded and extends to a bounded chain map
\be
\gamma: X(\Tc\Ac)\to \Ome\Tc\Ac\ .
\ee
\end{proposition}
{\it Proof:} Recall that the bornology $\Sg(\Om T\Ac)$ is generated by the collection of subsets $\bigcup_{n\ge 0}[n/2]!\, \Tt(\dd T)^n$, for any small $T\in\Sg(T\Ac)=\Sgan(\Om^+\Ac)$. We will consider a set of generators of $\Sgan(\Om^+\Ac)$ that will appear complicated at first sight, but it will simplify drastically the proof of boundedness. For any $S\in \Sg(\Ac)$, we put $T_n(S)=\St\odot(dSdS)^n\odot\St$, and consider the following subset of $\Om^+\Ac$:
$$
V(S)=(2S+S^2)\cup\bigcup_{N\ge 0} \sum_{n=0}^NT_n(S)\quad \subset \Om^+\Ac=T\Ac\ .
$$
We claim that $\Sgan(\Om^+\Ac)$ is generated by $\{V(S)|S\in\Sg(\Ac)\}$. Indeed, by definition $\Sgan(\Om^+\Ac)$ is generated by the set $\goth{U}=\{\bigcup_{n\ge 0} \St(dS)^{2n}|S\in\Sg(\Ac)\}$. Then for any $S\in\Sg(\Ac)$, $\bigcup_{n\ge 0} \St(dS)^{2n}\subset V(S)$. Conversely, given a small $S$ in $\Ac$, we have to show that $V(S)$ is contained in the circled convex hull of elements of $\goth{U}$. Let $n\in\nn$, one has
\beq
&&\St\odot(dSdS)^n\odot \St=(\St(dS)^{2n})\odot \St\subset\St(dS)^{2n}\St + (dS)^{2n+2} \non\\
&&\subset\St (dS)^{2n-1}d(S\St)+\St(dS)^{2n-2}d(S^2)dS+\ldots + (\St S)(dS)^{2n}+(dS)^{2n+2}\ ,\non
\eeq
where the latter sum contains $2n+2$ terms. Let $S'$ be the disk $(S\St)^{\Diamond}$, then we find
$$
\sum_{n=0}^NT_n(S)\subset \sum_{n=0}^{N+1}((n+1)\widetilde{S'}(dS')^{2n})^{\Diamond}\ .
$$
But the rescaling $U=4S'$ implies
$$
\sum_{n=0}^NT_n(S)\subset \sum_{n=0}^{N+1}\frac{1}{2^{2n+1}}(\widetilde{U}(dU)^{2n})^{\Diamond} \subset \left(\bigcup_{n\ge 0}\widetilde{U}(dU)^{2n}\right)^{\Diamond}\ ,
$$
because the sum $\sum_{n=0}^{N+1}1/2^{2n+1}$ is always less than $1$. This shows that $V(S)$ is contained in the disked hull of $\bigcup_{n\ge 0}\widetilde{U}(dU)^{2n}$. Consequently $V(S)$ is small for the bornology $\Sgan(T\Ac)$ as claimed, and $\{V(S)|S\in\Sg(\Ac)\}$ generates the analytic bornology of $T\Ac$.\\
We now show that $\phi$ is bounded. For a small $S\in\Sg(\Ac)$ let 
$$
\at_0\odot da_1da_2\odot\ldots \odot da_{2n-1}da_{2n}\odot \at_{2n+1}\ \in T_n(S)\subset \Om^+\Ac\ .
$$
The identity $\phi(x\odot y)=\phi(x)y+x\phi(y)+\dd x\dd y$ for any $x,y\in T\Ac$, as well as $\phi(a)=0$ $\forall a\in\Ac$, imply $\phi (da_1da_2)=-\dd a_1\dd a_2$, and more generally
\beq
\lefteqn{
\phi(\at_0\odot da_1da_2\odot\ldots \odot da_{2n-1}da_{2n}\odot \at_{2n+1}) = }\non\\
&& \dd \at_0\dd(da_1da_2\odot\ldots \odot da_{2n-1}da_{2n}\odot \at_{2n+1})\non\\
&& +\sum_{i=1}^n(-\at_0\odot da_1da_2\odot\ldots \dd a_{2i-1}\dd(a_{2i}\odot\ldots\odot \at_{2n+1})\non\\
&&+\at_0\odot da_1da_2\odot\ldots\dd(a_{2i-1}a_{2i})\dd(da_{2i+1}da_{2i+2}\ldots \odot\at_{2n+1})\non\\
&&- \at_0\odot da_1da_2\ldots\odot a_{2i-1}\dd a_{2i}\dd(da_{2i+1}da_{2i+2}\ldots \odot\at_{2n+1}))\ .\non
\eeq
This implies (with $T_n:=T_n(S)$ for any $n$)
$$
\phi(T_n)\subset \dd S\dd T_n+ \sum_{i=1}^n (T_{i-1}\dd(2S+S^2)\dd T_{n-i})^{\Diamond}
$$
and $\phi(\sum_{n=0}^NT_n)$ is contained in the disked hull of the sum $\dd S \dd(\sum_{n=0}^NT_n)+(\sum_{n-0}^NT_n)\dd( 2S+S^2)(\sum_{n=p}^NT_p)$. Furthermore, $\phi$ vanishes on $\Ac$ so that $\phi(2S+S^2)=0$, and with $V=V(S)$ one gets
\beq
\phi(V)&\subset& \left(\dd S \dd(\bigcup_N\sum_{n=0}^NT_n)+(\bigcup_N\sum_{n-0}^NT_n)\dd( 2S+S^2)(\bigcup_P\sum_{n=p}^PT_p)\right)^{\Diamond}\non\\
&\subset& (\dd V\dd V+V\dd V\dd V)^{\Diamond}\ = \ (\widetilde{V}\dd V\dd V)^{\Diamond}\ .\non
\eeq
Now, the bornology on $\Om T\Ac$ is generated by the sets $\bigcup_n [n/2]!\,\widetilde{V}(\dd V)^n$, for $V=V(S)$, $S\in\Sg(\Ac)$. One thus has
\beq
\phi(\widetilde{V}(\dd V)^n)&=& \nabla B(\widetilde{V}(\dd V)^n)\ \subset\ \nabla((n+1)(\dd V)^{n+1})^{\Diamond}\non\\
&\subset& (n+1)(\phi(V)(\dd V)^n)^{\Diamond}\ \subset\ (n+1)(\widetilde{V}(\dd V)^{n+2})^{\Diamond}\ .\non
\eeq
$(n+1)$ grows polynomially, so that by rescaling $V$, one can find a small set $V'\in\Sgan(\Om^+\Ac)$ such that $\phi(\bigcup_n [n/2]!\,\widetilde{V}(\dd V)^n)$ is contained in the disked hull of $\bigcup_n [1+n/2]!\,\widetilde{V'}(\dd V')^{n+2}$. Hence $\phi$ is {\it bounded} for the bornology of $\Om T\Ac$.\\
Next, since $\nabla(x_0\dd x_1\ldots\dd x_n)=x_0\phi(x_1)\dd x_2\ldots\dd x_x$ for any $x_i\in T\Ac$, $\nabla$ is also bounded. Let us now focus on $(1-\phi)^{-1}=\sum_{k=0}^{\infty}\phi^k$. For any $V=V(S)$, one has
$$
(1-\phi)^{-1}(\widetilde{V}(\dd V)^n)\subset\sum_{k=0}^{\infty}(n+1)(n+3)\ldots (n+2k+1)(\widetilde{V}(\dd V)^{n+2k})^{\Diamond}\ .
$$
In fact, this is a finite sum on all elements of $\widetilde{V}(\dd V)^n$. Elementary estimates on the function $n!$ show there is a constant number $\la$ such that $(n+1)(n+3)\ldots(n+2k+1)\le \la^{n+2k+1}\frac{[k+n/2]!}{[n/2]!}$. Hence $(1-\phi)^{-1}([n/2]!\widetilde{V}(\dd V)^n)$ is contained in the disked hull of $\sum_{k=0}^{\infty}[k+n/2]!\la\widetilde{V}(\la\dd V)^{n+2k}$, and by the rescaling $W=2\la V$, one gets
$$
(1-\phi)^{-1}([n/2]!\widetilde{V}(\dd V)^n)\subset \left(\bigcup_p [p/2]!\,\widetilde{W}(\dd W)^p\right)^{\Diamond}\ .
$$
Since $W$ does not depend on $n$, this shows that $(1-\phi)^{-1}$ is bounded.\\
It remains to study the morphism $\gamma:X(T\Ac)\to \Om T\Ac$. By definition, for any $x\in X_0(T\Ac)=T\Ac$, one has $\gamma(x)=(1-\phi)^{-1}(x)$, hence $\gamma$ is bounded on $X_0(T\Ac)$. Let now $\nat x\dd y\in X_1(T\Ac)=\Om^1T\Ac_{\nat}$. One has $\gamma(\nat x\dd y)=(1-\phi)^{-1}(x\dd y+b(x\phi(y)))$. If we use the bornological vector space isomorphism $\Om^1T\Ac_{\nat}\simeq \Tt\Ac\otimes\Ac$, it is sufficient to evaluate $\gamma$ on the element $\nat x\dd a \in \Tt\Ac\otimes\Ac$. Since $\phi(a)=0$, one has $\gamma(\nat x\dd a)=(1-\phi)^{-1}(x\dd a)$. We deduce that $\gamma$ is bounded, and also $Q=\gamma\pi$ and $h=(1-\phi)^{-1}\nabla(1-Q)$.\hfill\rule{1ex}{1ex}\\
 
\begin{corollary}\label{cgood}
The chain map $\gamma:X(\Tc\Ac)\to \Ome\Tc\Ac$ is an homotopy equivalence. Its class $[\gamma]$ in the bivariant entire cyclic cohomology $HE_0(\Ac,\Tc\Ac)$ is the inverse of $[\pi]\in HE_0(\Tc\Ac,\Ac)$.
\end{corollary}
{\it Proof:} By the universal properties of completions and proposition \ref{pequi} iii), one has $\pi\circ\gamma=\id$ on $X(\Tc\Ac)$ and $\gamma\circ\pi= \id +[b+B,h]$ on $\Ome\Tc\Ac$.\hfill\rule{1ex}{1ex}\\
\begin{corollary}\label{ccom}
Let $m:\Tc\Ac\to\Ac$ be the multiplication map, $v_{\Ac}:\Tc\Ac\to\Tc\Tc\Ac$ the canonical bounded homomorphism, and $\gamma,\pi$ as before. Let also $c:X(\Tc\cdot)\to \Ome(\cdot)$ be the bornological isomorphism (\ref{res}) and $P$ the spectral projection onto the $\kappa^2$-invariant forms. Then the following diagram of chain maps commutes up to homotopy:
\be
\xymatrix{
X(\Tc\Tc\Ac) \ar[rr]^{P\circ c} \ar@<-1ex>[d]_{X(m_*)} & & \Ome\Tc\Ac \ar[d]^{\Ome(m)} \ar@<1ex>[dll]^{\pi} \\
X(\Tc\Ac) \ar[u]_{X(v_{\Ac})} \ar[rr]_{P\circ c} \ar[urr]^{\gamma} & & \Ome\Ac } \label{cd}
\ee
Moreover, all the arrows are homotopy equivalences.
\end{corollary}
{\it Proof:} Let $\kappa$ be the Karoubi operator on $\Ome\Tc\Ac$. For any $x,y\in\Tc\Ac$ one has $\kappa(x)=x$ and $\kappa(x\dd y)=\dd y x$, hence the projection $\pi:\Ome\Tc\Ac\to X(\Tc\Ac)$ is $\kappa$-invariant: $\pi\circ\kappa=\pi$. It follows that $\pi$ is invariant under the spectral projection $P$, and $\pi\circ P\circ c=\pi\circ c$. Consider the composition of linear maps
$$
\begin{CD}
X(\Tc\Ac)  @>{X(v_{\Ac})}>> X(\Tc\Tc\Ac) @>{c}>>  \Ome\Tc\Ac @>{\pi}>> X(\Tc\Ac)\ .
\end{CD}
$$
A direct computation using the definitions shows that it is the identity on $X(\Tc\Ac)$, hence $\pi\circ P\circ c\circ X(v_{\Ac})=\Id_{X(\Tc\Ac)}$. Since $\pi$ and $P\circ c$ are homotopy equivalences, one deduces that $X(v_{\Ac})$ is also an homotopy equivalence. Next, the composition $\Tc\Ac\stackrel{v_{\Ac}}{\longrightarrow} \Tc\Tc\Ac \stackrel{m_*}{\longrightarrow}\Tc\Ac$ is the identity homomorphism of $\Tc\Ac$. Hence $X(m_*)\circ X(v_{\Ac})=\Id_{X(\Tc\Ac)}$, and $X(m_*)$ is an homotopy equivalence inverting $X(v_{\Ac})$. Since we know that $\gamma$ is the inverse of $\pi$, we are left with the commutative diagram up to homotopy
$$
\xymatrix{
X(\Tc\Tc\Ac) \ar[rr]^{P\circ c} \ar@<-1ex>[d]_{X(m_*)}& &\Ome\Tc\Ac \ar@<1ex>[dll]^{\pi} \\
X(\Tc\Ac) \ar[u]_{X(v_{\Ac})} \ar[urr]^{\gamma} & & }
$$
The bottom right corner of (\ref{cd}) follows from the functoriality of $\Ome(\cdot)$ and $X(\Tc\cdot)$, which implies $\Ome(m)\circ P\circ c=P\circ c\circ X(m_*)$. \hfill\rule{1ex}{1ex}

\section{Families of spectral triples}\label{bim}

\subsection{Definition}

Let $\Bc$ be a $\zz_2$-graded complete bornological algebra. Given a $\zz_2$-graded complete bornological vector space $\Hc$, we can consider the (graded) completed tensor product $\Ec=\Hc\hotimes\Bc$. Since the multiplication on $\Bc$ is bounded, the obvious right action of $\Bc$ on $\Hc\otimes \Bc$ extends to the completion $\Ec$. This turns $\Ec$ into a $\zz_2$-graded bornological right $\Bc$-module, i.e. the following bilinear map is bounded:
\beq
\Ec\times \Bc &\to& \Ec\non\\
(h\otimes b_1, b_2) &\mapsto& h\otimes b_1b_2\ .
\eeq
Denote by $\End_{\Bc}(\Ec)$ the $\zz_2$-graded algebra of bounded endomorphisms of $\Ec$, commuting with the action of $\Bc$. We always endow $\End_{\Bc}(\Ec)$ with the bornology of equibounded endomorphisms, so that it is a complete bornological algebra. 
\begin{definition}\label{dbim}
Let $\Ac$ and $\Bc$ be complete bornological algebras. We assume $\Ac$ is trivially graded and $\Bc$ is $\zz_2$-graded. Then a family of spectral triples over $\Bc$, or an unbounded $\Ac$-$\Bc$-bimodule, is a triple $(\Ec,\rho,D)$ corresponding to the following data:
\begin{itemize}
\item A $\zz_2$-graded complete bornological vector space $\Hc$ and the corresponding right $\Bc$-module $\Ec=\Hc\hotimes\Bc$.
\item A bounded homomorphism $\rho:\Ac\to \End_{\Bc}(\Ec)$ sending $\Ac$ to \emph{even degree} endomorphisms of $\Ec$. Hence $\Ec$ is a bornological left $\Ac$-module.
\item An unbounded endomorphism $D:\dom(D)\subset \Ec\to\Ec$ of odd degree, defined on a bornologically dense domain of $\Ec$ and commuting with the right action of $\Bc$:
\be
D\cdot(\xi b)=(D\cdot \xi)b\quad \forall \xi\in\Ec\ ,\ b\in\Bc\ .
\ee
$D$ is also called a \emph{Dirac operator}.
\item The commutator $[D,\rho(a)]$ extends to an element of $\End_{\Bc}(\Ec)$ for any $a\in\Ac$.
\item For any $t\in \rr_+$, the \emph{heat kernel} $\exp(-t D^2)$ is densely defined and extends to a bounded endomorphism of $\Ec$.
\end{itemize}
We denote by $\Psi(\Ac,\Bc)$ the set of such unbounded bimodules.
\end{definition}
It  is clear that this definition is an adaptation of the Baaj and Julg picture of unbounded Kasparov bimodules \cite{Bl}, with $C^*$-algebras replaced by bornological algebras. Remark however that we do not require the ``resolvent'' $(1+D^2)^{-1}$ to be a compact endomorphism as in Kasparov theory. Instead, we deal with the heat operator $\exp(-t D^2)$, and the compactness will be replaced by the so-called $\te$-summability condition (see definitions \ref{dwte} and \ref{dste}). Roughly speaking, $\te$-summability means that the heat kernel is a trace-class endomorphism for $t>0$, a well-defined notion in bornology.\\
Before proceeding further, let us mention some simple examples of unbounded bimodules, in the case of a trivially graded algebra $\Bc$:
\begin{example}\textup{
Homomorphisms: if $D=0$ and $\Hc=\cc^{n_+}\oplus \cc^{n_-}$ is a finite-dimensional $\zz_2$-graded space (with fine bornology), then the triple $(\Ec,\rho,D)$ reduces to a pair of bounded homomorphisms $\rho_{\pm}: \Ac\to M_{n_{\pm}}(\Bc)$. If moreover $\Ac=\cc$, the latter is equivalent to a pair of idempotents $e_{\pm}=\rho_{\pm}(1)\in M_{n_{\pm}}(\Bc)$. The difference ``$e_+-e_-$'' then describes an algebraic $K$-theory class of $\Bc$.}
\end{example}
\begin{example}\textup{
$\Bc=\cc$: then $\Psi(\Ac,\cc)$ is just the set of triples $(\Hc,\rho,D)$. If $\Hc$ is an Hilbert space, $D$ a selfadjoint unbounded operator and the heat kernel $\exp(-tD^2)$ is trace-class for any $t>0$, then $(\Hc,\rho,D)$ is a $\te$-summable spectral triple over $\Ac$.}
\end{example}
\begin{example}\textup{
$\Ac=\cc$ and $\rho(1)=1\in \End_{\Bc}(\Ec)$: the homomorphism $\cc\to \End_{\Bc}(\Ec)$ completely disappears. We view $\Ec=\Hc\hotimes\Bc$ as the space of sections of a trivial vector bundle over the non-commutative manifold $\Bc$, and $D$ represents a ``family of Dirac operators'' acting by endomorphisms on $\Ec$. This also describes a $K$-theory element of $\Bc$. More generally, if $\rho(1)=e\neq 1$ is any idempotent in $\End_{\Bc}(\Ec)$, then by virtue of the Serre-Swan theorem, $e\Ec$ is a twisted vector bundle over $\Bc$, and $(e,D)$ represents a twisted family of Dirac operators. Classically, such examples are provided by longitudinal elliptic operators on fibered manifolds or foliations \cite{B,C1}.}
\end{example}

\subsection{Higher bimodules and formal Bott periodicity}

We shall introduce the higher unbounded bimodules. Let $C_1=\cc\oplus\eps\cc$, be the one-dimensional complex Clifford algebra (with fine bornology). It is a $\zz_2$-graded algebra generated by the unit $1$ in degree zero and $\eps$ in degree one, with $\eps^2=1$. For any $n\ge 1$, the $n$-dimensional complex Clifford algebra is the graded tensor product $C_n=C_1^{\hotimes n}$, and by convention $C_0=\cc$. For any trivially graded complete bornological algebra $\Bc$, the completed tensor product $C_n\hotimes \Bc$ (which also coincides with the algebraic tensor product) is thus a $\zz_2$-graded complete bornological algebra.
\begin{definition}
Let $\Ac$ and $\Bc$ be trivially graded complete bornological algebras. For any $n\ge 0$, we set $\Psi_n(\Ac,\Bc):=\Psi(\Ac,C_n\hotimes \Bc)$.
\end{definition}
It is well-known that, due to the formal Bott periodicity $C_{n+2}\simeq M_2(C_n)$, only the first two cases $n=0$ and $n=1$ are relevant:
\vskip 2mm
\noindent $\bullet$ $n=0$: One has $\Psi_0(\Ac,\Bc)=\Psi(\Ac,\Bc)$. Let $(\Ec,\rho,D)$ be such a bimodule, with $\Ec=\Hc\hotimes \Bc$. The complete bornological space $\Hc$ is $\zz_2$-graded, hence it comes equipped with an involutive operator $\Gamma$, $\Gamma^2=1$, which splits $\Hc$ into two eigenspaces $\Hc_+$ and $\Hc_-$ of even and odd vectors respectively. Also the right $\Bc$-module $\Ec$ splits into two eigenspaces $\Ec_{\pm}=\Hc_{\pm}\hotimes \Bc$. We adopt the usual $2\times 2$ matrix notation
\be
\Ec=\left( \begin{array}{c}
          \Ec_+ \\
          \Ec_- \\
     \end{array} \right) \qquad
\Gamma=\left( \begin{array}{cc}
          1 & 0 \\
          0 & -1 \\
     \end{array} \right)\ .
\ee
The even (resp. odd) part of the $\zz_2$-graded algebra $\End_{\Bc}(\Ec)$ is represented by diagonal (resp. off-diagonal) matrices. By definition the homomorphism $\rho\to \End_{\Bc}(\Ec)$ commutes with $\Gamma$, whereas $D$ anticommutes. In matricial notations one thus has 
\be
\rho(a)=\left( \begin{array}{cc}
          \rho_+(a) & 0 \\
          0 & \rho_-(a) \\
     \end{array} \right)\qquad
D=\left( \begin{array}{cc}
          0 & D_- \\
          D_+ & 0 \\
     \end{array} \right) \label{dec0}
\ee
for any $a\in\Ac$.
\vskip 2mm
\noindent $\bullet$ $n=1$: Let $(\Ec,\rho,D)\in \Psi_1(\Ac,\Bc)=\Psi(\Ac,C_1\hotimes \Bc)$. One thus has $\Ec=\Hc\hotimes C_1\hotimes \Bc$ for some $\zz_2$-graded bornological space $\Hc=\Hc_+\oplus \Hc_-$. We may write the graded tensor product  $\Hc\hotimes C_1$ as the direct sum of its even and odd part:
\be
\Hc\hotimes C_1=(\Hc_+\oplus\Hc_-)\hotimes (\cc\oplus\eps\cc)=(\Hc_+\oplus\Hc_-\eps)\oplus(\Hc_+\eps\oplus\Hc_-)=\Kc\hotimes C_1\ ,
\ee
where $\Kc=\Hc_+\oplus \Hc_-\eps$ is a trivially graded bornological space. Hence the module $\Ec$ is the direct sum of two copies of $\Kc\hotimes\Bc$:
\be
\Ec=\left( \begin{array}{c}
          \Kc\hotimes\Bc \\
          \Kc\hotimes\Bc \\
     \end{array} \right) \ ,
\ee
whose right action of $C_1\hotimes\Bc$ is such that $\eps$ flips the two factors. It follows that any endomorphism $z\in\End_{C_1\hotimes\Bc}(\Ec)$ reads
\be
z=\left( \begin{array}{cc}
          x & y \\
          y & x \\
     \end{array} \right)\ ,
\ee
with $x,y\in \End_{\Bc}(\Kc\hotimes\Bc)$. As a consequence, there is a bounded homomorphism $\al:\Ac\to \End_{\Bc}(\Kc\hotimes\Bc)$ and an unbounded endomorphism $Q:\Kc\hotimes\Bc\to \Kc\hotimes\Bc$ such that
\be
\rho(a)=\left( \begin{array}{cc}
          \al(a) & 0 \\
          0 & \al(a) \\
     \end{array} \right)\qquad
D=\left( \begin{array}{cc}
          0 & Q \\
          Q & 0 \\
     \end{array} \right)\ .\label{dec1}
\ee
This is the general matricial form for an element of $\Psi_1(\Ac,\Bc)$.
\vskip 2mm
\noindent $\bullet$ $n\ge 2$: The study of an element  $(\Ec,\rho,D)\in\Psi_n(\Ac,\Bc)=\Psi(\Ac,C_n\hotimes\Bc)$ is analogous to the previous one for $\Psi_1(\Ac,\Bc)$. We can reduce $\Ec$ to a product $\Kc\hotimes C_n\hotimes\Bc $ for a certain trivially graded vector space $\Kc$, and consequently $\End_{C_n\hotimes\Bc}(\Ec)=\End_{\Bc}(\Kc\hotimes\Bc)\hotimes C_n$.\\ 
If $n=2k$ is even, then $C_{n}=M_2(M_{2^{k-1}}(\cc))$ as a $\zz_2$-graded algebra, with standard even/odd grading corresponding respectively to the diagonal/off-diagonal $2\times 2$ block matrices. It follows that $\End_{C_n\hotimes\Bc}(\Ec)=M_2(\End_{\Bc}(\Kc\hotimes\cc^{2^{k-1}}\hotimes\Bc))$ in $2\times 2$ matrix notation. The homomorphism $\rho$ and the Dirac operator thus decompose as in (\ref{dec0}). This shows that up to stabilization by matrices of arbitrary size, the elements of $\Psi_{2k}(\Ac,\Bc)$ correspond exactly to the elements of $\Psi_0(\Ac,\Bc)$.\\
If $n=2k+1$ is odd, one has $C_{n}=M_{2^k}(\cc)\hotimes C_1$ as a $\zz_2$-graded algebra, where $ M_{2^k}(\cc)$ is trivially graded and $C_1$ has its natural graduation. Consequently, $\End_{C_n\hotimes\Bc}(\Ec)=\End_{\Bc}(\Kc\hotimes\cc^{2^k}\hotimes\Bc)\hotimes C_1$. The homomorphism $\rho$ and the Dirac operator thus decompose as in (\ref{dec1}); hence up to stabilization by matrices, $\Psi_{2k+1}(\Ac,\Bc)$ corresponds to $\Psi_1(\Ac,\Bc)$.\\

All in all, due to formal Bott periodicity there are only two different sets of bimodules, the even ones $\Psi_0(\Ac,\Bc)$, and the odd ones $\Psi_1(\Ac,\Bc)$. In both cases, a $\zz_2$-graded module splits into the direct sum $\Ec=\Ec_+\oplus\Ec_-$, according to which the homomorphism $\rho$ has a diagonal form, and the Dirac operator $D$ is off-diagonal.

\subsection{Properties}

Let $\Ac$ and $\Bc$ be trivially graded complete bornological algebras. It is readily seen that $\Psi_*(\Ac,\Bc)$, $*=0,1$,  is a semigroup under direct orthogonal sum: $(\Ec,\rho,D)+(\Ec',\rho',D')=(\Ec\oplus \Ec',\rho\oplus \rho',D\oplus D')$. Since we don't deal with $C^*$-algebras, there is {\it a priori} no reason to find an interesting composition product $\Psi(\Ac,\Bc)\times \Psi(\Bc,\Cc)\to \Psi(\Ac,\Cc)$ as in the case of Kasparov's theory. Nevertheless, $\Psi$  is a bimodule over the category of bornological algebras in the following sense. If we let $\mor(\Ac,\Bc)\subset \Psi_0(\Ac,\Bc)$ be the set of bounded homomorphisms from $\Ac$ to $\Bc$, then there is a well-defined left product $\mor(\Ac,\Bc)\times \Psi_*(\Bc,\Cc)\to \Psi_*(\Ac,\Cc)$ given by
\be
\varphi\cdot (\Ec,\rho,D)= (\Ec,\rho\circ\varphi, D)\ ,
\ee
with $\Ec=\Hc\hotimes \Cc$. For the right product we must consider the {\it unitalizations} $\Bct$ and $\widetilde{\Cc}$. Let $(\Ec,\rho,D)\in\Psi_*(\Ac,\Bct)$ with $\Ec=\Hc\hotimes\Bct$, and consider a unital bounded homomorphism $\varphi:\Bct\to\Cct$. Then the right product $\Psi_*(\Ac,\Bct)\times \mor(\Bct,\Cct)\to \Psi_*(\Ac,\Cct)$ reads
\be
(\Ec,\rho,D)\cdot \varphi= (\Ec\hotimes_{\varphi}\Cct, \rho\otimes\Id, D\otimes\Id) \label{right}
\ee
where $\Ec\hotimes_{\varphi}\Cct$ is canonically isomorphic to $\Hc\hotimes\Cct$. Our aim is to construct a Chern character map
\be
\ch: \Psi_*^{\te}(\Ac,\widetilde{\Bc})\to HE_*(\Ac,\Bc)\ ,\quad *=0,1\ ,
\ee
with domain the strongly $\te$-summable unbounded $\Ac$-$\widetilde{\Bc}$-bimodules (see definitions \ref{dwte} and \ref{dste}), and range the bivariant entire cyclic cohomology of $\Ac$ and $\Bc$. This Chern character has to be additive, invariant under differentiable homotopies (definition \ref{homo}) and functorial with respect to $\Ac$ and $\Bc$, which means that the following diagram commutes:
\be
\begin{CD}
\mor(\Ac,\Bc) @. \times @. \Psi_*^{\te}(\Bc,\widetilde{\Cc}) @>>> \Psi_*^{\te}(\Ac,\widetilde{\Cc})\\
@VV{\ch}V     @.      @VV{\ch}V            @VV{\ch}V \\
HE_0(\Ac,\Bc) @. \times @. HE_*(\Bc,\Cc) @>>> HE_*(\Ac,\Cc)\\
\end{CD}
\ee
and similarly for the right product (\ref{right}).

\section{Algebra cochains and superconnections}\label{super}

Let $\Ac$ and $\Bc$ be trivially graded complete bornological algebras, and $\widetilde{\Bc}$ the unitalization of $\Bc$. To any unbounded bimodule $(\Ec,\rho,D)\in\Psi_*(\Ac,\Bct)$ verifying suitable $\te$-summability conditions, we will associate a bounded chain map $\chi(\Ec,\rho,D)$ from the $(b+B)$-complex of entire chains $\Ome\Ac$ to the $X$-complex $X(\Bc)$. This morphism, playing a central role in the bivariant Chern character, is obtained from the exponential of the curvature of a superconnection, as in the Bismut-Quillen approach to the family's index theorem \cite{B,Q1}. To do this, we adapt the theory of algebra cochains developed by Quillen \cite{Q2} to the bornological framework. For convenience, we postponed in appendix a self-contained account of Quillen's formalism. 

\subsection{Bar construction}

Let $\Ac$ be a complete bornological algebra, $\Om \Ac=\bigoplus_{n\ge 0}\Om^n\Ac$ the $(b,B)$-bicomplex of noncommutative forms over $\Ac$, with $\Om^n\Ac=\Act\hotimes\Ac^{\hotimes n}$. We use the \emph{bar complex} 
\be
\Bb(\Ac)=\bigoplus_{n\ge 0}\Bb_n(\Ac)\ ,
\ee
where $\Bb_n(\Ac)=\Ac^{\hotimes n}$ and $\Bb_0(\Ac)=\cc$. Recall (appendix) that $\Bb(\Ac)$ is a graded coassociative coalgebra. The coproduct $\Delta:\Bb(\Ac)\to\Bb(\Ac)\hotimes\Bb(\Ac)$ is given by
\be
\Delta(a_1\otimes\ldots\otimes a_n)=\sum_{i=0}^n (a_1\otimes\ldots\otimes a_i)\otimes(a_{i+1}\otimes\ldots\otimes a_n)\ ,
\ee
for any $a_j\in\Ac$. Furthermore, one has a boundary map $b':\Bb(\Ac)\to\Bb(\Ac)$ of degree $-1$:
\be
b'(a_1\otimes\ldots\otimes a_n)=\sum_{i=1}^{n-1}(-)^{i+1}a_1\otimes\ldots\otimes a_ia_{i+1}\otimes\ldots\otimes a_n
\ee
verifying ${b'}^2=0$ and $\Delta b'=(b'\otimes\id+\id\otimes b')\Delta$. Thus $\Bb(\Ac)$ is a graded differential coalgebra. There is an associated free bicomodule $\Om_1\Bb(\Ac)=\Bb(\Ac)\hotimes\Ac\hotimes\Bb(\Ac)$, with left and right bicomodule maps
\beq
\Delta_l=\Delta\otimes\id\otimes\id &:& \Om_1\Bb(\Ac)\to \Bb(\Ac)\hotimes\Om_1\Bb(\Ac)\ ,\non\\
\Delta_r=\id\otimes\id\otimes\Delta &:& \Om_1\Bb(\Ac)\to \Om_1\Bb(\Ac)\hotimes\Bb(\Ac)\ .
\eeq
$\Om_1\Bb(\Ac)$ is endowed with a differential $b'':\Om_1\Bb(\Ac)\to\Om_1\Bb(\Ac)$, ${b''}^2=0$, compatible with the bicomodule structure and $b'$ (see appendix). There is also a projection $\partial:\Om_1\Bb(\Ac)\to\Bb(\Ac)$ defined by
\be
\partial(a_1\otimes\ldots\otimes a_{i-1})\otimes a_i\otimes(a_{i+1}\otimes\ldots\otimes a_n)=a_1\otimes\ldots\otimes a_n\ .
\ee
It is a coderivation ($\Delta\d=(\id\otimes\d)\Delta_l+(\d\otimes\id)\Delta_r$) and a morphism of complexes ($\d b''=b'\d$). The last operator we need is the injection $\nat: \Om\Ac\to \Om_1\Bb(\Act)$:
\be
\nat(\at_0da_1\ldots da_n)=\sum_{i=0}^n(-)^{n(i+1)}(a_{i+1}\otimes\ldots\otimes a_n)\otimes \at_0\otimes(a_1\otimes\ldots\otimes a_i)\ ,
\ee
for any $\at_0$ in the unitalization $\Act=\cc\oplus\Ac$ and $a_j\in\Ac$. Then $\nat$ is a \emph{cotrace}, see appendix.\\

We now endow the bar complex and its associated bicomodule with new bornologies satisfying the \emph{entire growth condition}. Let $\Sge(\Bb(\Ac))$ be the bornology generated by the sets $\bigcup_{n\ge 0}[n/2]!\, S^{\otimes n}$ for any $S\in\Sg(\Ac)$. We denote by $\Bbe(\Ac)$ the completion of $\Bb(\Ac)$ with respect to this bornology. Also, let $\Sge(\Om_1\Bb(\Ac))$ be the bornology generated by $\bigcup_{n,p\ge 0}[(n+p)/2]!\, (S^{\otimes n})\otimes S\otimes(S^{\otimes p})$, $S\in\Sg(\Ac)$. The corresponding completion $\Om_1\Bbe(\Ac)$ identifies with $\Bbe(\Ac)\hotimes\Ac\hotimes\Bbe(\Ac)$. 
\begin{lemma}
All the maps $\Delta,\Delta_{l,r},b',b'',\d,\nat$ are bounded for the entire bornology and thus extend to the completions $\Bbe(\Ac)$, $\Om_1\Bbe(\Ac)$ and $\Ome\Ac$.
\end{lemma}
{\it Proof:} It is a direct consequence of the definitions. Let us for example check the boundedness of the coproduct $\Delta$ on $\Bb(\Ac)$. For any small $S\in\Sg(\Ac)$ one has
\beq
\lefteqn{\Delta([n/2]!\, S^{\otimes n})\subset [n/2]!\sum_{i=0}^n(S^{\otimes i})\otimes(S^{\otimes(n-i)})} \non\\
&&\subset \sum_{i=0}^n\frac{[n/2]!}{[i/2]![(n-i)/2]!}\left(([i/2]!S^{\otimes i})\otimes([(n-i)/2]!S^{\otimes(n-i)})\right)^{\Diamond}\ .\non
\eeq
But there is a constant $\la$ such that $\frac{[n/2]!}{[i/2]![(n-i)/2]!}\le \la^n$ for any $n$ and $i\le n$, so that $\Delta([n/2]!\, S^{\otimes n})$ is contained in
$$
\la^n(n+1)\left(\sum_{i=0}^n\frac{1}{n+1}([i/2]!S^{\otimes i})\otimes([(n-i)/2]!S^{\otimes(n-i)})\right)^{\Diamond}\ .
$$
The sum over $i$ lies in the circled convex hull of the set $(\bigcup_m[m/2]!S^{\otimes m})\otimes(\bigcup_p[p/2]!S^{\otimes p})$, hence by rescaling appropriately $S$, one can find a small $T\in\Sg(\Ac)$ such that 
\be
\Delta(\bigcup_n[n/2]!\, S^{\otimes n})\subset \left((\bigcup_m[m/2]!T^{\otimes m})\otimes(\bigcup_p[p/2]!T^{\otimes p})\right)^{\Diamond}\ ,
\ee
and the conclusion follows. The other operators are treated similarly.\hfill\rule{1ex}{1ex}\\

In particular, the completed bar complex is a $\zz_2$-graded differential coalgebra, and $\Om_1\Bbe(\Ac)$ is a $\Bbe(\Ac)$-bicomodule. All the results of the appendix extend also to these completions. 

\subsection{The bimodule $\Omd\Ec$}

Let $\Ac$ and $\Bc$ be complete bornological algebras, $\Bct$ the unitalization of $\Bc$, and consider an unbounded bimodule $(\Ec,\rho,D)\in \Psi_*(\Ac,\Bct)$. One thus has $\Ec=\Hc\hotimes\Bct$ for some complete $\zz_2$-graded bornological space $\Hc$; $\rho$ is a bounded homomorphism from $\Ac$ to the even part of $\End_{\Bct}(\Ec)$ and $D:\Ec\to\Ec$ is an unbounded endomorphism with dense domain. Let $\Omtd\Bc$ be the unitalization of the complete DG algebra $\Omd\Bc$. Recall that the latter is the completion of $\Om \Bc$ with respect to the de Rham-Karoubi bornology. By definition the unit $1\in\Omtd\Bc$ verifies $d1=0$. Since $\Ec$ is a bornological right $\Bct$-module and $\Omtd\Bc$ a left $\Bct$-module, we can form the completed tensor product over $\Bct$:
\be
\Omd\Ec:= \Ec\hotimes_{\Bct}\Omtd\Bc\ .
\ee
Since $\Bct$ is unital, $\Omd\Ec$ identifies with $\Hc\hotimes \Omtd\Bc$. The space $\Omd\Ec$ is naturally a $\zz_2$-graded complete bornological right $\Omtd\Bc$-module, endowed with a (bounded) differential $d$ induced by $d(h\otimes\om):=(-)^{|h|}h\otimes d\om$ for any $h\otimes\om\in\Hc\hotimes\Omtd\Bc$. Here $|h|$ is the degree of the homogeneous element $h\in\Hc$.\\ 

Let $\Lc=\End_{\Omtd\Bc}(\Omd\Ec)$ be the $\zz_2$-graded complete bornological algebra of bounded endomorphisms of $\Omd\Ec$. It has a differential induced by the differential on $\Omd\Ec$: for any $x\in\Lc$, one has $dx=d\circ x-(-)^{|x|}x\circ d$. $\Lc$ has a unit $1_{\Lc}$ corresponding to the identity endomorphism of $\Omd\Ec$, satisfying $d1_{\Lc}=0$; hence $\Lc$ is a complete unital DG algebra. \\
Any endomorphism $y\in\End_{\Bct}(\Ec)$ gives rise to an endomorphism of $\Omd\Ec$ by
\be
y\cdot(\xi\otimes\om)=(y\cdot \xi)\otimes\om\quad\forall \xi\otimes\om\in \Ec\hotimes_{\Bct}\Omtd\Bc\ ,
\ee
whence a bounded homomorphism $\End_{\Bct}(\Ec)\to \Lc$. Composing $\rho$ with this map yields a bounded representation of $\Ac$ into $\Lc$, hence $\Omd\Ec$ becomes a complete bornological $\Ac$-$\Omtd\Bc$-bimodule. In the subsequent constructions we will need to extend the homomorphism $\rho:\Ac\to \Lc$ to the unitalization $\Act$ by setting $\rho(1_{\Act})=1_{\Lc}$. \\
Next, the endomorphism $D:\Ec\to\Ec$ being unbounded, it may fail to extend to the completion of $\Ec\otimes_{\Bct}\Omtd\Bc$. Therefore we have to assume that $D$ is a densely defined unbounded operator on $\Omd\Ec$, commuting with the right action of $\Omtd\Bc$. \\

\subsection{Trace-class endomorphisms}

Let $\Ec=\Hc\hotimes\Bct$ be a $\zz_2$-graded right $\Bct$-module, and consider the complete bornological space $\Ec'$ consisting of bounded right $\Bct$-module maps $\Ec\to\Bct$. In other words, the action of $\Ec'$ on $\Ec$ is given by a bounded bilinear braket $\langle\ ,\ \rangle:\Ec'\times\Ec\to\Bct$ satisfying
\be
\langle v,\xi b\rangle=\langle v,\xi\rangle b\quad \forall v\in\Ec'\ ,\ \xi\in\Ec\ ,\ b\in\Bct\ .
\ee
Then $\Ec'$ is naturally a left $\Bct$-module: for any $v\in\Ec'$ and $b\in\Bct$, the product $bv\in\Ec'$ is defined by $\langle bv,\xi\rangle =b\langle v,\xi\rangle$, $\forall \xi\in\Ec$. 
\begin{definition}
The algebra of \emph{trace-class} endomorphisms of $\Ec$ is the $\zz_2$-graded complete bornological algebra $\ell^1(\Ec)=\Ec\hotimes_{\Bct}\Ec'$.
\end{definition}
The product on $\ell^1(\Ec)$ is induced by the braket: 
\be
(\xi_1\otimes v_1)\cdot(\xi_2\otimes v_2)=\xi_1\langle v_1,\xi_2\rangle\otimes v_2\quad \forall \xi_i\in\Ec\ ,\ v_i\in\Ec'\ .
\ee
Also, $\ell^1(\Ec)$ acts on $\Ec$ by bounded endomorphisms: one has a bounded bilinear map $\ell^1(\Ec)\times \Ec \to \Ec$ sending $(\xi_1\otimes v, \xi_2)$ to $\xi_1\langle v,\xi_2\rangle$, compatible with the right action of $\Bct$, whence a canonical bounded homomorphism $\ell^1(\Ec)\to \End_{\Bct}(\Ec)$. This map may fail to be injective in general. Finally, $\ell^1(\Ec)$ is a $\End_{\Bct}(\Ec)$-bimodule, the left multiplication $\End_{\Bct}(\Ec)\times \ell^1(\Ec)\to\ell^1(\Ec)$ corresponding to
\be
(x,\xi\otimes v)\mapsto x(\xi)\otimes v\quad 
\ee
for any $x\in \End_{\Bct}(\Ec)$ , $ \xi\in\Ec$, $ v\in\Ec'$, and the right multiplication $\ell^1(\Ec)\times\End_{\Bct}(\Ec)\to\ell^1(\Ec)$ sends $(\xi\otimes v,x)$ to $\xi\otimes (v\circ x)$.\\

Let us now turn to \emph{partial supertraces}. We first define a map $\Tr:\Ec\otimes \Ec'\to\Bct$ by 
\be
\Tr((h\otimes b)\otimes v)=(-)^{|h||v|}b\langle v, h\otimes 1_{\Bct}\rangle\quad \forall \ h\otimes b\in\Ec\ ,\ v\in\Ec'\ .
\ee
Note the sign appearing in the r.h.s depending on the degrees of the graded elements $h\in\Hc$ and $v\in\Ec'$. Moreover this map is well-defined on $\Ec\otimes_{\Bct}\Ec'$ and bounded, hence it extends to a bounded map on the completion
\be
\Tr: \ell^1(\Ec)\to \Bct\ .
\ee
$\Tr$ is a partial supertrace on $\ell^1(\Ec)$ viewed as a $\End_{\Bct}(\Ec)$-bimodule. This means that, if we compose by the universal trace $\nat:\Bct\to\Bct_{\nat}=(\Bct/[\ ,\ ])_{\mathrm{completed}}$ (see appendix), the resulting bounded map $\nat\Tr:\ell^1(\Ec)\to \Bct_{\nat}$ vanishes on the supercommutators $[\ell^1(\Ec),\End_{\Bct}(\Ec)]$.\\

The same discussion holds for the right $\Omtd\Bc$-module $\Omd\Ec$, with some improvements due to the presence of the differential $d$. Here the space $(\Omd\Ec)'$ of bounded right $\Omtd\Bc$-module maps from $\Omd\Ec$ to $\Omtd\Bc$ is a graded (complete) left $\Omtd\Bc$-module, also with a differential: for $v\in(\Omd\Ec)'$, one puts $dv=d\circ v-(-)^{|v|}v\circ d$. Consequently the set of trace-class endomorphisms $\ell^1(\Omd\Ec):=\Omd\Ec\hotimes_{\Omtd\Bc}(\Omd\Ec)'$ is a DG algebra (in general non-unital), the differential corresponding to
\be
d(\xi\otimes v)=d\xi\otimes v+(-)^{|\xi|}\xi\otimes dv\quad\forall \xi\in\Omd\Ec\ ,\ v\in(\Omd\Ec)'\ ,
\ee
and it is well-defined on $\ell^1(\Omd\Ec)$ because $d(\xi\otimes\om v)=d(\xi\om\otimes v)$ for any $\om\in\Omtd\Bc$. It is also easy to show that the natural bounded map $\ell^1(\Omd\Ec)\to \End_{\Omtd\Bc}(\Omd\Ec)=\Lc$ is a DG algebra (and $\Lc$-bimodule) morphism, and that the partial supertrace $\Tr: \ell^1(\Omd\Ec)\to \Omtd\Bc$ commutes with $d$.

\subsection{Superconnections}

We are ready now to introduce the fondamental chain map $\chi$. Let $(\Ec,\rho,D)\in\Psi_*(\Ac,\Bct)$ be an unbounded bimodule. As above, $\Bbe(\Act)$ denotes the completed bar coalgebra of the unitalization of $\Ac$, with coproduct $\Delta:\Bbe(\Act)\to \Bbe(\Act)\hotimes\Bbe(\Act)$, and $\Lc=\End_{\Omtd\Bc}(\Omd\Ec)$ is the unital DG algebra of bounded endomorphisms on $\Omd\Ec$, with product $m:\Lc\hotimes\Lc\to\Lc$. The space of bounded linear maps
\be
\Rc=\hom(\Bbe(\Act),\Lc)
\ee
is a $\zz_2$-graded complete bornological algebra for the convolution product $fg=m\circ(f\otimes g)\circ\Delta$, $\forall f,g\in\Rc$. The differentials $d$ on $\Lc$ and $b'$ on $\Bbe(\Act)$ induce two anticommuting differentials on $\Rc$,
\be
df=d\circ f\ ,\quad \delta f=-(-)^{|f|}f\circ b'\ ,\quad \forall f\in\Rc\ ,
\ee
which moreover satisfy the Leibniz rule for the convolution product. Associated to the $\Bbe(\Act)$-bicomodule $\Om_1\Bbe(\Act)$ is the $\zz_2$-graded $\Rc$-bimodule (see appendix)
\be
\Mc=\hom(\Om_1\Bbe(\Act),\Lc)\ ,
\ee
the left and right multiplication maps being respectively given by $f\gamma=m\circ(f\otimes\gamma)\circ\Delta_l$ and $\gamma f=m\circ(\gamma\otimes f)\circ\Delta_r$, for any $f\in\Rc$, $\gamma\in\Mc$. $\Mc$ is also endowed with two anticommuting differentials $d\gamma=d\circ\gamma$ and $\delta\gamma=-(-)^{|\gamma|}\gamma\circ b''$, compatible with the $\Rc$-bimodule maps. Last but not least, the transposed of the canonical coderivation $\d:\Om_1\Bbe(\Act)\to\Bbe(\Act)$ yields a bounded derivation $\d:\Rc\to\Mc$, commuting with $d$ and $\delta$.\\

Let us now consider the bounded homomorphism $\rho:\Ac\to\End_{\Bct}(\Ec)$. We know that it gives rise to a bounded homomorphism from $\Ac$ to $\Lc$, and we extend it to $\Act$ by imposing $\rho(1_{\Act})=1_{\Lc}$. Next, since the projection $\Bb(\Act)\to \Bb_1(\Act)=\Act$ is clearly bounded for the entire bornology, it extends to a bounded map $\Bbe(\Act)\to \Act$. Then composing with $\rho:\Act\to\Lc$, we obtain a linear map of degree 1, which we also denote by $\rho$:
\be
\rho\in\hom(\Bbe(\Act),\Lc)=\Rc\ ,\quad |\rho|=1\ . 
\ee
We go back to the right $\Omtd\Bc$-module $\Omd\Ec$. The action of the algebra of endomorphisms $\Lc=\End_{\Omtd\Bc}(\Omd\Ec)$ yields a bounded map $m':\Lc\hotimes \Omd\Ec\to\Omd\Ec$. We form the left $\Rc$-module
\be
\Fc=\hom(\Bbe(\Act),\Omd\Ec)\ .
\ee
The module map $\Rc\times\Fc\to\Fc$ comes from the convolution product $f\cdot\xi=m'\circ(f\otimes\xi)\circ\Delta$, $\forall f\in\Rc$, $\xi\in\Fc$. It is immediate to check the compatibility of this action with the product on $\Rc$: the coassociativity of $\Bbe(\Act)$ implies $f\cdot(g\cdot\xi)=(fg)\cdot\xi$. The differentials $d$ on $\Omd\Ec$ and $b'$ on $\Bbe(\Act)$ imply as above that $\Fc$ is a bidifferential $\Rc$-bimodule: one has $d\xi=d\circ\xi$ and $\delta\xi=-(-)^{|\xi|}\xi\circ b'$ for any $\xi\in\Fc$, and these differentials are compatible with the ones on $\Rc$, i.e. $d(f\cdot\xi)=df\cdot\xi +(-)^{|f|}f\cdot d\xi$ and $\delta(f\cdot\xi)=\delta f\cdot\xi +(-)^{|f|}f\cdot \delta\xi$ for any $f\in\Rc$.\\
The last ingredient we have is the Dirac operator $D$ acting on $\Omd\Ec$ as an unbounded endomorphism of odd degree. We assume the induced unbounded linear map $D:\Fc\to\Fc$ has dense domain. Remark that if $D$ were bounded, it could be considered as a bounded homomorphism from $\cc=\Bb_0(\Act)$ to $\Lc$ and thus would define an element of $\Rc$. The map $D:\Fc\to\Fc$ would correpond to the left action of this element. The unboundedness of $D$ however prevents us to consider it as an element of $\Rc$.\\

We now introduce a \emph{superconnection} $\Dc:\Fc\to\Fc$
\be
\Dc=\delta -d +\rho+D\ ,
\ee
where $\rho$ is the odd element of $\Rc$ induced by the unital homomorphism $\rho:\Act\to\Lc$ and $D$ is the odd unbounded operator on $\Fc$. For any $\at_1,\at_2\in\Act$ one has $\delta\rho(\at_1,\at_2)=\rho b'(\at_1,\at_2)=\rho(\at_1\at_2)$ and $\rho^2(\at_1,\at_2)=-\rho(\at_1)\rho(\at_2)$. Since $\rho$ is an homomorphism, it follows that $\delta\rho+\rho^2=0$. Furthermore $D$ can be viewed as a 0-cochain on the bar complex, hence $\delta D=D\circ b'=0$. This implies that the curvature of $\Dc$ reads
\beq
\Dc^2&=&(\delta - d)(\rho +D)+(\rho+D)^2\non\\
&=& -d(\rho+D)+[D,\rho]+D^2\ .\label{curv}
\eeq
Because of the terms $D^2$ and $dD$, the curvature is an unbounded operator of even degree acting on $\Fc$. In the following, we want the \emph{heat kernel} $\exp (-t\Dc^2)$ to be a bounded operator on $\Fc$, hence defining an element of $\Rc$. We first have to precise the meaning of the heat kernel of $\Dc^2$. Using a Duhamel-type expansion, we write for any $t\in\rr_+$
\be
\exp(-t\Dc^2)=\sum_{n\ge 0}(-t)^n\int_{\Delta_n}ds_1\ldots ds_n\, e^{-ts_0D^2}\Te e^{-ts_1D^2}\ldots \Te e^{-ts_nD^2}\ ,\label{du}
\ee
where $\Delta_n$ is the $n$-simplex $\{(s_0,\ldots,s_n)\in [0,1]^{n+1}|\sum_i s_i=1\}$, and $\Te=-d(\rho+D)+[D,\rho]$. By hypothesis (definition \ref{dbim}), the heat kernel $\exp(-uD^2)$ is a bounded endomorphism of $\Ec$ for any $u\in\rr_+$, hence defines an even element of $\Rc$. Also, the commutator $[D,\rho]$ takes its values in $\End_{\Bct}(\Ec)$ and thus lies in $\Rc$. However $\Te$ may not be in $\Rc$ because $dD$ is not necessarily a bounded operator on $\Fc$. Therefore, we must impose $\exp(-uD^2)$ to be a \emph{regulator}, so that each term of the Duhamel expansion is bounded (hence in $\Rc$), and that the series itself converges bornologically. This is part of the content of the following definition.
\begin{definition}[Weak $\te$-summability]\label{dwte}
Let $\Ac$ and $\Bc$ be complete bornological algebras. An unbounded bimodule $(\Ec,\rho,D)\in\Psi_*(\Ac,\Bct)$ is called \emph{weakly $\te$-summable} iff the following conditions hold:\\
i) Boundedness condition: the  heat kernel $\exp(-t\Dc^2)$, given by the power series (\ref{du}), converges bornologically to an element of $\Rc=\hom(\Bbe(\Act),\Lc)$ for any $t\in\rr_+$.\\
ii) Trace-class condition: the natural homomorphism $\ell^1(\Omd\Ec)\to\Lc$ is injective, and for any $t>0$, the heat kernel lies in $\hom(\Bbe(\Act),\ell^1(\Omd\Ec))$.
\end{definition}

Recall there is a derivation of degree zero $\d:\Rc\to\Mc$. Thus for $\rho\in\Rc$, $\d\rho\in\Mc$ is odd. Assuming the $\te$-summability condition \ref{dwte}, we form the following odd element of $\Mc$:
\be
\mu= \int_0^1 dt\, e^{-t\Dc^2}\partial\rho \,e^{(t-1)\Dc^2}\ :\ \Om_1\Bbe(\Act)\to\ell^1(\Omd\Ec)\ .
\ee
Then composing $\mu$ by the cotrace $\nat:\Ome \Ac\to \Om_1\Bbe(\Act)$ yields an entire cochain on the $(b+B)$-complex of $\Ac$, namely $\mu\nat \in \hom(\Ome \Ac, \ell^1(\Omd\Ec))$.
\begin{proposition}\label{p41}
The bounded map $\mu\nat:\Ome \Ac\to \ell^1(\Omd\Ec)$ satisfies the Bianchi identity
\be
\mu\nat (b+B) +[\mu,\rho+D]\nat =d\mu\nat\ . \label{bian}
\ee
\end{proposition}
{\it Proof:} One has $\partial(\Dc^2)=\partial\Dc\Dc+\Dc\partial\Dc=[\Dc,\partial\rho]$, thus
$$
[\Dc,\mu]= \int_0^1dt\, e^{-t\Dc^2}[\Dc,\partial\rho] \,e^{(t-1)\Dc^2}
= \int_0^1dt\, e^{-t\Dc^2}\partial(\Dc^2) \,e^{(t-1)\Dc^2}
= -\partial e^{-\Dc^2}\ ,
$$
which is equivalent to $(\delta-d)\mu + [\rho+D,\mu]=-\partial\exp(-\Dc^2)$. Thus composing with the cotrace $\nat$, one gets (recall $\mu$ is odd)
$$
\delta\mu\nat+\partial e^{-\Dc^2}\nat+[\mu,\rho+D]\nat=d\mu\nat\ .
$$
Now from lemma \ref{A1} one has $\delta\mu\nat=\mu\nat b$. Moreover, $\exp(-t\Dc^2)$ is an even element of $\Rc$. It vanishes if one of its arguments is equal to the unit $1\in\Act$, because it involves the commutator $[D,\rho]$ and the differential $d\rho$. Thus lemma \ref{A2} implies
$$
\mu\nat B=\int_0^1dt\, e^{-t\Dc^2}\partial\rho \,e^{(t-1)\Dc^2}\nat B=\partial e^{-\Dc^2}\nat\ ,
$$
and the conclusion follows. \hfill\rule{1ex}{1ex}\\

The next step is to compose $\mu\nat$ with a partial supertrace $\tau$ on the superalgebra $\ell^1(\Omd\Ec)$. Of course we have the canonical map $\Tr:\ell^1(\Omd\Ec)\to\Omtd\Bc$, but what we need is a little bit more complicated, depending on the parity of the unbounded bimodule $(\Ec,\rho,D)\in\Psi_*(\Ac,\Bct)$:\\

\noindent a) $(\Ec,\rho,D)\in\Psi_0(\Ac,\Bct)$: we have $\Ec=\Hc\hotimes\Bct$ for a certain $\zz_2$-graded complete bornological vector space $\Hc=\Hc_+\oplus\Hc_-$, and $\Omd\Ec=\Hc\hotimes\Omtd\Bc$ is given its natural $\zz_2$-graduation. We simply take $\tau$ as the \emph{even} partial supertrace $\Tr$:
\beq
\tau=\Tr &:& \ell^1(\Omd\Ec) \to \Omtd\Bc \non\\
&&(h\otimes\om)\otimes v \mapsto \pm \om\langle v,h\otimes 1_{\Omtd\Bc}\rangle
\eeq
for any $h\in\Hc$, $\om\in\Omtd\Bc$, and $v\in(\Omd\Ec)'$. The sign $\pm$ depends on the parity of $|h|(|\om|+|v|)$. The bounded map $\tau\mu\nat:\Ome\Ac\to \Omtd\Bc$ is thus an {\it even} entire cochain, i.e. it sends an even (resp. odd) entire chain on $\Ac$ to an even (resp. odd) chain in the de Rham-Karoubi completion of forms over $\Bc$.\\

\noindent b) $(\Ec,\rho,D)\in\Psi_1(\Ac,\Bct)$: in this case, $\Ec=\Kc\hotimes C_1\hotimes\Bct$ for a {\it trivially graded} bornological space $\Kc$, and $\Omd\Ec=\Kc\hotimes C_1\hotimes\Omtd\Bc$. The operators $\rho$ and $D$, acting by endomorphisms on $\Omd\Ec$, commute with the right action of the Clifford algebra $C_1$. Since the map $\mu\nat:\Ome\Ac\to \ell^1(\Omd\Ec)$ is made out of $\rho$ and $D$, its range must lie in the trace-class endomorphisms of $\Omd\Ec$ commuting with $C_1$. This subalgebra of $\ell^1(\Omd\Ec)$ identifies with $\ell^1(\Kc\hotimes\Omtd\Bc)\hotimes C_1$. Therefore, we choose the partial supertrace $\tau:\ell^1(\Kc\hotimes\Omtd\Bc)\hotimes C_1\to \Omtd\Bc$ to be the tensor product of the canonical (even) partial supertrace $\Tr:\ell^1(\Kc\hotimes\Omtd\Bc)\to\Omtd\Bc$ with an {\it odd} supertrace $\zeta:C_1\to\cc$. The latter is {\it unique} up to a multiplication factor, because the universal supercommutator quotient space ${C_1}_{\nat}:=C_1/[C_1,C_1]$ is one-dimensional. This normalization factor can be fixed uniquely by imposing the compatibility of the bivariant Chern character with suspension an Bott periodicity. This is done in section \ref{bott}. One finds $\zeta(\eps)=\sqrt{2i}$ and $\zeta(1)=0$. Thus the partial supertrace $\tau$ is odd and reads
\beq
\tau=\Tr\otimes\zeta &:& \ell^1(\Kc\hotimes\Omtd\Bc)\hotimes C_1\to \Omtd\Bc \label{tr}\\
&& x+\eps y \mapsto \sqrt{2i} \Tr(y)\qquad \forall x,y\in\ell^1(\Kc\hotimes\Omtd\Bc)\ .\non
\eeq
The bounded map $\tau\mu\nat:\Ome\Ac\to \Omtd\Bc$ is then an {\it odd} entire cochain, i.e. it sends an even (resp. odd) entire chain on $\Ac$ to an odd (resp. even) chain on $\Bc$.\\

For any $n\ge 0$, let $p_n:\Om\Bc\to \Om^n\Bc$ be the natural projection. It is bounded for the de Rham-Karoubi bornology $\Sgd(\Om \Bc)$, hence extends to a bounded map on the (unital) completion $p_n:\Omtd\Bc\to\Om^n\Bc$. We denote by $\Om^1\Bc_{\nat}$ the completion of the commutator quotient space $\Om^1\Bc/[\Bc,\Om^1\Bc]$, and by $\nat:\Om^1\Bc\to\Om^1\Bc_{\nat}$ the bounded map induced by projection. We set $\chi_0=p_0\tau\mu\nat:\Ome\Ac\to \Bc$ and $\chi_1=\nat p_1\tau\mu\nat:\Ome\Ac\to \Om^1\Bc_{\nat}$. These are the components of a bounded map from the space of entire chains on $\Ac$ to the $X$-complex of $\Bc$:
\be
\chi(\Ec,\rho,D)\in\hom(\Ome\Ac,X(\Bc))\ .
\ee
$\chi$ has the same parity as the unbounded bimodule $(\Ec,\rho,D)$. The components $\chi_0$ and $\chi_1$ can be expressed explicitly through the map $\mu_0:\Om_1\Bbe(\Act)\to\ell^1(\Omd\Ec)$ defined by
\be
\mu_0=\int_0^1 dt\, e^{-t\te}\partial\rho \,e^{(t-1)\te}\ ,
\ee
with $\te=D^2+[D,\rho]$. The exponentials of $\te$ have a Duhamel expansion
\be
e^{-t\te}=\sum_{n\ge 0}(-t)^n\int_{\Delta_n} ds_1\ldots ds_n e^{-s_0tD^2}[D,\rho] e^{-s_1tD^2}\ldots [D,\rho] e^{-s_ntD^2}\ .
\ee
The map $\tau\mu_0\nat$ is valued in $\Bct$ and $\chi_0$ is its projection onto $\Bc$. On the other hand, the composition $\nat p_1\tau:\ell^1(\Omd\Ec)\to \Om^1\Bc_{\nat}$ is a trace, which implies that $\nat p_1\tau\cdot\nat$ is a trace on the $\Rc$-bimodule $\Mc$ (see appendix A4). One thus has
\be
\nat p_1\tau\mu\nat=\nat p_1\tau\int_0^1dt\,e^{-t\Dc^2}\partial\rho \,e^{(t-1)\Dc^2}\nat = \nat p_1\tau\partial\rho e^{-\Dc^2}\nat\ ,
\ee
hence
\be
\chi_1= \nat \tau\partial\rho \int_0^1 dt\, e^{-t\te} d(\rho+D)\,e^{(t-1)\te}\nat = \nat\tau\mu_0 d(\rho+D)\nat\ .
\ee
\begin{proposition}\label{p42}
$\chi$ is a morphism from the $(b+B)$-complex of entire chains on $\Ac$ to the $X$-complex of $\Bc$, i.e. $\chi_0\circ (b+B) = \pm\bb\circ\chi_1$ and $\chi_1\circ (b+B)= \pm\nat d\circ\chi_0$, the sign $\pm$ depending on the parity of $\chi$.
\end{proposition}
{\it Proof:} i) $\chi_0(b+B) = \pm\bb\chi_1$: Composing eq. (\ref{bian}) with the trace $\tau:\ell^1(\Omd\Ec)\to\Omtd\Bc$ yields the equality of linear maps $\tau\mu\nat(b+B)+\tau[\mu,\rho+D]\nat=\tau d\mu\nat$, and projecting the range of these maps onto $\Bct$, one gets 
$$
\tau\mu_0\nat(b+B)+\tau[\mu_0,\rho+D]\nat=0\ .
$$
We shall use the left and right bicomodule maps
\beq
\Delta_l &:& \Om_1\Bbe(\Ac)\to \Bbe(\Ac)\hotimes\Om_1\Bbe(\Ac)\ ,\non\\
\Delta_r &:& \Om_1\Bbe(\Ac)\to \Om_1\Bbe(\Ac)\hotimes\Bbe(\Ac)\ ,\non
\eeq
as well as the graded flip $\si:\Om_1\Bbe(\Ac)\hotimes\Bbe(\Ac)\rightleftarrows\Bbe(\Ac)\hotimes\Om_1\Bbe(\Ac)$ exchanging the two factors (with signs). The bounded map $\nat:\Ome\Ac\to\Om_1\Bbe(\Act)$ is a cotrace, which means $\si\Delta_r\nat =\Delta_l\nat$ and $\si\Delta_l\nat=\Delta_r\nat$. Let $m:\Lc\hotimes\Lc\to \Lc$ denote the multiplication and $\si':\Lc\otimes\Lc\rightleftarrows \Lc\otimes\Lc$ the graded flip. Then if we treat formally $dD$ as an element of $\Lc$, we can write
$$
\chi_1=\nat\tau\mu_0d(\rho+D)\nat=\nat\tau m(\mu_0\otimes d(\rho+D))\Delta_r\nat\ .
$$
Now let $x\in\ell^1(\Ec)$ and $y\in\End_{\Bct}(\Ec)$. The tracial properties of $\tau$ imply $\bb\nat\tau(xdy)=(-)^{|\tau|+|x|}\tau([x,y])$ or equivalently
$$
\bb\nat\tau(xdy)=(-)^{|\tau|+|x|}\tau m(x\otimes y-\si'(x\otimes y))\ .
$$
So we have
\beq
\bb\chi_1&=&-(-)^{|\tau|}\tau m(\mu_0\otimes (\rho+D)-\si'(\mu_0\otimes(\rho+D)))\Delta_r\nat \non\\
&=& -(-)^{|\tau|}\tau\mu_0(\rho+D)\nat +(-)^{|\tau|}\tau m\si'(\mu_0\otimes(\rho+D))\si^2\Delta_r\nat\non
\eeq
because $\si^2=\Id$, and since $\si'(\mu_0\otimes(\rho+D))\si=-(\rho+D)\otimes\mu_0$, one gets
\beq
\bb\chi_1&=&-(-)^{|\tau|}\tau\mu_0(\rho+D)\nat -(-)^{|\tau|}\tau m((\rho+D)\otimes\mu_0 )\Delta_l\nat\non\\
&=& -(-)^{|\tau|}\tau[\mu_0,\rho+D]\nat\ =\ (-)^{|\tau|}\tau\mu_0\nat(b+B)\ .\non
\eeq
But the range of $\bb$ lies in $\Bc\subset\Bct$, so that $\tau\mu_0\nat(b+B)$ takes its values in $\Bc$, hence $\tau\mu_0\nat(b+B)=\chi_0(b+B)$, whence the result.\\
ii) $\chi_1(b+B)= \pm\nat d\chi_0$: Projecting the range of eq. (\ref{bian}) onto $\Om^1\Bc_{\nat}$ yields $\nat\tau\mu_0d(\rho+D)\nat(b+B) =\nat\tau d\mu_0\nat$ because $\nat\tau\cdot\nat$ is a trace on $\Mc$. Furthermore, we know that the canonical supertrace $\Tr$ on $\ell^1(\Omd\Ec)$ commutes with the differential $d$, hence $d\tau=(-)^{|\tau|}\tau d$ and $\nat\tau d\mu_0\nat= (-)^{|\tau|}\nat d \tau\mu_0\nat$. The fact that $\nat d$ vanishes on the unit $1\in\Bct$ yields $\chi_1(b+B)=(-)^{|\tau|}\nat d\chi_0$ as required. \hfill\rule{1ex}{1ex}\\

The space of bounded linear maps $\hom(\Ome\Ac,X(\Bc))$ is a $\zz_2$-graded complete bornological complex, the differential of a map $\varphi$ corresponding to the graded commutator $(\nat d,\bb)\circ \varphi -(-)^{|\varphi|}\varphi\circ (b+B)$. Hence the cocycles of $\hom(\Ome\Ac,X(\Bc))$ are the bounded chain maps between $\Ome\Ac$ and $X(\Bc)$, and $\chi(\Ec,\rho,D)$ is a cocycle whose degree coincides with the parity of the $\te$-summable unbounded bimodule $(\Ec,\rho,D)\in\Psi_*(\Ac,\Bct)$.

\subsection{Homotopy invariance}

We have to show that the cohomology class of the cocycle $\chi\in\hom(\Ome\Ac,X(\Bc))$ is invariant under suitable homotopies on the set of $\te$-summable unbounded bimodules. From the construction above, it is clear that the correct notion of homotopy is obtained by {\it suspension}. Let $\cinf[0,1]$ be the algebra of smooth complex-valued functions on $[0,1]$, such that all derivatives of order $\ge 1$ vanish at the endpoints, while the functions themselves take arbitrary values at $0$ and $1$. We endow this algebra with the usual Fr\'echet topology, generated by the countable family of norms
\be
||f||_n=\sum_{i=0}^n\frac{1}{i!}\sup_{x\in[0,1]}|f^{(i)}(x)|\quad \forall f\in\cinf[0,1]\ ,\ n\in\nn\ .
\ee
The Fr\'echet topology generates the bounded bornology $\goth{Bound}(\cinf[0,1])$, a subset being small iff it is bounded for all norms. This turns $\cinf[0,1]$ into a {\it complete bornological algebra}. Given any complete bornological space $\Vc$, we define its suspension as the completed tensor product
\be
\Vc[0,1]:=\Vc\hotimes\cinf[0,1]\ .
\ee
For any $t\in[0,1]$, there is a bounded evaluation map $\ev_t:\cinf[0,1]\to\cc$ sending a function $f$ to its value $f(t)$. This extends for any complete bornological space $\Vc$ to a bounded evaluation map $\ev_t=\Id\hotimes \ev_t:\Vc[0,1]\to \Vc$.\\

Let $(\Ec,\rho,D)\in\Psi_*(\Ac,\Bct)$, with $\Ec=\Hc\hotimes\Bct$. The suspension $\Ec[0,1]$ is a right module over the complete algebra $\Bct[0,1]$, for the product
\be
(\xi\otimes f)\cdot(b\otimes g)=\xi b\otimes fg\quad \forall\ \xi\in\Ec\ ,\ b\in\Bct\ ,\ f,g\in\cinf[0,1]\ .
\ee
Consider the canonical bounded map $\iota:\Ec\to\Ec[0,1]$ given by $\iota(\xi)=\xi\otimes 1$ for any $\xi\in\Ec$, where $1$ stands for the constant function $1\in\cinf[0,1]$. Given any (possibly unbounded) endomorphism $Q:\Ec[0,1]\to\Ec[0,1]$ commuting with the right action of $\Bct[0,1]$, we define the evaluation of $Q$ at $t$ as the endomorphism $\ev_t(Q)$ of $\Ec$ corresponding to the composition
\be
{\ev}_t(Q)\ :\ \Ec \stackrel{\iota}{\to} \Ec[0,1]\stackrel{Q}{\longrightarrow} \Ec[0,1] \stackrel{\ev_t}{\longrightarrow} \Ec\ .
\ee
\begin{definition}\label{homo}
Let $\Ac$ and $\Bc$ be complete bornological algebras. Two unbounded bimodules $(\Ec_0,\rho_0,D_0)$ and $(\Ec_1,\rho_1,D_1)$ in $\Psi_*(\Ac,\Bct)$ are differentiably homotopic iff $\Ec_0=\Ec_1=\Hc\hotimes\Bct$ and there is an unbounded bimodule $(\Ec,\rho,D)\in\Psi_*(\Ac,\Bct[0,1])$ such that $\Ec=\Hc\hotimes\Bct[0,1]$ and $\ev_t(\rho)=\rho_t$, $\ev_t(D)=D_t$ for $t=0,1$. Differentiable homotopy is an equivalence relation. A similar definition holds for  $\te$-summable bimodules, where the interpolating bimodule $(\Ec,\rho,D)$ has also to be $\te$-summable.
\end{definition}

\begin{proposition}\label{p43}
Let $(\Ec,\rho,D)\in\Psi_*(\Ac,\Bct)$ be a $\te$-summable bimodule. The cohomology class of the cocycle $\chi(\Ec,\rho,D)$ in $H_*(\hom(\Ome \Ac,X(\Bc)))$ is invariant with respect to differentiable homotopies of $\te$-summable bimodules.
\end{proposition}
{\it Proof:} Let $(\Ec_0,\rho_0,D_0)$ and $(\Ec_1,\rho_1,D_1)$ be homotopic $\te$-summable unbounded $\Ac$-$\Bct$-modules. By definition there is an interpolating bimodule $(\Ec,\rho,D)\in\Psi_*(\Ac,\Bct[0,1])$. One thus has $\Ec=\Hc\hotimes\Bct[0,1]$ for a given complete bornological space $\Hc$. Let $\Om^*[0,1]$ be the (graded) commutative differential algebra of de Rham forms on $[0,1]$ with its Fr\'echet topology. We endow $\Om^*[0,1]$ with the bounded bornology, and note $d_t$ the (bounded) de Rham coboundary. Consider the unital complete bornological DG algebra $\Omtd\Bc\hotimes\Om^*[0,1]$, endowed with the total differential $d+d_t$. We shall mimic the construction of the map $\chi$ before, with the right $\Omtd\Bc\hotimes\Om^*[0,1]$-module
$$
\Omd\Ec:= \Hc\hotimes\Omtd\Bc\hotimes\Om^*[0,1]\ .
$$
Then $\rho$ and $D$ lift to endomorphisms of $\Omd\Ec$ as before, and we consider the superconnection $\Dc=\delta-(d+d_t)+\rho+D$ acting on $\Fc=\hom(\Bbe(\Act),\Omd\Ec)$. In this way, one gets a bounded map
$$
\mu= \int_0^1 dt\, e^{-t\Dc^2}\partial\rho \,e^{(t-1)\Dc^2}\ :\ \Om_1\Bbe(\Act)\to\ell^1(\Omd\Ec)\ .
$$
With the cotrace $\nat:\Ome\Ac\to\Om_1\Bbe(\Act)$ and the partial trace $\tau:\ell^1(\Omd\Ec)\to\Omtd\Bc\hotimes\Om^*[0,1]$, the analogue of proposition \ref{p41} yields
\be
\tau\mu\nat (b+B) +\tau[\mu,\rho+D]\nat =(-)^{|\tau|}(d+d_t)\tau\mu\nat\ .\label{ga}
\ee
For any $n\in\nn$ and $k=0,1$, we let $p_{n,k}$ be the natural bounded map from $\Omtd\Bc\hotimes\Om^*[0,1]$ to the tensor product $\Om^n\Bc\hotimes\Om^k[0,1]$. Composing equation (\ref{ga}) with $p_{0,k}$ implies
$$
p_{0,k}\tau\mu\nat (b+B) +p_{0,k}\tau[\mu,\rho+D]\nat =(-)^{|\tau|}p_{0,k}(d+d_t)\tau\mu\nat=(-)^{|\tau|}d_t p_{0,k-1}\tau\mu\nat\ .
$$
Next, the bounded map $\nat:\Om^1\Bc\hotimes\Om^k[0,1]\to\Om^1\Bc_{\nat}\hotimes\Om^k[0,1]$ is a trace because the algebra $\Om^*[0,1]$ is graded commutative. Thus composing (\ref{ga}) with $\nat p_{1,k}$ yields
$$ 
\nat p_{1,k}\tau\mu\nat(b+B)=(-)^{|\tau|}\nat p_{1,k}(d+d_t)\tau\mu\nat=(-)^{|\tau|}\nat d p_{0,k}\tau\mu\nat+(-)^{|\tau|}\nat d_t p_{1,k-1}\tau\mu\nat\ .
$$
Moreover, since $\Om^*[0,1]$ is graded commutative, the same computation as in the proof of proposition \ref{p42} shows that $p_{0,k}\tau[\mu,\rho+D]\nat=-(-)^{|\tau|}\bb\nat p_{1,k}\tau\mu\nat$, hence we get a couple of equations
$$
\left\{
\begin{array}{l}
 p_{0,k}\tau\mu\nat(b+B)-(-)^{|\tau|}\bb\nat p_{1,k}\tau\mu\nat = (-)^{|\tau|}d_t p_{0,k-1}\tau\mu\nat\ ,\\
 \nat p_{1,k}\tau\mu\nat(b+B)-(-)^{|\tau|}\nat d p_{0,k}\tau\mu\nat =(-)^{|\tau|} \nat d_t p_{1,k-1}\tau\mu\nat\ .
\end{array} \right.
$$
For $k=0$, we introduce the notations $\chi_0=p_{0,0}\tau\mu\nat$ and $\chi_1=\nat p_{1,0}\tau\mu\nat$. They are the components of a bounded map $\chi:\Ome\Ac\to X(\Bc)\hotimes\cinf[0,1]$, and the above equations yield the cocycle condition
$$
\left\{
\begin{array}{l}
\chi_0(b+B)-(-)^{|\tau|}\bb\chi_1=0\ ,\\
\chi_1(b+B)-(-)^{|\tau|}\nat d\chi_0=0\ .
\end{array}\right.
$$
If we compose $\chi$ with the evaluation map $\ev_t:X(\Bc)\hotimes\cinf[0,1]\to X(\Bc)$, we recover $\chi(\Ec_0,\rho_0,D_0)$ for $t=0$ and $\chi(\Ec_1,\rho_1,D_1)$ for $t=1$. Next, for $k=1$, define the Chern-Simons transgressions $cs_0=p_{0,1}\tau\mu\nat$ and $cs_1=\nat p_{1,1}\tau\mu\nat$. They form the components of a bounded map $cs:\Ome\Ac\to X(\Bc)\hotimes \Om^1[0,1]$, satisfying
\be
\left\{
\begin{array}{l}
cs_0(b+B)-(-)^{|\tau|}\bb\, cs_1=(-)^{|\tau|}d_t \chi_0\ ,\\
 cs_1(b+B)-(-)^{|\tau|}\nat d\, cs_0=(-)^{|\tau|}d_t\chi_1\ .
\end{array}
\right. \label{trans}
\ee
Now, remark that the integration map of one-forms $\int:\Om^1[0,1]\to \cc$ is bounded and extends to an integration map $\int :X(\Bc)\hotimes\Om^1[0,1]\to X(\Bc)$. Furthermore, for any $x\in X(\Bc)\hotimes\cinf[0,1]$, one has
$$
\int d_tx= {\ev}_1(x)-{\ev}_0(x)\in X(\Bc)\ .
$$
Thus integrating (\ref{trans}) shows that the difference $\chi(\Ec_0,\rho_0,D_0)-\chi(\Ec_1,\rho_1,D_1)$ is the coboundary of $\int cs$ in the complex $\hom(\Ome\Ac,X(\Bc))$, whence the result. \hfill\rule{1ex}{1ex}\\

Let us speak about functoriality. We know that for any complete bornological algebras $\Ac_1,\Ac_2,\Bc_1,\Bc_2$, there is a left product $\mor(\Ac_1,\Ac_2)\times \Psi_*(\Ac_2,\Bct_1)\to \Psi_*(\Ac_1,\Bct_1)$ given by $\varphi\cdot (\Ec,\rho,D)= (\Ec,\rho\circ\varphi, D)$ for any bounded homomorphism $\varphi:\Ac_1\to\Ac_2$, and a right product $\Psi_*(\Ac_2,\Bct_1)\times \mor(\Bct_1,\Bct_2)\to \Psi_*(\Ac_2,\Bct_2)$ given by $(\Ec,\rho,D)\cdot \psi= (\Ec\hotimes_{\psi}\Bct_2, \rho\otimes\Id, D\otimes\Id)$ where $\psi:\Bct_1\to\Bct_2$ is a unital bounded homomorphism. Using the explicit construction of $\chi$ in terms of $\rho$ and $D$, one sees that it is functorial, i.e. the following diagram commutes:
\be
\xymatrix{
\Ome\Ac_1 \ar[rr]^{\chi(\varphi\cdot(\Ec,\rho,D))} \ar[d]_{\Ome(\varphi)} & & X(\Bc_1)  \ar[d]^{X(\psi)} \\
\Ome\Ac_2 \ar[rr]_{\chi((\Ec,\rho,D)\cdot\psi)} \ar[urr]|{\chi(\Ec,\rho,D)} & & X(\Bc_2)}
\ee
We collect the preceding results in a theorem:
\begin{theorem}\label{tchi}
Let $\Ac$ and $\Bc$ be complete bornological algebras. To any unbounded $\te$-summable bimodule $(\Ec,\rho,D)\in\Psi_*(\Ac,\Bct)$, we associate a bounded chain map $\chi(\Ec,\rho,D) :\Ome\Ac\to X(\Bc)$ of the same parity. Its associated cohomology class in $H_*(\Ome\Ac,X(\Bc))$ is invariant under differentiable homotopies and functorial in $\Ac$ and $\Bc$. \hfill\rule{1ex}{1ex}
\end{theorem}
\begin{remark}\label{r41}\textup{
In particular if $D=0$ then $\mu_0=\d\rho$ and the two components of $\chi(\Ec,\rho,0)$ reduce to $\chi_0 = p\tau\partial\rho\nat$, where $p$ is the projection $\Bct\to\Bc$, and $\chi_1=\nat \tau\partial\rho d\rho\nat$. One sees that $\chi_0$ and $\chi_1$ are respectively a zero-cochain and a one-cochain on the $(b+B)$-complex of entire chains over $\Ac$, explicitly 
\be
\chi_0(a)=p\tau\rho(a)\ ,\qquad \chi_1(a_0da_1)=\nat\tau \rho(a_0)d\rho(a_1)\ ,
\ee
for any $a,a_0,a_1\in\Ac$. }
\end{remark}
\begin{remark}\textup{
For any $\te$-summable unbounded bimodule, the composition of $\tau\mu\nat$ by the universal trace $\nat:\Omtd\Bc\to\Omtd\Bc_{\nat}$, yields a bounded map from the $(b+B)$-complex of entire chains over $\Ac$, to the (unitalized) entire de Rham-Karoubi complex of $\Bc$:
\be
\nat\tau\mu\nat\in \hom(\Ome\Ac,\Omtd\Bc_{\nat})\ .
\ee
Proposition \ref{p41} implies that it is in fact a cocycle: $\nat \tau\mu\nat(b+B)-(-)^{|\tau|}\nat d\tau\mu\nat=0$. This cocycle was considered in a dual context for example in \cite{Go} (without superconnection and in periodic theory rather than in entire cyclic theory), and we know that it can be adapted to compute the action of unbounded representatives of $K\!K(\Ac,\Bc)$ classes on cyclic cocycles over $\Bc$ (see \cite{C1} p. 434). }
\end{remark}
\begin{remark}\textup{
In fact the construction of $\chi$ works as well if $\rho:\Ac\to\End_{\Bct}(\Ec)$ is simply a bounded linear map and not necessarily an homomorphism. In that case, the curvature $\delta\rho+\rho^2$ of $\rho$ does not vanish and has to be included in formula (\ref{curv}). One thus obtains some generalisations of the cocycles constructed by Quillen in \cite{Q2,Q3}. In the sequel we will always consider that $\rho$ is an homomorphism.}
\end{remark}

\section{The bivariant Chern character}\label{biv}

We are ready to construct a bivariant Chern character for unbounded bimodules satisfying some strong $\te$-summability conditions. Given two complete bornological algebras $\Ac$ and $\Bc$, our goal is to lift an element $(\Ec,\rho,D)\in\Psi_*(\Ac,\Bct)$ to a $\Tc\Ac$-$\Tct\Bc$-bimodule and construct the corresponding bounded chain map $\chi:\Ome\Tc\Ac\to X(\Tc\Bc)$. Then composing $\chi$ with the homotopy equivalence $\gamma:X(\Tc\Ac)\to\Ome\Tc\Ac$ given by the Goodwillie theorem, we obtain a bounded chain map $\chi\gamma\in\hom(X(\Tc\Ac),X(\Tc\Bc))$ whose class in the bivariant entire cyclic cohomology $HE_*(\Ac,\Bc)$ is the bivariant Chern character of $(\Ec,\rho,D)$. \\

So we fix an unbounded bimodule $(\Ec,\rho,D)\in\Psi_*(\Ac,\Bct)$, $\Ec=\Hc\hotimes\Bct$. Recall that $\Oman\Bc$ is the completion of $\Om \Bc$ for the analytic bornology $\Sgan(\Om\Bc)$ generated by the sets $\bigcup_{n\ge 0}\St (dS)^n$, for any $S\in\Sg(\Bc)$. It is a complete bornological DG algebra for the product of forms and the differential $d$. We let $\Omtan\Bc$ be its unitalization, with $d1=0$. Since the latter is a left $\Bct$-module, we can introduce the analytic right $\Omtan\Bc$-module and its even/odd form part:
\be
\Oman\Ec:= \Ec\hotimes_{\Bct}\Omtan\Bc\ ,\qquad \Oman^{\pm}\Ec:=\Ec\hotimes_{\Bct}\Omtan^{\pm}\Bc\ .
\ee
As a complete $\zz_2$-graded bornological vector space, $\Oman\Ec$ is isomorphic to $\Hc\hotimes\Omtan\Bc$, and has naturally a bounded differential induced by $d$. Now endow the subspace $\Omtan^+\Bc$ of {\it even} forms with the (bounded) Fedosov product
\be
\om_1\odot\om_2=\om_1\om_2-d\om_1d\om_2\quad \forall \om_1,\om_2\in\Omtan^+\Bc\ .
\ee
The unit $1\in\Omtan^+\Bc$ is also the unit for the Fedosov product, and the correspondence (\ref{cor}) shows that the associative algebra $(\Omtan^+\Bc,\odot)$ is isomorphic to the unitalized analytic tensor algebra $\Tct\Bc$. Then, we can endow the $\zz_2$-graded space $\Oman^+\Ec=\Hc\hotimes\Omtan^+\Bc$ with a right action of this Fedosov algebra $(\Omtan^+\Bc,\odot)\simeq\Tct\Bc$:
\beq
\odot&:& \Oman^+\Ec\times \Omtan^+\Bc \to \Oman^+\Ec \\
&& (\xi,\om) \mapsto \xi\odot\om:= \xi\om- (-)^{|\xi|}d\xi d\om\ ,\non
\eeq
where $d\xi\in \Oman^-\Ec$ and $d\om\in\Oman^-\Bc$. It is easy to check that $\Oman^+\Ec$ is a right $\Tct\Bc$-module: $(\xi\odot\om_1)\odot \om_2=\xi\odot(\om_1\odot\om_2)$ for any $\om_1,\om_2\in\Omtan^+\Bc=\Tct\Bc$, and as such it is isomorphic to the tensor product $\Hc\hotimes\Tct\Bc$. Hence we have just lifted the $\Bct$-module $\Ec$ to a $\Tct\Bc$-module.\\

Let $\End_{\Tct\Bc}(\Oman^+\Ec)$ be the complete bornological algebra of bounded endomorphisms of $\Oman^+\Ec$ commuting with $\Tct\Bc$. From the left representation $\rho:\Ac\to\End_{\Bct}(\Ec)$, we want to construct a bounded homomorphism $\rho_*:\Tc\Ac\to\End_{\Tct\Bc}(\Oman^+\Ec)$. First, we have a bounded linear map (not an homomorphism) $\rho_*:\Ac\to\End_{\Tct\Bc}(\Oman^+\Ec)$ given by a Fedosov-type action:
\be
\rho_*(a)\odot\xi:= \rho(a)\xi- d\rho(a) d\xi\ ,\quad \forall a\in\Ac\ ,\ \xi\in\Oman^+\Ec\ ,
\ee
where $\rho(a)$ and $d\rho(a)$ are viewed as elements of the DG algebra $\End_{\Omtan\Bc}(\Oman\Ec)$, while $\xi,d\xi$ are elements of $\Oman\Ec$. One has $\rho_*(a)\odot(\xi\odot\om)=(\rho_*(a)\odot\xi)\odot\om$ for any $\om\in\Tct\Bc$, hence $\rho_*(a)$ is indeed an endomorphism of $\Oman^+\Ec$. This induces a representation (=homomorphism) of the non-completed tensor algebra $\rho_*:T\Ac\to\End_{\Tct\Bc}(\Oman^+\Ec)$ by
\be
\rho_*(a_1\otimes\ldots\otimes a_n)\odot\xi=\rho_*(a_1)\odot\ldots\odot\rho_*(a_n)\odot\xi\ ,\quad \forall a_1\otimes\ldots\otimes a_n\in T\Ac\ ,\ \xi\in\Oman^+\Ec\ .
\ee
Under the identification $T\Ac\simeq(\Om^+\Ac,\odot)$, the above action reads
\beq
\lefteqn{\rho_*(a_0da_1\ldots da_{2n})\odot\xi=(\rho(a_0)d\rho(a_1)\ldots d\rho(a_{2n}))\odot\xi}\\
&&\qquad\qquad =\rho(a_0)d\rho(a_1)\ldots d\rho(a_{2n})\xi- d\rho(a_0)d\rho(a_1)\ldots d\rho(a_{2n})d\xi\non
\eeq
for any $a_0da_1\ldots da_{2n}\in\Om^+\Ac$, where $\rho(a_0)d\rho(a_1)\ldots d\rho(a_{2n})$ is viewed as an element of the DG algebra $\End_{\Omtan\Bc}(\Oman\Ec)$. In general, {\it we don't know if the representation $\rho_*$ is bounded} for the analytic bornology on $T\Ac$. This is true, for example, when $\Hc$ is a Banach space endowed with the bounded bornology. In the following, we always assume that $\rho_*$ is bounded (this will be part of the strong $\te$-summability assumption below), and consequently extends to the desired bounded representation of the completion $\Tc\Ac$ in $\End_{\Tct\Bc}(\Oman^+\Ec)$.\\

Let us now deal with the Dirac operator $D$. It is an odd, unbounded endomorphism of $\Ec$, an extends to an unbounded endomorphism (with dense domain) of the right $\Tct\Bc$-module $\Oman\Ec$. Once again, we deform its restriction on $\Oman^+\Ec$ into a Fedosov-type action:
\be
D\odot\xi:= D\xi+dD d\xi\ ,\quad \forall \xi\in\Oman^+\Ec\ ,\label{defo}
\ee
so that $D\odot(\xi\odot\om)=(D\odot\xi)\odot\om$ for any $\om\in\Omtan^+\Bc$. In this way, $D$ defines an unbounded endomorphism of the right $\Tct\Bc$-module $\Oman^+\Ec$. Note that the sign $+$ in front of $dD$ in eq. (\ref{defo}) is due to the odd degree of $D$.\\

What we have obtained so far is the following. Starting from a bimodule $(\Ec,\rho,D)\in\Psi_*(\Ac,\Bct)$, with $\Ec=\Hc\hotimes\Bct$, we constructed the right $\Tct\Bc$-module $\Oman^+\Ec=\Hc\hotimes\Tct\Bc$, endowed with a bounded left representation $\rho_*:\Tc\Ac\to\End_{\Tct\Bc}(\Oman^+\Ec)$, and with an odd, unbounded Dirac endomorphism $D$. It is natural to wonder if this lifted bimodule defines an element of $\Psi_*(\Tc\Ac,\Tct\Bc)$. In general this may be false, because of the following reasons:\\
a) For any $x\in\Tc\Ac$, the commutator for the Fedosov action $[D,\rho_*(x)]_{\odot}:=D\odot\rho_*(x)-\rho_*(x)\odot D$ may not act by a {\it bounded} endomorphism on $\Oman^+\Ec$.\\
b) The heat kernel {\it for the Fedosov product}, given by the formal power series
\be
\exp_{\odot}(-t D^{\odot 2}):=\sum_{n\ge 0}\frac{(-t)^n}{n!}(D^{\odot 2})^{\odot n}
\ee
may not be a bounded endomorphism of $\Oman^+\Ec$. In order to understand what we mean exactly by this exponential, we state the following lemma:
\begin{lemma}\label{lexp}
Let $H$ be any (possibly unbounded) even endomorphism of $\Oman^+\Ec$ acting by a Fedosov-type deformation. Then the formal power series $\sum_n\frac{1}{n!}H^{\odot n}$ may be rewritten as a Duhamel-type expansion
\be
\exp_{\odot}H= \sum_{n\ge 0}(-)^n\int_{\Delta_n}ds_1\ldots ds_n \, e^{s_0H}dHd(e^{s_1H})dH\ldots d(e^{s_{n-1}H})dHd(e^{s_nH})\ ,\label{exp}
\ee
where $e^{s_iH}$ is the exponential for the usual product of endomorphisms. We call $\exp_{\odot}H$ the \emph{Fedosov exponential} of $H$.
\end{lemma}
{\it Proof:} We establish a first-order differential equation for the Fedosov exponential. For any $t\in[0,1]$ and $\xi\in\Oman^+\Ec$, one has
$$
\frac{d}{dt}(\exp_{\odot}(tH)\odot\xi)=H\odot\exp_{\odot}(tH)\odot\xi=(H\exp_{\odot}(tH)-dHd\exp_{\odot}(tH))\odot\xi\ .
$$
Thus $\frac{d}{dt}\exp_{\odot}(tH)=H\exp_{\odot}(tH)-dHd\exp_{\odot}(tH)$. A well-known trick of perturbative quantum mechanics is to introduce the \emph{interaction scheme} $I(t):=e^{-tH}\exp_{\odot}(tH)$, for which the evolution equation reads
$$
\frac{d}{dt}I(t)=-e^{-tH}dHd\exp_{\odot}(tH)\ .
$$
Using the fact that $I(0)=1$, the solution is expressed in integral form
$$
I(t)=1-\int_0^tds\, e^{-sH}dHd\exp_{\odot}(sH)\ ,
$$
or equivalently
$$
\exp_{\odot}(tH)=e^{tH}-\int_0^tds \, e^{(t-s)H}dHd\exp_{\odot}(sH)\ .
$$
The perturbative resolution of this equation gives rise to the result.\hfill\rule{1ex}{1ex}\\

Substituting $H$ by $-tD^{\odot 2}=-t(D^2+dDdD)$ in (\ref{exp}) gives a power series of differential forms involving the heat operator $\exp(-sD^2)$, which by hypothesis is a bounded endomorphism of $\Ec$ playing the role of a regulator, together with some derivatives $d\exp(-uD^2)$, $d(D^2)$ and $dDdD$. The obtained formula is really the definition of $\exp_{\odot}(-tD^{\odot 2})$. The bornological convergence of this series in $\End_{\Tct\Bc}(\Oman^+\Ec)$ is part of the strong $\te$-summability assumption below. This being understood, we can perform the construction of the previous section with $\Ac$, $\Bct$, $\Ec$ replaced by $\Tc\Ac$, $\Tct\Bc$, $\Oman^+\Ec$ respectively, and get a bounded chain map
\be
\chi(\Oman^+\Ec,\rho_*,D): \Ome\Tc\Ac\to X(\Tc\Bc)\ .
\ee
Let us recall briefly the main steps. We first form the right $\Omtd\Tc\Bc$-module $\Omd\Oman^+\Ec$, and denote by $(\Lc,\dd)$ the DG algebra $\End_{\Omtd\Tc\Bc}(\Omd\Oman^+\Ec)$. Then, using the completed bar complex $\Bbe(\Tct\Ac)$ and its associated bicomodule $\Om_1\Bbe(\Tct\Ac)$, we consider the algebra $\Rc=\hom(\Bbe(\Tct\Ac),\Lc)$ and its associated $\Rc$-bimodule $\Mc=\hom(\Om_1\Bbe(\Tct\Ac),\Lc)$; then the left $\Rc$-module $\Fc=\hom(\Bbe(\Tct\Ac),\Omd\Oman^+\Ec)$ endowed with two differentials $\dd,\delta$; and finally the superconnection $\Dc=\delta-\dd+\rho_*+D:\Fc\to\Fc$. Since we want the heat operator $\exp(-t\Dc^2)$ to define a trace-class element of $\Rc$, the lifted $\Tc\Ac$-$\Tct\Bc$-bimodule $(\Oman^+\Ec,\rho_*,D)$ must be weakly $\te$-summable. This leads to the strong version of $\te$-summability:
\begin{definition}[Strong $\te$-summability]\label{dste}
Let $\Ac$ and $\Bc$ be complete bornological algebras. An unbounded bimodule $(\Ec,\rho,D)\in\Psi_*(\Ac,\Bct)$ is called \emph{strongly $\te$-summable} iff the following conditions hold:\\
i) The homomorphism $\rho_*:T\Ac\to \End_{\Tct\Bc}(\Oman^+\Ec)$ is bounded for the analytic bornology on $T\Ac$, and thus extends to a bounded representation of $\Tc\Ac$ into $\End_{\Tct\Bc}(\Oman^+\Ec)$. This turns the lift $(\Oman^+\Ec,\rho_*,D)$ into a $\Tc\Ac$-$\Tct\Bc$-bimodule.\\
ii) The thus obtained lift $(\Oman^+\Ec,\rho_*,D)$, though not necessarily in $\Psi_*(\Tc\Ac,\Tct\Bc)$, nevertheless verifies the weak $\te$-summability condition as stated in definition \ref{dwte}.\\ 
We denote by $\Psi_*^{\te}(\Ac,\Bct)$ the abelian semigroup of strongly $\te$-summable bimodules. Two strongly $\te$-summable bimodules are homotopic iff their lifts are homotopic.
\end{definition} 
If the strong $\te$-summability conditions are satisfied, then from the odd element of $\Mc$
\be
\mu=\int_0^1 dt\, e^{-t\Dc^2}\d\rho e^{(t-1)\Dc^2}\in\hom(\Om_1\Bbe(\Tct\Ac),\ell^1(\Omd\Oman^+\Ec))\ ,
\ee
one gets the two components $\chi_0=p_0\tau\mu\nat$ and $\chi_1=\nat p_1\tau\mu\nat$ of the cocycle $\chi(\Oman^+\Ec,\rho_*,D)\in\hom(\Ome\Tc\Ac,X(\Tc\Bc))$. Then composing $\chi$ with the Goodwillie equivalence $\gamma\in\hom(X(\Tc\Ac),\Ome\Tc\Ac)$ of section \ref{good} yields a bivariant entire cyclic cohomology class $[\chi\circ\gamma]\in HE_*(\Ac,\Bc)$. This is the bivariant Chern character of $(\Ec,\rho,D)$. Thus we are led to the following theorem:
\begin{theorem}
Let $\Ac$ and $\Bc$ be complete bornological algebras. There is a bivariant Chern character map 
\be
\ch: \Psi_*^{\te}(\Ac,\Bct)\to HE_*(\Ac,\Bc)\ ,\quad *=0,1\ ,
\ee
sending a strongly $\te$-summable bimodule $(\Ec,\rho,D)$ to the bivariant entire cyclic cohomology class $\ch(\Ec,\rho,D):=[\chi(\Oman^+\Ec,\rho_*,D)\circ\gamma]$. The Chern character is additive, invariant for differentiable homotopies inside $\Psi_*^{\te}(\Ac,\Bct)$, and functorial in both variables.
\end{theorem}
{\it Proof:} This is a consequence of theorem \ref{tchi} applied to the lifted bimodule $(\Oman^+\Ec,\rho_*,D)$ and the fact that the Goodwillie map $\gamma$ is obviously functorial with respect to $\Ac$.\hfill\rule{1ex}{1ex}\\
\begin{remark}\label{r51}\textup{
The strong $\te$-summability condition \ref{dste} should not be taken too seriously. In concrete applications, it is sufficient to verify {\it a posteriori} that the composition map $\chi\circ\gamma:X(\Tc\Ac)\to X(\Tc\Bc)$ is bounded. On the other hand, when dealing with commutative algebras, one can replace the universal DG algebra of noncommutative forms $\Om \Bc$ by the smaller (graded) commutative algebra of de Rham forms over $\Bc$, and similarly for the module $\Om\Ec$. In these circumstances, the $\te$-summability conditions are much less restrictive, and give rise to a bivariant Chern character in ordinary de Rham cohomology, which is satisfactory in many concrete geometrical examples. Also, in some situations it is not necessary to unitalize the algebra $\Bc$, and the construction of the Chern character can be performed on $\Psi_*^{\te}(\Ac,\Bc)$. The following section provides commutative examples illustrating these particular cases with the study of Bott elements.}
\end{remark}

\section{Examples}\label{ex}

Let us have a look at some examples related to $K$-theory. It will illustrate our bivariant Chern character on the two extremal cases $\Psi_*(\cc,\Ac)$ and $\Psi_*(\Ac,\cc)$, describing respectively the $K$-theory and $K$-homology of a complete bornological algebra $\Ac$.

\subsection{Index pairing and the JLO cocycle}

Let $\Ac$ and $\Bc$ be complete bornological algebras. First of all, a bounded homomorphism $\rho:\Ac\to\Bc\subset \Bct$ must be considered as the fundamental example of even $\Ac$-$\Bct$-bimodule. In this case one chooses $\Ec=\Bct$ with trivial graduation, and the Dirac operator is equal to zero, hence we get an unbounded bimodule $(\Bct,\rho,0)\in \Psi_0(\Ac,\Bct)$. 
\begin{proposition}
For any bounded homomorphism $\rho:\Ac\to\Bc$, the bivariant Chern character of $(\Bct,\rho,0)\in \Psi_0(\Ac,\Bct)$ is equal to the class $\ch(\rho)\in HE_0(\Ac,\Bc)$ of the chain map
\be
X(\rho_*): X(\Tc\Ac)\to X(\Tc\Bc)
\ee
induced by the bounded homomorphism $\rho_*:\Tc\Ac\to\Tc\Bc$.
\end{proposition}
{\it Proof:} $\ch(\Bct,\rho,0)$ is the cohomology class of the composition of chain maps $X(\Tc\Ac)\stackrel{\gamma}{\to}\Ome\Tc\Ac\stackrel{\chi}{\to} X(\Tc\Bc)$, where $\chi$ is constructed as follows. With $\Ec=\Bct$ one has $\Oman^+\Ec=\Bct\hotimes_{\Bct}\Omtan^+\Bc\simeq\Tct\Bc$. Furthermore, the homomorphism $\rho_*:\Tc\Ac\to \End_{\Tct\Bc}(\Oman^+\Ec)\simeq\Tct\Bc$ is simply given by $\rho_*(a_1\otimes\ldots\otimes a_n)=\rho(a_1)\otimes\ldots\otimes\rho(a_n)$. Hence the lift of $(\Bct,\rho,0)$ corresponds to the $\Tc\Ac$-$\Tct\Bc$-bimodule $(\Tct\Bc,\rho_*,0)$. Thus by remark \ref{r41}, the two components of the morphism $\chi:\Ome\Tc\Ac\to X(\Tc\Bc)$ are respectively a $0$-cochain and a $1$-cochain on the $(b+B)$-complex of universal forms over $\Tc\Ac$:
$$
\chi_0(x)=\rho_*(x)\ ,\forall x\in\Tc\Ac\ ,\quad\chi_1(x\dd y)=\nat\rho_*(x)\dd\rho_*(y)\ ,\forall x\dd y\in\Om^1\Tc\Ac\ .
$$
Hence $\chi$ vanishes on any differential form over $\Tc\Ac$ of degree $\ge 2$. From the explicit expression of the Goodwillie equivalence $\gamma$, we can compute easily the composition $\chi\gamma: X(\Tc\Ac)\to X(\Tc\Bc)$:
\beq
&&x\stackrel{\gamma}{\longmapsto}x+\mathrm{degree\ge 2} \stackrel{\chi_0}{\longmapsto} \rho_*(x)\ ,\non\\
&&\nat x\dd y\stackrel{\gamma}{\longmapsto}x\dd y+b(x\phi(y))+\mathrm{degree\ge 3}\stackrel{\chi_1}{\longmapsto}\nat \rho_*(x)\dd\rho_*(y)\ ,\non
\eeq
for any $x,y\in\Tc\Ac$. This is precisely the morphism of complexes $X(\rho_*)$.\hfill\rule{1ex}{1ex}\\

We focus on the algebra $\cc$. One knows \cite{Me} that $X(\Tc\cc)$ is homotopic to $X(\cc):\cc\rightleftarrows 0$. The generator of $HE_0(\cc)=\cc$ is represented by the following even cycle $\hat{e}\in X_0(\Tc\cc)\simeq\Oman^+\cc$. Denoting by $e$ the unit of $\cc$, then
\be
\hat{e}:=e+\sum_{n\ge 1}\frac{(2n)!}{(n!)^2}(e-\frac{1}{2})(dede)^n\quad\in\Oman^+\cc=\Tc\cc
\ee
is an idempotent: $\hat{e}^2=\hat{e}$ in $\Tc\cc$, which implies $\nat \dd \hat{e}=0$. Thus $\hat{e}$ indeed defines an even cycle of $X(\Tc\cc)$.\\
Now let $\Ac$ be a complete bornological algebra, and fix an integer $N\in\nn$. Let $\Hc=\Hc_+\oplus\Hc_-$ be the $\zz_2$-graded complete bornological space such that $\Hc_+=\Hc_-=\cc^N$, an consider the right $\Act$-module $\Ec=\Hc\hotimes\Act$. The algebra of endomorphisms $\End_{\Act}(\Ec)$ identifies with the $\zz_2$-graded matrix algebra $M_2(M_N(\Act))$, the graduation corresponding to the decomposition into diagonal/off-diagonal matrices as usual. Any pair of idempotents $e_{\pm}=1+u_{\pm}\in M_N(\Act)$, with $u_{\pm}\in M_N(\Ac)$, is described by an even bounded homomorphism $\rho=\left( \begin{array}{cc}
          \rho_+ & 0 \\
          0 & \rho_- \\
     \end{array} \right)$ from $\cc$ to $M_2(M_N(\Act))$, with $\rho_{\pm}(e)=e_{\pm}$, $e\in\cc$. This gives rise to a bimodule $(\Ec,\rho,0)\in\Psi_0(\cc,\Act)$. The pair $e_{\pm}$ is called degenerate if $e_+=e_-$. The set of (differentiable) homotopy classes of such bimodules modulo degenerates is the $K$-theory group $K_0(\Ac)$. The bivariant Chern character yields a well-defined additive map $K_0(\Ac)\to HE_0(\cc,\Ac)\simeq HE_0(\Ac)$ which coincides with the usual Chern character on $K$-theory \cite{CQ1}:
\begin{proposition}
Let $\Ac$ be a complete bornological algebra, $e_{\pm}=1+u_{\pm}$ a pair of idempotents with $u_{\pm}\in M_N(\Ac)$, and $(\Ec,\rho,0)$ the corresponding bimodule in $\Psi_0(\cc,\Act)$ representing the $K$-theory element $[e_+]-[e_-]\in K_0(\Ac)$. Then the Chern character $\ch(\Ec,\rho,0)\in HE_0(\Ac)$ is represented by the entire chain $\ch(e_+)-\ch(e_-)\in\Tc\Ac$, with
\be
\ch(e_{\pm})=\tr (e_{\pm})+\sum_{n\ge 1}\frac{(2n)!}{(n!)^2}\tr((e_{\pm}-\frac{1}{2})(de_{\pm}de_{\pm})^n)\quad\in\Oman^+\Ac=\Tc\Ac\ ,
\ee
where $\tr$ is the usual trace on $N\times N$ matrices. The difference $\ch(e_+)-\ch(e_-)$ is well-defined on $K_0(\Ac)$ because it vanishes on degenerates and the bivariant Chern character is homotopy invariant.
\end{proposition}
{\it Proof:} One has $\Ec=\Hc\hotimes\Act$ with $\Hc=\cc^N\oplus\cc^N$. Thus $\Oman^+\Ec$ is equal to $\Hc\hotimes\Tct\Ac$ and the homomorphism $\rho:\cc\to\End_{\Act}(\Ec)=M_{2N}(\Act)$ lifts to an homomorphism $\rho_*:\Tc\cc\to \End_{\Tct\Ac}(\Oman^+\Ec)=M_{2N}(\Tct\Ac)$. Then by remark \ref{r41}, $\ch(\Ec,\rho,0)$ is the image of the generator $\hat{e}\in\Tc\cc$ under the composition of bounded maps
$$
\Tc\cc\stackrel{\rho_*}{\longrightarrow}M_{2N}(\Tct\Ac)\stackrel{{\tr}_s}{\longrightarrow}\Tct\Ac\stackrel{p}{\longrightarrow} \Tc\Ac\ ,
$$
where $\tr_s$ is the usual supertrace on supersymmetric $2N\times 2N$ matrices, and $p$ is the projection. Since $e_{\pm}=1+u_{\pm}$ with $u_{\pm}\in M_N(\Ac)$, we have $\tr_s(\rho(e))=\tr(e_+)-\tr(e_-)=\tr(u_+)-\tr(u_-)\in\Ac$, hence
\beq
\ch(\Ec,\rho,0)&=&{\tr}_s( \rho(e))+\sum_{n\ge 1}\frac{(2n)!}{(n!)^2}{\tr}_s((\rho(e)-\frac{1}{2})(d\rho(e)d\rho(e))^n)\non\\
&=&\ch(e_+)-\ch(e_-)\quad \in\Tc\Ac\non
\eeq
as claimed.\hfill\rule{1ex}{1ex}\\

Let $\mathrm{Idem}(\Ac)$ be the set of idempotents in the inductive limit of matrix algebras $M_{\infty}(\Ac)=\limind M_N(\Ac)$. Then we have  a pairing $\mathrm{Idem}(\Ac)\times \Psi_*(\Ac,\Bct)\to \Psi_*(\cc,\Bct)$ given by left composition with the homomorphisms $\cc\to M_N(\Ac)$ corresponding to the idempotents. The functorial properties of the Chern character
\be
\begin{CD}
\mathrm{Idem}(\Ac) @. \times @. \Psi_*^{\te}(\Ac,\Bct) @>>> \Psi_*^{\te}(\cc,\Bct)\\
 @VV{\ch}V @.        @VV{\ch}V      @VV{\ch}V \\
HE_0(\Ac) @. \times @. HE_*(\Ac,\Bct) @>>> HE_*(\Bct)\\
\end{CD} \label{kt}
\ee
show that we can compute in homology the pairing between $K$-theory and bivariant modules. In particular when $\Bc=\cc$, the unbounded $\Ac$-$\cc$-bimodules essentially describe the spectral triples over $\Ac$. Recall that a spectral triple is given by a $\zz_2$-graded Hilbert space, a representation $\rho$ of $\Ac$ into the algebra $\End(\Hc)$ of bounded operators, and an unbounded selfadjoint odd operator $D$. Extending the left actions of $\rho$ and $D$ to $\Ec:=\Hc\hotimes \cct$, we get a bimodule $(\Ec,\rho,D)\in \Psi_*(\Ac,\cct)$ describing the above spectral triple. Its bivariant Chern character is an entire cyclic cohomology class in $HE_*(\Ac,\cc)\simeq HE^*(\Ac)$. Our construction represents $\ch(\Ec,\rho,D)$ as a bounded cocycle on $X(\Tc\Ac)$. Exploiting the homotopy equivalence between $X(\Tc\Ac)$ and $\Ome\Ac$, we may also represent the Chern character as an entire cocycle on the $(b+B)$-complex of $\Ac$. When doing this, we recover exactly the JLO cocycle \cite{JLO}:
\begin{proposition}
Let $\Ac$ be a complete bornological algebra, and $(\Ec,\rho,D)\in\Psi_*(\Ac,\cct)$ a $\te$-summable spectral triple, with $\Ec=\Hc\hotimes\cct$. Then the Chern character of $(\Ec,\rho,D)$ is represented by the entire cocycle $\chi(\Ec,\rho,D):\Ome\Ac\to\cc$ on the $(b+B)$-complex of $\Ac$, corresponding to the JLO formula:\\
i) In the even case $\Hc=\Hc_+\oplus\Hc_-$, $\rho=\left( \begin{array}{cc}
          \rho_+ & 0 \\
          0 & \rho_- \\
     \end{array} \right) $ and $D=\left( \begin{array}{cc}
          0 & D_- \\
          D_+ & 0 \\
     \end{array} \right)$. Then $\chi$ is the even entire cochain
\beq
\lefteqn{ \chi(\Ec,\rho,D)(a_0da_1\ldots da_{2n})=\int_{\Delta_{2n}}ds_1\ldots ds_{2n}\times}\label{even}\\
&&\times{\Tr}_s(\rho(a_0)e^{-s_0D^2}[D,\rho(a_1)]e^{-s_1D^2}\ldots [D,\rho(a_{2n})]e^{-s_{2n}D^2})\ ,\non
\eeq
for any $n\in\nn$ and $a_i\in\Ac$, where ${\Tr}_s$ is the supertrace of operators on the $\zz_2$-graded Hilbert space $\Hc$.\\
ii) In the odd case $\Hc$ is the sum of two copies of a trivially graded Hilbert space $\Kc$. One has $\rho=\left( \begin{array}{cc}
          \al & 0 \\
          0 & \al \\
     \end{array} \right)$ for a given bounded homomorphism $\al:\Ac\to\End(\Kc)$, and $D=\left( \begin{array}{cc}
          0 & Q \\
          Q & 0 \\
     \end{array} \right)$ for an unbounded operator $Q$. Then $\chi$ is the odd entire cochain
\beq
\lefteqn{\chi(\Ec,\rho,D)(a_0da_1\ldots da_{2n+1})=-\sqrt{2i}\int_{\Delta_{2n+1}}ds_1\ldots ds_{2n+1}\times}\label{odd}\\
&&\times\Tr(\rho(a_0)e^{-s_0Q^2}[Q,\rho(a_1)]e^{-s_1Q^2}\ldots [Q,\rho(a_{2n+1})]e^{-s_{2n+1}Q^2})\ ,\non
\eeq
for any $n\in\nn$ and $a_i\in\Ac$, where $\Tr$ is the trace of operators on $\Kc$.
\end{proposition}
{\it Proof:} The isomorphism $HE_*(\Ac,\cc)\simeq HE^*(\Ac)$ is obtained as follows. Given a bounded chain map $f:X(\Tc\Ac)\to X(\Tc\cc)$ representing a bivariant entire cyclic cohomology class, we associate the bounded cocycle $X(m)\circ f:X(\Tc\Ac)\to X(\cc)\simeq \cc$ obtained after composition with the projection morphism $X(m):X(\Tc\cc)\to X(\cc)$ coming from the multiplication map $m:\Tc\cc\to\cc$. The functoriality of the construction $\chi$ (theorem \ref{tchi}) yields a commutative square
$$
\xymatrix{
\Ome\Tc\Ac  \ar[rr]^{\chi(\Oman^+\Ec,\rho_*,D)} \ar[d]_{\Ome(m)} & & X(\Tc\cc) \ar[d]^{X(m)} \\
\Ome\Ac \ar[rr]^{\chi(\Ec,\rho,D)} & &  \cc
}
$$
Combining this square with the commutative diagram (\ref{cd}) of corollary \ref{ccom} shows that under the homotopy equivalence $P\circ c:X(\Tc\Ac)\stackrel{\sim}{\longrightarrow} \Ome\Ac$, the Chern character of the spectral triple $\ch(\Ec,\rho,D)\in HE^*(\Ac)$ indeed corresponds to the class of the $(b+B)$-cocycle $\chi(\Ec,\rho,D)$. Since $X_1(\cc)=0$, the only remaining component of $\chi$ is $\chi_0=\tau\mu_0\nat$, where $\tau$ is the even/odd supertrace of endomorphisms on $\Hc$, depending on the parity of the spectral triple. Since $\tau\cdot \nat$ is a (super)trace, one has
$$
\chi_0=\tau\int_0^1dt\, e^{-t\te}\d\rho e^{(t-1)\te}\nat=\tau\d\rho e^{-\te}\nat\ ,
$$
with $\te=D^2+[D,\rho]$. As usual $\exp(-\te)$ is given by a Duhamel expansion
$$
e^{-\te}=\sum_{n\ge 0}(-)^n\int_{\Delta_{n}}ds_1\ldots ds_{n}\,e^{-s_0D^2}[D,\rho]e^{-s_1D^2}\ldots [D,\rho]e^{-s_{n}D^2}\ .
$$
i) Even case: Then $\Hc=\Hc_+\oplus\Hc_-$ and $\tau={\Tr}_s$. Also, $\rho$ is a diagonal matrix whereas $D$ is off-diagonal, so that $\tau$ selects only even powers of $D$, and therefore $\chi_0$ is an even entire cocycle on $\Ome\Ac$. Equation (\ref{even}) follows.\\
ii) Odd case: $\Hc=\Kc\oplus\Kc$ and $D=\eps Q$, with $\eps=\left( \begin{array}{cc}
          0 & 1 \\
          1 & 0 \\
     \end{array} \right)$. For any $x,y\in\End(\Kc)$, one has $\tau(x+\eps y)=\sqrt{2i}\Tr(y)$, hence $\tau$ selects only odd powers of $D$; $\chi_0$ is thus an odd entire cocycle over $\Ac$, whence (\ref{odd}).\hfill\rule{1ex}{1ex}\\

This shows in particular that for any idempotent $e\in\mathrm{Idem}(A)$, and any $\te$-summable even spectral triple $(\Ec,\rho,D)$, the coupling of the Chern characters calculates the index pairing between $K$-theory and $K$-homology:
\begin{proposition}
Let $\Ac$ be a complete bornological algebra, $e\in M_{\infty}(\Ac)$ an idempotent and $(\Ec,\rho,D)$ a $\te$-summable even spectral triple over $\Ac$. Then the composition of the Chern characters by the map $HE_0(\Ac)\times HE^0(\Ac)\to \cc$ computes the index pairing
\be
\langle \ch(e),\ch(\Ec,\rho,D) \rangle = \mathrm{index}(eDe)
\ee
\end{proposition}
{\it Proof:} One has $e\in M_N(\Ac)$, and $\Ec=\Hc\hotimes\cct$ for a certain Hilbert space $\Hc$. The proposition is easily proved by exploiting the homotopy invariance of the Chern character with respect to $D$. For notational simplicity we still denote by $e$ the idempotent $\rho(e)\in\End(\Hc\hotimes\cc^N)$. For $t\in [0,1]$ the operator
$$
D_t= D+t([D,e]e-e[D,e])
$$
is a bounded perturbation of $D$, connecting homotopically $D$ to $D_1$. One has $[D_1,e]=0$, which shows that we can reduce to the situation where $D$ commutes with $e$. The Chern character of $e$ in the $(b+B)$-complex of entire chains over $\Ac$ is obtained from the $X$-complex cycle $\hat{e}$ by means of the rescaling factor of equation (\ref{res}):
$$
\ch(e)=\tr (e)+\sum_{n\ge 1}(-)^n\frac{(2n)!}{n!}\tr((e-\frac{1}{2})(dede)^n)\ .
$$
Since we assume $[D,e]=0$, equation (\ref{even}) implies that $\chi(\Ec,\rho,D)$ vanishes on any term involving a (strictly) positive power of $dede$, hence
$$
\chi(\Ec,\rho,D)(\ch(e))={\Tr}_s(e\,e^{-D^2})\ .
$$
This is the McKean-Singer formula computing the index of $D$ relative to $e$, whose proof involves spectral theory in a simple way \cite{Gi}.\hfill \rule{1ex}{1ex}

\subsection{The Bott class }\label{bott}

We now examine the generators of the $K$-theory and $K$-homology of the $n$-dimensional real vector space. This will illustrate some features of commutative algebras, and explain the role of the somewhat mysterious Fedosov exponential (lemma \ref{lexp}). Let $\Sc(\rr^n)$ be the commutative algebra of smooth rapidly decreasing functions on $\rr^n$. We denote by $\{x_1,\ldots ,x_n\}$ the canonical coordinate system, giving to $\rr^n$ its canonical orientation. $\Sc(\rr^n)$ is a Fr\'echet algebra for the locally convex topology given by the countable family of seminorms
\be
||a||_{\al,\delta}=\sup_{x\in\rr^n}|x^{\al}\d^{\delta}a(x)|\quad \forall a\in\Sc(\rr^n)\ ,
\ee
where $\al=(\al_1,\ldots,\al_n)$ and $\delta=(\delta_1,\ldots,\delta_n)$ are collections of positive integers such that $x^{\al}=(x_1)^{\al_1}\ldots(x_n)^{\al_n}$ and $\d^{\delta}$ is the partial differentiation operator $\frac{\d^{|\delta|}}{\d {x_1}^{\delta_1}\ldots \d {x_n}^{\delta_n}}$. We endow $\Sc(\rr^n)$ with the bounded bornology.\\

For any $n\in\nn$, we shall first construct a spectral triple over $\Sc(\rr^n)$. It comes from the Dirac operator acting on sections of the trivial spinor bundle over $\rr^n$. So let $C_n$ be the $n$-dimensional complex Clifford algebra. It is generated by a basis $\gamma^1,\ldots,\gamma^n$ of $\cc^n$, subject to the anticommutation relations $\{\gamma^{\mu},\gamma^{\nu}\}=2\delta^{\mu\nu}$. We let $S_n$ be the complex spinor representation of $C_n$ endowed with the fine bornology. Then the fundamental class of $\rr^n$ in $K$-homology is represented by the following bornological spectral triple: $\Hc_n:=S_n\hotimes\Sc(\rr^n)$ is the complete bornological space consisting of rapidly decreasing sections of the trivial spinor bundle; the representation $\rho:\Sc(\rr^n)\to \End(\Hc_n)$ is given by (left) multiplication; and for any real parameter $t>0$, the usual Dirac operator $D_t=i\sqrt{t}\gamma^{\mu}\frac{\d}{\d x^{\mu}}$ acts as an unbounded endomorphism of $\Hc_n$. If $\Ec_n$ denotes the right $\cct$-module $\Hc_n\hotimes\cct$, then $(\Ec_n,\rho,D_t)\in \Psi_*(\Sc(\rr^n),\cct)$ is a $\te$-summable spectral triple with parity equal to $n$ mod 2 (for $n$ odd, replace $\Ec_n$ by two copies of $\Hc_n\hotimes\cct$, $\rho$ by $\Id_2\otimes\rho$ and $D_t$ by $\eps D_t$).\\
In the subsequent calculations we will need some explicit properties of the matrix representation of the Clifford algebra into $S_n$. This depends on the parity of $n$:\\
i) $n=2k$: Then the spinor representation $S_n$ is a $2^k$-dimensional $\zz_2$-graded space. The generators $\gamma^{\mu}$ are {\it odd} operators represented by hermitian matrices. The grading operator is the element of highest degree in $C_n$:
\be
\Gamma=(-i)^k \gamma^1\ldots\gamma^n\ ,\quad (\Gamma)^2=1\ .
\ee
The supertrace of linear operators on $S_n$ is ${\tr}_s=\tr(\Gamma\cdot)$. Then for any $j<n$ one has $\tr_s(\gamma^1\ldots\gamma^j)=0$, while $\tr_s(\gamma^1\ldots\gamma^n)=(2i)^k$.\\
ii) $n=2k+1$: Then $S_n$ is a trivially graded $2^k$-dimensional space. We represent also the generators $\gamma^{\mu}$ by hermitian matrices. Then for any $j<n$ one has $\tr(\gamma^1\ldots\gamma^j)=0$, and $\tr(\gamma^1\ldots\gamma^n)=(2i)^k$.

\begin{proposition}
Let $n$ be a positive integer. The Chern character of the fundamental class $(\Ec_n,\rho,D_t)\in\Psi_{n+2\zz}(\Sc(\rr^n),\cct)$ in $HE^{n+2\zz}(\Sc(\rr^n))$ retracts, when $t\to 0$, to an $n$-dimensional $(b+B)$-cocycle $\chi^n:\Om^n\Sc(\rr^n)\to\cc$. It corresponds to the fundamental class of $\rr^n$ in cyclic cohomology:
\be
\chi^n(a_0da_1\ldots da_n)=\frac{1}{n!(2\pi i)^{n/2}}\int_{\rr^n}a_0da_1\wedge\ldots\wedge da_n\ ,
\ee
for any $a_i\in\Sc(\rr^n)$.
\end{proposition}
{\it Proof:} It is a well-known fact that when the Dirac operator acts on the sections of a spinor bundle on a manifold $M$, the JLO cocycle retracts  on local expressions involving the $\widehat{A}$-genus of $M$ at the limit $t\to 0$ \cite{CM93}. In our case, $\rr^n$ is a flat manifold so that the computation is particularly simple, and can be performed by means of the asymptotic symbol calculus for example as in \cite{CM95}. We don't give the details here because it is a classical result.\hfill\rule{1ex}{1ex}\\

In fact the retracted $(b+B)$-cocycle $\chi^n$ is also a cyclic $n$-cocycle, that is, a closed graded trace of degree $n$ on the DG algebra $\Om\Sc(\rr^n)$. It follows that $\chi^n$ is invariant under the Karoubi operator $\kappa$ on $\Ome\Sc(\rr^n)$, thus it vanishes on the contractible subspace $P^{\bot}\Ome\Sc(\rr^n)$ (see section \ref{X}). It follows that $\chi^n$ is also a cocycle on the $X$-complex $X(\Tc\Sc(\rr^n))\simeq(\Oman\Sc(\rr^n),\nat \dd,\bb)$. Taking into account the rescaling factor $(-)^n[n/2]!$ of equation (\ref{res}) when passing from $\Oman\Sc(\rr^n)$ to $\Ome\Sc(\rr^n)$, we deduce the expression of the fundamental class as an $X$-complex cocycle:
\begin{corollary}\label{cdir}
Let $n\in\nn$, and $(\Ec_n,\rho,D_t)$ be the fundamental $K$-homology class of $\rr^n$. \\
i) If $n$ is even, then the Chern character of $(\Ec_n,\rho,D_t)$ in $HE^0(\Sc(\rr^n))$ is represented by the following trace on the algebra $\Tc\Sc(\rr^n)\simeq(\Oman^+\Sc(\rr^n),\odot)$:
\be
\ch(\Ec_n,\rho,D_t)(a_0da_1\ldots da_n)=\frac{[n/2]!}{n!(2\pi i)^{n/2}}\int_{\rr^n}a_0da_1\wedge\ldots\wedge da_n
\ee
for any $a_i\in\Sc(\rr^n)$.\\
ii) If $n$ is odd, then the Chern character of $(\Ec_n,\rho,D_t)$ in $HE^1(\Sc(\rr^n))$ is represented by the following one-cocycle on $\Om^1\Tc\Sc(\rr^n)_{\nat}\simeq\Oman^-\Sc(\rr^n)$:
\be
\ch(\Ec_n,\rho,D_t)(\nat a_0da_1\ldots da_{n-1}\dd a_n)=-\frac{[n/2]!}{n!(2\pi i)^{n/2}}\int_{\rr^n}a_0da_1\wedge\ldots\wedge da_{n-1}\wedge da_n
\ee
for any $a_i\in\Sc(\rr^n)$.\hfill\rule{1ex}{1ex}
\end{corollary}
One sees that there is a simplification here, due to the fact that $\Sc(\rr^n)$ is a commutative algebra: the cocycle factors through de Rham cohomology. Hence we can deal with the ordinary exterior algebra of differential forms $\Om^*(\rr^n)$ instead of the universal DG algebra of non-commutative forms $\Om\Sc(\rr^n)$. We endow $\Om^*(\rr^n)$ with its usual Fr\'echet topology and the associated bounded bornology. Then the universal property of $\Om\Sc(\rr^n)$ implies that there is a unique bounded DG algebra morphism $\Om\Sc(\rr^n)\to\Om^*(\rr^n)$ extending the identity map $\Sc(\rr^n)\to\Sc(\rr^n)$. The image of $a_0da_1\ldots da_k$ is equal to $a_0da_1\wedge\ldots\wedge da_k$ if $k\le n$ and zero otherwise. As a consequence, this DG morphism is also bounded for the analytic bornology on $\Om\Sc(\rr^n)$ and thus extends to a bounded DG morphism $\Oman\Sc(\rr^n)\to \Om^*(\rr^n)$. We endow the even part $\Om^+(\rr^n)$ with the Fedosov product
\be
\om_1\odot\om_2:=\om_1\om_2- d\om_1\wedge d\om_2\quad \forall \om_1,\om_2\in\Om^+(\rr^n)\ .
\ee
Then $T_n:=(\Om^+(\rr^n),\odot)$ is an associative (non-commutative!) complete bornological algebra, and we get a canonical bounded homomorphism $(\Oman^+\Sc(\rr^n),\odot)\to(\Om^+(\rr^n),\odot)$, or equivalently $\Tc\Sc(\rr^n)\to T_n$. This yields a bounded chain map $X(\Tc\Sc(\rr^n))\to X(T_n)$. Now, the fundamental class of $\rr^n$ gives rise to a bounded cocycle $[\rr^n]: X(T_n)\to\cc$:
\be
[\rr^n](x)=\int_{\rr^n}x\ ,\quad [\rr^n]( x\dd y)=\int_{\rr^n}x\wedge dy\ ,\quad \forall x,y\in T_n\ .
\ee
It is easily checked that $[\rr^n]$ vanishes on the commutators $[T_n,\Om^1T_n]$, hence is well-defined on $\Om^1{T_n}_{\nat}=X_1(T_n)$. Consequently, the Chern character of $(\Ec_n,\rho,D)\in\Psi_{n+2\zz}(\Sc(\rr^n),\cct)$ factors through $X(T_n)$:
\be 
\ch(\Ec_n,\rho,D_t): X(\Tc\Sc(\rr^n))\to X(T_n)\stackrel{\hat{\chi}^n}{\longrightarrow}\cc\ ,
\ee
where $\hat{\chi}^n$ is the cocycle $(-)^n\frac{[n/2]!}{n!(2\pi i)^{n/2}}[\rr^n]$. \\

We now construct the Bott generator of the $K$-theory of $\rr^n$. It will be represented by a bimodule $\beta_n\in\Psi_{n+2\zz}(\cc,\Sc(\rr^n))$. Since we deal with the exterior algebra of ordinary differential forms and its Fedosov deformation $T_n$, we don't need to consider the unitalization of $\Sc(\rr^n)$. This property also is a advantage of the commutative case. Let again $\Hc_n=S_n\hotimes\Sc(\rr^n)$ be the space of rapidly decreasing sections of the trivial spinor bundle over $\rr^n$, considered this time as a right $\Sc(\rr^n)$-module. There is an ovious homomorphism $\al:\cc\to\End_{\Sc(\rr^n)}(\Hc_n)$, sending the unit $e\in\cc$ to the identity endomorphism. For any real parameter $\la>0$, we introduce an unbounded Dirac operator $Q_{\la}$ acting on $\Hc_n$ by Clifford multiplication with respect to the vector $x$:
\be
(Q_{\la}\xi)(x)= \sqrt{\la}x_{\mu}\gamma^{\mu}\cdot\xi(x)\quad \forall \xi\in\Hc_n\ ,\ x\in\rr^n\ .
\ee
Note that $Q_{\la}$ is the Fourier transform of the previous Dirac operator $D_{\la^{-1}}$. Then the unbounded $\cc$-$\Sc(\rr^n)$-bimodule $\beta_n=(\Hc_n,\al,Q_{\la})$ represents the Bott generator of $\rr^n$. Its Chern character $\ch(\beta_n)\in HE_{n+2\zz}(\Sc(\rr^n))$ is represented by entire chains over $\Sc(\rr^n)$. We want to evaluate $\ch(\beta_n)$ on the fundamental class of $\rr^n$, so that only its image in $X(T_n)$ is important. All the construction of the bivariant Chern character then transpose immediately to the situation where the universal DG algebra $\Om\Sc(\rr^n)$ is replaced by $\Om^*(\rr^n)$.
\begin{proposition}\label{pbott}
For any $n\in\nn$, let $\beta_n=(\Hc_n,\al,Q_{\la})\in\Psi_{n+2\zz}(\cc,\Sc(\rr^n))$ be the Bott generator. Its Chern character is represented by the following cycle in the $X$-complex of the Fedosov algebra $T_n=(\Om^+(\rr^n),\odot)$:\\
i) $n$ even:
\be
\ch(\beta_n)=\frac{n!}{(n/2)!}(2i\la)^{n/2}e^{-\la x^2}dx_1\wedge\ldots \wedge dx_n\ \in T_n\ .
\ee
ii) $n=2k+1$:
\be
\ch(\beta_n)=-\frac{(2k)!}{k!}(2i\la)^{n/2}\sum_{j=1}^n\nat e^{-\la x^2}dx_{j+1}\wedge\ldots \wedge dx_{j-1}\dd x_j\ \in \Om^1{T_n}_{\nat}\ .
\ee
These are top-degree differential forms with gaussian shape over $\rr^n$.
\end{proposition}
{\it Proof:} For $n$ even we set $\Ec_n=\Hc_n$, $D=Q_{\la}$ and $\rho=\al$. For $n$ odd we set $\Ec_n=\Hc_n\oplus\Hc_n$, $D=\eps Q_{\la}$ and $\rho=\Id_2\otimes\al$, where $\eps=\left( \begin{array}{cc}
          0 & 1 \\
          1 & 0 \\
     \end{array} \right)$ is the odd generator of the one-dimensional Clifford algebra $C_1$. Then for any $n$, the unbounded bimodule $(\Ec_n,\rho,D)\in\Psi_{n+2\zz}(\cc,\Sc(\rr^n))$ represents the Bott element, and $\ch(\beta_n)$ is the image of the generator $\hat{e}\in HE_0(\cc)$ under the composition of chain maps
$$
X(\Tc\cc)\stackrel{\gamma}{\longrightarrow}\Ome\Tc\cc\stackrel{\chi}{\longrightarrow} X(\Tc\Sc(\rr^n))\to X(T_n)\ ,
$$
where $\gamma$ is the Goodwillie equivalence and $\chi=\chi(\Oman^+\Ec_n,\rho_*,D)$ is the core of the bivariant Chern character. We know that the generator of $HE_0(\cc)$ is represented by the idempotent
$$
\hat{e}=e+\sum_{k\ge 1}\frac{(2k)!}{(k!)^2}(e-\frac{1}{2})(dede)^k\quad\in\Oman^+\cc=\Tc\cc\ ,
$$
where $e$ is the unit of $\cc$. We claim that its image $\gamma(\hat{e})$ in $\Ome\Tc\cc$ has the same homology class as the $(b+B)$ entire cycle
$$
\hat{f}=\hat{e}+\sum_{k\ge 1}(-)^k\frac{(2k)!}{k!}(\hat{e}-\frac{1}{2})(d\hat{e}d\hat{e})^k\quad\in\Ome\Tc\cc\ .
$$
Indeed, the projection $\pi:\Ome\Tc\cc\to X(\Tc\cc)$ maps $\hat{f}$ to $\hat{e}$, and corollary \ref{cgood} shows that $\pi$ and $\gamma$ are inverse homotopy equivalences. Thus $\ch(\beta_n)$ is the image of $\hat{f}$ under $\chi(\Oman^+\Ec_n,\rho_*,D)$ projected to $X(T_n)$. The construction of the bivariant Chern character carries over to the situation where the universal DG algebra $\Om\Sc(\rr^n)$ is replaced by $\Om^*(\rr^n)$. We thus consider the right $T_n$-module $\Om^+\Ec_n:=S_n\hotimes T_n$. There is a unique bounded homomorphism $\rho_*:\Tc\cc\to \End_{T_n}(\Om^+\Ec_n)$ extending $\rho:\cc\to \End_{\Sc(\rr^n)}(\Ec_n)$. By definition one has $\rho(e)=1$ and $d1=0$, so that $\rho_*(e(dede)^k)=1(d1d1)^k=0$ whenever $k\ge 1$, and $\rho_*(\hat{e})=1$. Next, the chain map $\chi(\Om^+\Ec_n,\rho_*,D):\Ome\Tc\cc\to X(T_n)$ has two components: the $T_n$-valued $\chi_0=\tau\mu_0\nat$, and the $\Om^1{T_n}_{\nat}$-valued $\chi_1=\nat\tau\mu_0 \dd(D+\rho_*)\nat$, with
$$
\mu_0=\int_0^1dt\, \exp_{\odot}(-t\te)\d\rho_*\exp_{\odot}((t-1)\te)\ ,\quad \te=D^{\odot 2}+[D,\rho_*]_{\odot}\ ,
$$
and $\tau: \End_{T_n}(\Om^+\Ec_n)\to T_n$ is the supertrace. Recall that the exponentials and commutators are taken with respect to the Fedosov product on $(\Om^+(\rr^n),\odot)= T_n$. Since $\rho_*(\hat{e})=1$, the Fedosov commutator $[D,\rho_*(\hat{e})]_{\odot}$ vanishes, and we simply have
\beq
\tau\mu_0\nat(\hat{f})&=&\int_0^1dt\,\tau e_{\odot}^{-tD^{\odot 2}}\d\rho_* e_{\odot}^{(t-1)D^{\odot 2}}\nat(\hat{e})=\tau e_{\odot}^{-D^{\odot 2}}\ ,\non\\
\nat\tau\mu_0\dd(D+\rho_*)\nat(\hat{f})&=&\int_0^1dt\nat\tau e_{\odot}^{-tD^{\odot 2}}\d\rho_* e_{\odot}^{(t-1)D^{\odot 2}}\dd(D+\rho_*)\nat(\hat{f})\non\\
&=&\nat\tau e_{\odot}^{-D^{\odot 2}}\dd D\ .\non
\eeq
Let us now compute the Fedosov exponential $\exp_{\odot}(-D^{\odot 2})$ in terms of differential forms. One has $D^{\odot 2}=D^2 +dDdD$, where the laplacian $D^2$ is a scalar function of $x\in\rr^n$: indeed if $n$ is even, the matrices $\gamma^{\mu}$ are odd for the $\zz_2$-graduation of $S_n$ and
$$
D^2=Q_{\la}^2=\la x_{\mu}\gamma^{\mu}x_{\nu}\gamma^{\nu}=\frac{1}{2}\la x_{\mu}x_{\nu}\{\gamma^{\mu},\gamma^{\nu}\}=\la x^2\ .
$$
If $n$ is odd, then $S_n$ is trivially graded, and the product $\eps\gamma^{\mu}$ is odd: 
$$
D^2=(\eps Q_{\la})^2=\la \eps x_{\mu}\gamma^{\mu}\eps x_{\nu}\gamma^{\nu}=\frac{\eps^2}{2}x_{\mu}x_{\nu}\{\gamma^{\mu},\gamma^{\nu}\}=\la x^2\ .
$$
Next, with $H=-D^{\odot 2}$, lemma \ref{lexp} implies that the Fedosov exponential is the following differential form on $\rr^n$:
$$
\exp_{\odot}H= \sum_{k\ge 0}(-)^k\int_{\Delta_k}ds_1\ldots ds_k \, e^{s_0H}dHde^{s_1H}dH\ldots de^{s_{k-1}H}dHde^{s_kH}\ .
$$
But $dH=-d(D^2)=-\la d(x^2)$ and $d\exp(sH)=sdH\exp(sH)$, hence $d(x^2)$ always appears by pairs in the expression above. This means that all the terms corresponding to $k\ge 1$ vanish, and
$$
\exp_{\odot}H= e^H= e^{-D^2-dDdD}=e^{-D^2}\sum_{k\ge 0}\frac{(-)^k}{k!}(dDdD)^k\ ,
$$
because the scalar $D^2$ commutes (for the ordinary product of differential forms) with $dDdD$. For $n$ even one has $dDdD=\la dx_{\mu}\gamma^{\mu}dx_{\nu}\gamma^{\nu}=-\la dx_{\mu}dx_{\nu}\gamma^{\mu}\gamma^{\nu}$ because $\gamma^{\mu}$ and $dx_{\nu}$ are odd, and for $n$ odd $dDdD=\la dx_{\mu}\eps\gamma^{\mu}dx_{\nu}\eps\gamma^{\nu}=-\la dx_{\mu}dx_{\nu}\gamma^{\mu}\gamma^{\nu}$. Thus in any case, the Fedosov exponential reads
$$
\exp_{\odot}(-D^{\odot 2})=e^{-\la x^2}\sum_{k\ge 0}\frac{\la^k}{k!}dx_{\mu_1}\ldots dx_{\mu_{2k}}\gamma^{\mu_1}\ldots\gamma^{\mu_{2k}}\ .
$$
Let us now compute the Chern character of the Bott element:\\
i) $n$ even: then $\tau$ is equal to the supertrace $\tr_s$ on the $\zz_2$-graded spinor representation $S_n$, and $\ch(\beta_n)=\tr_s\exp_{\odot}(-D^{\odot 2})$. Thus
$$
\ch(\beta_n)= e^{-\la x^2}\sum_{k\ge 0}\frac{\la^k}{k!}dx_{\mu_1}\ldots dx_{\mu_{2k}}{\tr}_s(\gamma^{\mu_1}\ldots\gamma^{\mu_{2k}})\ .
$$
However, if $2k<n$, then the supertrace over the $\gamma$-matrices vanishes, and if $2k>n$, the differential form $dx_{\mu_1}\ldots dx_{\mu_{2k}}$ is identically zero. Hence only the term $2k=n$ remains:
\beq
\ch(\beta_n)&=&\frac{\la^{n/2}}{(n/2)!}e^{-\la x^2}dx_{\mu_1}\ldots dx_{\mu_n}{\tr}_s(\gamma^{\mu_1}\ldots\gamma^{\mu_n})\non\\
&=& \frac{\la^{n/2}}{(n/2)!}n! e^{-\la x^2}dx_1\ldots dx_n{\tr}_s(\gamma^1\ldots\gamma^n)\non\\
&=& \frac{n!}{(n/2)!}(2i\la)^{n/2} e^{-\la x^2}dx_1\ldots dx_n\ .\non
\eeq
ii) $n$ odd: then $D=\eps Q_{\la}$ and $\tau(x+\eps y)=\sqrt{2i}\,\tr(y)$ for any endomorphisms $x,y$ of the trivially graded spinor representation $S_n$. Thus
$$
\ch(\beta_n)=\nat\tau e_{\odot}^{-D^{\odot 2}}\dd D=\nat\tau e_{\odot}^{-D^{\odot 2}}\dd(\eps Q_{\la})=-\sqrt{2i}\,\nat\tr(e_{\odot}^{-D^{\odot 2}}\dd Q_{\la})\ ,
$$
because $\eps$ anticommutes with $\dd$. One has
\beq
\ch(\beta_n)&=& -\sqrt{2i}\, \nat\tr( e^{-\la x^2}\sum_{k\ge 0}\frac{\la^k}{k!}dx_{\mu_1}\ldots dx_{\mu_{2k}}\gamma^{\mu_1}\ldots\gamma^{\mu_{2k}}\dd(\sqrt{\la}x_{\nu}\gamma^{\nu}))\non\\
&=& -\sqrt{2i}\sum_{k\ge 0}\frac{\la^{n/2}}{k!}\tr(\gamma^{\mu_1}\ldots\gamma^{\mu_{2k}}\gamma^{\nu})\,\nat e^{-\la x^2}dx_{\mu_1}\ldots dx_{\mu_{2k}}\dd x_{\nu}\ .\non
\eeq
With the same argument as in the even case, only the term $2k+1=n$ remains and
\beq
\ch(\beta_n)&=& -\sqrt{2i}\frac{\la^{n/2}}{k!}\tr(\gamma^{\mu_1}\ldots\gamma^{\mu_n})\,\nat e^{-\la x^2}dx_{\mu_1}\ldots dx_{\mu_{n-1}}\dd x_{\mu_n}\non\\
&=& -\sqrt{2i}\frac{\la^{n/2}}{k!}(2k)!\tr(\gamma^1\ldots\gamma^n)\sum_{j=1}^n\nat  e^{-\la x^2}dx_{j+1}\ldots dx_{j-1}\dd x_j\non\\
&=& -\frac{(2k)!}{k!}(2i\la)^{n/2}\sum_{j=1}^n\nat e^{-\la x^2}dx_{j+1}\ldots dx_{j-1}\dd x_j\ .\non
\eeq
The proof is complete.\hfill\rule{1ex}{1ex}
\begin{corollary}
Let $n\in\nn$. Then the pairing between the Chern characters of the Bott element $\beta_n\in\Psi_{n+2\zz}(\cc,\Sc(\rr^n))$ and the Dirac spectral triple $(\Ec_n,\rho,D_t)\in\Psi_{n+2\zz}(\Sc(\rr^n),\cct)$ is normalized:
\be
\langle \ch(\beta_n), \ch(\Ec,\rho,D_t)\rangle =1\ .
\ee
\end{corollary}
{\it Proof:} It is a consequence of corollary \ref{cdir} and proposition \ref{pbott}. \hfill\rule{1ex}{1ex}\\

This explains the normalization factor $\sqrt{2i}$ appearing in the definition (\ref{tr}) of the canonical trace $\tau$ for the Chern character on $\Psi_1$. It is interesting also to note that this factor is the only one compatible with the {\it external} product on $K$-homology $K\!K(\Ac,\cc)\times K\!K(\Bc,\cc)\to K\!K(\Ac\hotimes \Bc,\cc)$, see \cite{C1} p.295.

\appendix

\renewcommand{\theequation}{\Alph{section}.\arabic{equation}}

\section{Appendix}

In this appendix we adapt Quillen's formalism of algebra cochains \cite{Q2} to the bornological framework. All the results presented here are straightforwardly obtained from Quillen's paper by replacing arbitrary algebras by complete bornological algebras, tensor products by completed tensor products and linear maps by bounded linear maps.

\subsection{Bar construction}

Let $\Ac$ be an associative complete bornological algebra. The bar construction of $\Ac$ is the graded space
\be
\Bb(\Ac)=\bigoplus_{n\ge 0}\Bb_n(\Ac)\ ,
\ee
where $\Bb_n(\Ac)=\Ac^{\hotimes n}$ is localized in degree $n$. $\Bb(\Ac)$ endowed with the direct sum bornology is a complete bornological space. The decomposable element $a_1\otimes...\otimes a_n$ of $\Ac^{\hotimes n}$ will be written $(a_1,...,a_n)$. $\Bb(\Ac)$ is naturally a bornological coassociative coalgebra, with bounded coproduct $\Delta: \Bb(\Ac)\to \Bb(\Ac)\hotimes\Bb(\Ac)$ given by
\be
\Delta(a_1,...,a_n)= \sum_{i=0}^{n}(a_1,...,a_i)\otimes (a_{i+1},...,a_n)
\ee
and counit $\eta: \Bb(\Ac)\to\cc$ corresponding to the projection onto $\Ac^{\hotimes 0}=\cc$. On $\Bb(\Ac)$ is defined a bounded differential $b'$ of degree $-1$:
\be
b'(a_1,...,a_n)=\sum_{i=1}^{n-1} (-)^{i-1} (a_1,...,a_ia_{i+1},...,a_n)\ ,\label{118}
\ee
with $b'=0$ for $n=0,1$. One readily verifies that ${b'}^2=0$ and that the coproduct and counit are  morphisms of (graded) complexes, i.e. $\Delta b'= (b'\otimes 1 + 1\otimes b') \Delta$ and $ \eta b'=b'\eta =0$, taking care of the signs occuring when graded symbols are permuted, for instance
$$
(1\otimes b') ((a_1,...,a_i)\otimes (a_{i+1},...,a_n))=(-)^i(a_1,...,a_i)\otimes b'(a_{i+1},...,a_n)
$$
according to the respective degrees of $(a_1,...,a_i)$ and $b'$. This turns $\Bb(\Ac)$ into a differential graded (DG) complete bornological coalgebra.\\

Next we consider the free bicomodule over $\Bb(\Ac)=\Bb$
\be
\Om_1\Bb =\Bb\hotimes \Ac\hotimes\Bb\ .
\ee
The generic element $(a_1,...,a_{i-1})\otimes a_i\otimes(a_{i+1},...,a_n)$ of $\Om_1\Bb$ will be written $(a_1,...,a_{i-1}|a_i|a_{i+1},...,a_n)$. The left and right comodule maps $\Delta_l: \Om_1\Bb\to \Bb\hotimes\Om_1\Bb$ and $\Delta_r: \Om_1\Bb\to \Om_1\Bb\hotimes\Bb$ are bounded and given by
\beq
\lefteqn{ \Delta_l(a_1,...,a_{i-1}|a_i|a_{i+1},...,a_n) = }\non\\
&& \sum_{j=0}^{i-1} (a_1,...,a_j)\otimes (a_{j+1},...,a_{i-1}| a_i|a_{i+1},...,a_n)\ ,
\eeq
and similarly for $\Delta_r$. $\Om_1\Bb$ also has a grading over the integers,
\be
(\Om_1\Bb)_n= \bigoplus_{i=1}^n \Bb_{i-1}\hotimes \Ac \hotimes\Bb_{n-i}\ \mbox{for}\ n\ge 1\ ,\quad (\Om_1\Bb)_0=0\ ,
\ee
just counting the number of arguments in $\Ac$. There is a bounded differential $b''$ of degree $-1$
\beq
\lefteqn{ b''(a_1,...,a_{i-1}|a_i|a_{i+1},...,a_n)=(b'(a_1,...,a_{i-1})| a_i|a_{i+1},...,a_n)}\non\\
&& +(-)^i(a_1,...,a_{i-2}|a_{i-1}a_i|a_{i+1},...,a_n)+(-)^{i+1}(a_1,...,a_{i-1}|a_ia_{i+1}|a_{i+2},...,a_n)\non\\
&&+(-)^i(a_1,...,a_{i-1}|a_i| b'(a_{i+1},...,a_n))\ .
\eeq
One has ${b''}^2=0$, and $\Delta_{l,r}$ are morphisms of (graded) complexes, i.e. $\Delta_l b'' =(b'\otimes 1 + 1\otimes b'')\Delta_l$ and similarly for $\Delta_r$. The last operator we will consider is the obvious bounded map $\partial: \Om_1\Bb\to \Bb$ induced by
\be
\partial (a_1,...,a_{i-1}|a_i|a_{i+1},...,a_n)=(a_1,...,a_n)\ . \label{op}
\ee
It is a coderivation: $\Delta\partial = (1\otimes\partial)\Delta_l+(\partial\otimes 1)\Delta_r$, and a morphism of complexes: $\partial b''=b'\partial$, of degree zero with respect to gradings. 

\subsection{Algebra cochains}

Let $\Lc$ be a complete bornological $\zz_2$-graded algebra with unit $1$ and differential $d$. The space of bounded linear maps
\be
\Rc=\hom(\Bb(\Ac),\Lc)
\ee
endowed with the bornology of equibounded maps is complete, and splits into the even/odd subspaces coming from the $\zz_2$-gradings of $\Bb$ and $\Lc$. We denote by $|f|$ the degree of an homogeneous element $f\in \Rc$. Since $\Bb(\Ac)$ is a coalgebra (coproduct $\Delta$), and $\Lc$ is an algebra (product $m:\Lc\hotimes \Lc\to \Lc$), $\Rc$ is naturally endowed with a complete bornological algebra structure given by the convolution product $fg = m(f\otimes g)\Delta$, $\forall f,g\in \Rc$. Explicitly on a $n$-chain one has
\be
(fg)(a_1,...,a_n)= \sum_{i=0}^n (-)^{|g|i} f(a_1,...,a_i)g(a_{i+1},...,a_n)\ .
\ee
Note the sign $(-)^{|g|i}$ occuring when the chain $(a_1,...,a_i)$ crosses $g$. The differentials $b'$ and $d$ induce two bounded differentials of odd degree on $\Rc$:
\be
df =d\circ f\ , \qquad \delta f= -(-)^{|f|}f\circ b'\ ,\qquad d\delta+\delta d=\delta^2=d^2=0\ .\label{127}
\ee
$d$ and $\delta$ are derivations with respect to the convolution product. Thus $\Rc$ is a bidifferential $\zz_2$-graded (complete bornological) algebra, with unit $1\eta:\Bb(\Ac)\to \Lc$.\\

To the bicomodule $\Om_1\Bb(\Ac)$ it corresponds by duality a graded $\Rc$-bimodule
\be
\Mc=\hom(\Om_1\Bb(\Ac),\Lc)
\ee
with the bounded left multiplication $\Rc\hotimes \Mc\to \Mc$ given by $f\gamma = m(f\otimes\gamma)\Delta_l$, $\forall f\in \Rc$, $\gamma\in \Mc$, and similarly the bounded right multiplication reads $\gamma f=m(\gamma\otimes f)\Delta_r$. Explicitly, the product evaluated on an element of $\Om_1\Bb(\Ac)$ is 
\beq
\lefteqn{ (f\gamma)(a_1,...,a_{i-1}|a_i|a_{i+1},...,a_n)= }\non\\ 
&& \sum_{j=0}^{i-1}(-)^{|\gamma|j} f(a_1,...,a_j)\gamma(a_{j+1},...,a_{i-1}|a_i|a_{i+1},...,a_n)\ .
\eeq
As before $b''$ induces by duality a bounded differential of odd degree on $\Mc$,
\be
\delta\gamma = -(-)^{|\gamma|}\gamma\circ b''\ ,
\ee
compatible with the $\Rc$-bimodule structure:
\be
\delta(f\gamma)=\delta f\gamma+(-)^{|f|}f\delta\gamma\ ,\quad
\delta(\gamma f)=\delta \gamma f+(-)^{|\gamma|}\gamma\delta f\ .
\ee
This differential together with $d$ implies that $\Mc$ is a bidifferential graded (complete bornological) bimodule. Last but not least, transposing the operator (\ref{op}) yields a bounded derivation $\partial: \Rc\to \Mc$ commuting with $\delta$ and $d$.

\subsection{Noncommutative differential forms}

Let $\Act=\Ac\oplus \cc$ be the complete bornological algebra obtained from $\Ac$ by adjoining a unit $1$ (even if $\Ac$ is already unital). The space of noncommutative forms is the complete bornological space $\Om \Ac=\bigoplus_{n\ge 0}\Om^n\Ac$ with $\Om^n\Ac=\Act\hotimes \Ac^{\hotimes n}$ for $n\ge 1$ and $\Om^0\Ac=\Ac$. The element $a_0\otimes...\otimes a_n\in \Om^n\Ac$ (resp. $1\otimes a_1...\otimes a_n$) is denoted by $a_0da_1...da_n$ (resp. $da_1...da_n$). Then $\Om \Ac$ is a (non-unital) complete bornological DG algebra when specifying the differential
\be
d(a_0da_1...da_n)=da_0da_1...da_n\ ,\qquad d(da_1...da_n)=0\ ,\qquad d^2=0\ ,
\ee
verifying the Leibniz rule with respect to the ordinary product on differential forms. The Hochschild operator $b: \Om^n\Ac\to \Om^{n-1}\Ac$ is the bounded map defined by $b(\om da)=(-)^{|\om|}[\om,a]$ for any $\om\in\Om \Ac$ and $a\in \Ac$, and $b(a)=0$. From this one gets the Karoubi operator $\kappa=1-(bd+db)$ and Connes' boundary $B=(1+\kappa+...+\kappa^n)d$ on $\Om^n\Ac$, both bounded, verifying $B^2=b^2=bB+Bb=0$ and $B\kappa=\kappa B=B$. Thus $(\Om \Ac,b,B)$ becomes a complete bornological bicomplex.\\

We now can use the bar construction for $\Act$ in order to get cochains on the bicomplex $\Om \Ac$. First consider the bounded injection $\nat: \Om^n\Ac\to(\Om_1\Bb(\Act))_{n+1}$
\be
\nat(\at_0 da_1...da_n)= \sum_{i=0}^n (-)^{n(i+1)} (a_{i+1},...,a_n|\at_0|a_1,...,a_i)\ .\label{135}
\ee
Then by direct computation one checks that $\nat b=b''\nat$. Let $\Lc$ be a unital complete bornological $\zz_2$-graded algebra, and consider the associated algebra and bimodule $\Rc=\hom(\Bb(\Act),\Lc)$ and $\Mc=\hom(\Om_1\Bb(\Act),\Lc)$. Then composing $\nat$ with an element $\gamma$ of $\Mc$ we get a bounded cochain $\gamma\nat \in \hom(\Om \Ac, \Lc)$. The following lemma relates the Hochschild operator $b$ on $\Om \Ac$ with the differential $\delta$ on $\Mc$.
\begin{lemma} \label{A1}
For any $\gamma\in \Mc$ one has $\delta\gamma \nat = -(-)^{|\gamma|}\gamma\nat b$ in $\hom(\Om \Ac,\Lc)$.
\end{lemma}
{\it Proof:} By direct computation one checks that $\nat b=b''\nat$, and then immediately $\delta\gamma\nat =-(-)^{|\gamma|}\gamma b''\nat=-(-)^{|\gamma|}\gamma\nat b$. \hfill\rule{1ex}{1ex}\\

It remains to relate the operator $B$ to the derivation $\partial : \Rc\to \Mc$. For this we have to consider a bounded linear map $\rho:\Act\to \Lc^0$ with values in the even part of $\Lc$, and preserving the unit: $\rho(1)=1$. We view it as an element of $\hom(\Bb_1(\Act),\Lc)\subset \Rc$ of degree $|\rho|=1$. Then the following lemma holds:
\begin{lemma}\label{A2}
Let $\rho:\Act\to \Lc^0$ be a bounded unital linear map (not necessarily an homomorphism), and let $f,g$ be two elements of $\Rc$ vanishing if one of their arguments is equal to $1\in\Act$. Then one has
\be
\partial(fg)\nat = (-)^{|g|} f\,\partial\rho\,g\nat B
\ee
in $\hom(\Om \Ac,\Lc)$.
\end{lemma}
{\it Proof:} We may suppose $f\in\hom(\Bb_p(\Act),\Lc)$ and $g\in\hom(\Bb_q(\Act),\Lc)$ with $p+q=n+1$. One has
$$
f\partial\rho g\nat B(\at_0da_1...da_n)=\sum_{i=0}^n (-)^{n(i+1)} f\partial\rho g\nat(da_{i+1}...da_nda_0...da_i)\ ,
$$
with $\at_0\in\Act$ and $a_0$ its projection on $\Ac$. We compute
\beq
f\partial\rho g\nat(da_0...da_n)&=&\sum_{i=0}^{n+1}(-)^{(n+1)(i+1)}(f\partial\rho g)(a_i,...,a_n|1|a_0,...,a_{i-1})\non\\
&=& (-)^{(n+1)(q+1)}(f\partial\rho g)(a_q,...,a_n|1|a_0,...,a_{q-1})\non\\
&=& (-)^{(n+1)(q+1)}(-)^{|g|+p}(fg)(a_q,...,a_n,a_0,...,a_{q-1})\non\\
&=& (-)^{|g|+nq}(fg)(a_q,...,a_n,a_0,...,a_{q-1})\non
\eeq
where we retained only the term corresponding to $i=q$ and used the fact that $\rho(1)=1$. Similarly for any $0\le i\le n$ one has
$$
f\partial\rho g\nat(da_{i+1}...da_nda_0...da_i)=(-)^{|g|+nq}(fg)(a_{i+q+1},...,a_i,a_{i+1},...,a_{i+q})\ ,
$$
where the indices of the $a$'s are defined modulo $n+1$. Thus
\beq
f\partial\rho g\nat B(\at_0da_1...da_n)&=& \sum_{i=0}^n (-)^{n(i+1+q)+|g|}(fg)  (a_{i+q+1},...,a_{i+q})\non\\
&=& (-)^{|g|}\sum_{i=0}^n(-)^{n(i+1)}(fg)(a_{i+1},...,a_n,a_0,...,a_i)\non
\eeq
by reindexing $i+q\to i$. On the other hand 
\beq
\partial(fg)\nat (\at_0da_1...da_n)&=& \sum_{i=0}^n(-)^{n(i+1)}\partial(fg)(a_{i+1},...,a_n|\at_0|a_1,...,a_i)\non\\
&=& \sum_{i=0}^n(-)^{n(i+1)} (fg)(a_{i+1},...,a_n,\at_0,a_1,...,a_i)\non\\
&=& \sum_{i=0}^n(-)^{n(i+1)} (fg)(a_{i+1},...,a_n,a_0,a_1,...,a_i)\non
\eeq
since $f,g$ are supposed to vanish on $1$, and the conclusion follows. \hfill\rule{1ex}{1ex}

\subsection{Traces}

Let $\Lc$, $\Ac$ be as above. Let $\Vc$ be a complete bornological vector space and $\tau: \Lc\to \Vc$ a bounded trace, i.e. a bounded linear map vanishing on the graded commutators $[\Lc,\Lc]$. Another way to specify this is to consider the permutation map $\si: \Lc\hotimes \Lc\to \Lc\hotimes \Lc$ which flips the two factors:
\be
\si(x\otimes y )=(-)^{|x||y|}y\otimes x
\ee
according to their respective degrees. Then $\tau$ is a trace if and only if $\tau m \si=\tau m$, where $m: \Lc\hotimes \Lc\to \Lc$ is the multiplication map. An essential example is the universal trace
\be
\nat: \Lc\to \Lc_{\nat}=(\Lc/[\Lc,\Lc])_{\mathrm{completed}}\ .
\ee
Its universal property stems from the fact that any trace $\tau$ factors through $\nat$. \\

At the dual level, the injection $\nat: \Om \Ac\to \Om_1\Bb(\Act)$ is a bounded cotrace. Indeed if we introduce the map $\si: \Om_1\Bb(\Act)\hotimes \Bb(\Act)\rightleftarrows \Bb(\Act)\hotimes \Om_1\Bb(\Act)$ which permutes the two factors (with signs), then one has $\Delta_l\nat = \si \Delta_r\nat$ and $\si\Delta_l\nat =\Delta_r\nat$.\\

We now put traces and cotraces together. For any bounded trace $\tau$ on $\Lc$, the map from $\Mc=\hom(\Om_1\Bb(\Act),\Lc)$ to $\hom(\Om \Ac, \Vc)$ which sends $\gamma$ to $\tau \gamma\nat$ is a trace on the $\Rc$-bimodule $\Mc$, that is, it vanishes on the graded commutators $[\Rc,\Mc]$. Indeed for any $f\in \Rc$ and $\gamma\in \Mc$, one has $\tau (\gamma f)\nat = \tau m (\gamma\otimes f)\Delta_r\nat = (-)^{|\gamma||f|}\tau m\si (f\otimes\gamma)\si\Delta_r\nat = (-)^{|\gamma||f|}\tau m(f\otimes\gamma)\Delta_l\nat = (-)^{|\gamma||f|}\tau (f\gamma)\nat$.

\end{document}